\documentclass{article}
\usepackage[utf8]{inputenc}

\usepackage[
backend=bibtex,
url=false,isbn=false,doi=false,
date=year,
hyperref=auto,
style=alphabetic,
natbib=true,
sorting=nty,
firstinits=true,
maxnames=2, maxbibnames=10,
]{biblatex}

\bibliography{references.bib} 

\usepackage{hyperref}
\hypersetup{
    bookmarks=true,         
    unicode=false,          
    pdftoolbar=true,        
    pdfmenubar=true,        
    pdffitwindow=false,     
    pdfstartview={FitH},    
    pdfauthor={Reinhard Heckel},     
    pdfsubject={Subject},   
    pdfcreator={Reinhard Heckel},   
    pdfproducer={Producer}, 
    pdfnewwindow=true,      
    colorlinks=true,       
    linkcolor=red,          
    citecolor=green,        
    filecolor=magenta,      
    urlcolor=cyan           
}

\usepackage[utf8]{inputenc} 
\usepackage[T1]{fontenc}    
\usepackage{hyperref}       
\usepackage{url}            
\usepackage{booktabs}       
\usepackage{amsfonts}       
\usepackage{nicefrac}       
\usepackage{microtype}      
\usepackage[ruled,vlined]{algorithm2e}
\usepackage{rays_defs_18}

\usepackage{graphicx}
\usepackage{epstopdf}
\usepackage{caption}
\usepackage{subcaption}
\usepackage{amssymb}
\usepackage{bbm}
\usepackage[normalem]{ulem}
\usepackage{enumitem}
\usepackage{float}

\usepackage{amsmath}
\usepackage{blkarray}
\usepackage{tikz,pgfplots}
\usetikzlibrary{pgfplots.statistics}
\usetikzlibrary{calc}
\usepackage{pgfplotstable}
\usepgfplotslibrary{colormaps}
\usetikzlibrary{quotes,angles}
\usepackage{textpos}
\usepackage{pgfplotstable}
\usepackage{array}
\usepackage{authblk}
\usepackage{multirow}
\usepgfplotslibrary{fillbetween}
\usepackage[export]{adjustbox}
\usetikzlibrary{matrix,shapes,arrows,positioning}

\usepackage{threeparttable}
\usepackage{comment}
\usepackage{booktabs}
\usepackage{colortbl} 
\usepackage{xcolor} 
\usepackage{xfrac}
\usepackage{caption} 
\pgfplotsset{compat=1.14}
\usepgfplotslibrary{groupplots}
\usepgfplotslibrary{dateplot}
\usepackage[a4paper, total={6in, 10in}]{geometry}

\DeclareMathAlphabet{\mathcal}{OMS}{cmsy}{m}{n}
\SetMathAlphabet{\mathcal}{bold}{OMS}{cmsy}{b}{n}
\providecommand{\abs}[1]{\lvert#1\rvert}

\usetikzlibrary{backgrounds,
    calc,
    fadings,
    shadows,
    shapes.arrows,
    shapes.symbols
}

\usepackage{xparse}
\edef\RestoreEndlinechar{%
    \endlinechar=\the\endlinechar\relax
}
\usepackage{smartdiagram}
\RestoreEndlinechar

\tikzset{bubble node/.append style={
        draw=none, opacity=0.5
    }
}

\usepackage{adjustbox}
\usepackage{booktabs}
\usepackage{relsize,stackengine}

\usepackage{ragged2e,tabularx}
\newcolumntype{L}[1]{>{\raggedright\let\newline\\\arraybackslash\hspace{0pt}}p{#1}}

\definecolor{steelblue}{rgb}{0.27, 0.51, 0.71}
\definecolor{amber}{rgb}{1.0, 0.49, 0.0}
\definecolor{caribbeangreen}{rgb}{0.0, 0.8, 0.6}
\definecolor{ash}{rgb}{0.7, 0.75, 0.71}
\definecolor{brightpink}{rgb}{1.0, 0.0, 0.5}
\definecolor{fgreen}{rgb}{0.13, 0.55, 0.13}

\newcommand{\steelb}[1]{\textcolor{steelblue!70!black}{#1}}

\usepackage{bm}
\newcommand\vtheta{{\bm \theta}}

\long\def\comment#1{}

\begin{document}


\begin{center}

{\bf{\LARGE{
Measuring Robustness in Deep Learning Based Compressive Sensing
}}}

\vspace*{.2in}

{\large{
\begin{tabular}{cccc}
Mohammad Zalbagi Darestani$^{\ast}$, 
Akshay S. Chaudhari$^{\ddagger}$,
and 
Reinhard Heckel$^{\dagger,\ast}$
\end{tabular}
}}

\vspace*{.05in}

\begin{tabular}{c}
$^\ast$Dept. of Electrical and Computer Engineering, Rice University\\
$^{\ddagger}$Dept. of Radiology and Dept. of Biomedical Data Science, Stanford University \\
$^\dagger$Dept. of Electrical and Computer Engineering, Technical University of Munich
\end{tabular}

\vspace*{.1in}

\today

\vspace*{.1in}

\end{center}

\begin{abstract}
Deep neural networks give state-of-the-art accuracy for reconstructing images from few and noisy measurements, a problem arising for example in accelerated magnetic resonance imaging (MRI).
However, recent works have raised concerns that deep-learning-based image reconstruction methods are sensitive to perturbations and are less robust than traditional methods:
Neural networks (i) may be sensitive to small, yet adversarially-selected perturbations, (ii) may perform poorly under distribution shifts, and (iii) may fail to recover small but important features in an image.
In order to understand the sensitivity to such perturbations,
in this work, we measure the robustness of different approaches for image reconstruction including trained and un-trained neural networks as well as traditional sparsity-based methods. 
We find, contrary to prior works, that both trained and un-trained methods are vulnerable to adversarial perturbations. Moreover, both trained and un-trained methods tuned for a particular dataset suffer very similarly from distribution shifts. Finally, we demonstrate that an image reconstruction method that achieves higher reconstruction quality, also performs better in terms of accurately recovering fine details. Our results indicate that the state-of-the-art deep-learning-based image reconstruction methods provide improved performance than traditional methods without compromising robustness.
\end{abstract}

\section{Introduction}\label{sec:intro}

Neural networks outperform traditional (e.g., sparsity-based) methods in a variety of image reconstruction tasks across common metrics of image quality. For instance, consider the compressive sensing problem arising in magnetic resonance imaging (MRI), where the goal is to reconstruct a diagnostic quality image using linear under-sampled measurements for accelerating the MRI scans. 
A large body of literature shows that neural networks can enable higher reconstruction quality and faster reconstruction computations for MRI compared to clinically utilized traditional sparsity-based reconstruction methods~\cite{hammernik2018learning,zbontar2018fastmri,knoll2020advancing,sriram2020end,wang2019pyramid,schlemper2017deep,putzky2019invert,arora2020untrained,darestani2020can}. 

However, recent works have raised robustness concerns about neural-network-based image recovery~\citep{cohen2018distribution,huang2018some,antun2020instabilities,gottschling2020troublesome,genzel2020rob}.
Specifically, 
\citet{antun2020instabilities} 
demonstrated that small, adversarially-selected perturbations in the under-sampled measurements may result in significant reconstruction artifacts. 
Based on those findings, \citet{antun2020instabilities} and \citet{gottschling2020troublesome} concluded that despite providing worse reconstruction quality than neural networks, traditional sparsity-based compressive sensing methods tend to be more robust to adversarial perturbations. 

Moreover, the results from the fastMRI challenge, a competition for improving the performance of image reconstruction systems for accelerated MRI, have shown that while all top performing deep networks yield high scores according to various image quality metrics, there is also a potential to miss small, clinically relevant pathologies~\citep[Fig.~3]{knoll2020advancing}, which may lead to false-negative diagnoses. 

These findings, however, do not address the question whether the identified robustness concerns are specific to training a network for image reconstruction, or whether un-trained neural networks and traditional un-trained sparsity-based reconstruction methods are similarly sensitive to perturbations. 

In addition, it is unknown whether neural networks for image reconstruction are sensitive to various distribution shifts, which can commonly arise in clinical practice.
Natural distribution shifts have shown to degrade performance in learning-based systems, for example for classification problems~\citep{recht2019imagenet,taori2020measuring,miller_linearfits_2021,Yadav_Bottou_2019} and for question answering models in natural language processing~\cite{Miller_Krauth_Recht_Schmidt_2020}).


\tikzstyle{block} = 
    [
        rectangle
      , draw
      , text width=12em
      , text centered
      , rounded corners
      , minimum height=2em
      ]

\tikzstyle{block2} = 
    [
        rectangle
      , draw
      , text width=6.5em
      , text centered
      , rounded corners
      , minimum height=2em
      ]

\tikzstyle{line} = 
    [
        draw
     , -latex'
     ]
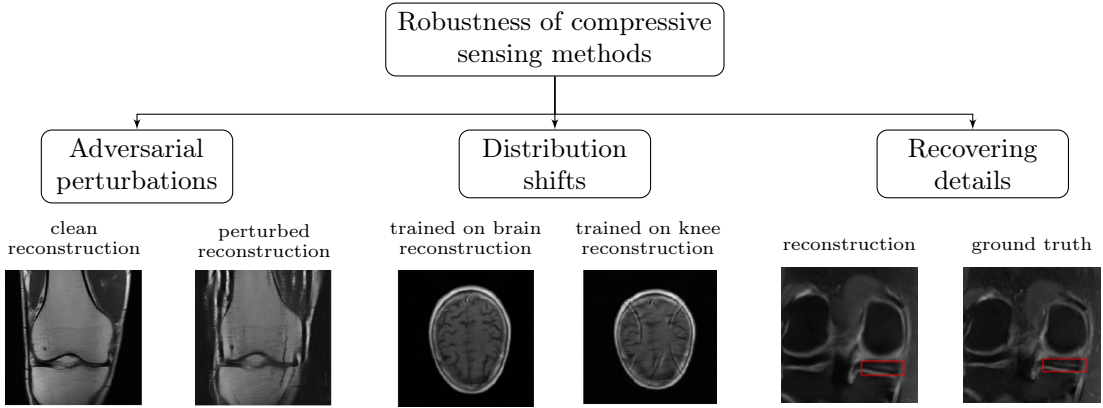
\begin{figure*}
\centering
\begin{tikzpicture}       
  \matrix (mat) [matrix of nodes, nodes=block, column sep=10mm, row sep=20pt] 
{
& & &\node [block] (main) {Robustness of compressive \\ sensing methods};  & & & \\ 
&\node [block2] (adv) {Adversarial \\ perturbations};   & & \node [block2] (details)  {Distribution \\ shifts}; & & \node [block2] (dist)  {Recovering \\ details}; &     \\
&&
&&
&&
\\ 
};  
\path[line] (main.south)   -- +(0,-0.5) -| node [pos=0.3, above] {} (adv.north);
\path[line] (main.south)   -- +(0,-0.5) -| node [pos=0.3, above] {} (details.north);
\path[line] (main.south)   -- +(0,-0.5) -| node [pos=0.3, above] {} (dist.north);
\node[above left,label={[label distance=-0.75pt, align=center]90:\scriptsize clean \\[-1ex] \scriptsize reconstruction }] (adv.south) at +(-5.3,-3.8) {\includegraphics[width=1.8cm,height=1.8cm]{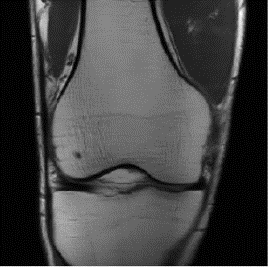}};
\node[above left,label={[label distance=-1.75pt, align=center]90:\scriptsize perturbed \\[-1ex] \scriptsize reconstruction}] (adv.south) at +(-2.8,-3.8) {\includegraphics[width=1.8cm,height=1.8cm]{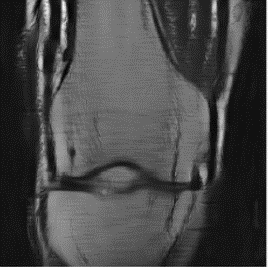}};
\node[above left,label={[label distance=-0.75pt, align=center]90:\scriptsize trained on brain\\[-1ex] \scriptsize reconstruction}] (adv.south) at +(-0.15,-3.8) {\includegraphics[width=1.8cm,height=1.8cm]{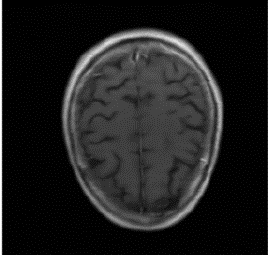}};
\node[above left,label={[label distance=-0.75pt, align=center]90:\scriptsize trained on knee\\[-1ex] \scriptsize reconstruction}] (adv.south) at +(2.25,-3.8) {\includegraphics[width=1.8cm,height=1.8cm]{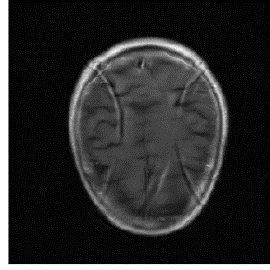}};
\node[above left,label={[label distance=-0.75pt, align=center]90:\scriptsize reconstruction}] (adv.south) at +(4.9,-3.75) {\includegraphics[width=1.8cm,height=1.8cm,trim={1cm 1cm 1cm 1cm},clip]{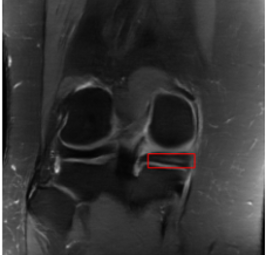}};
\node[above left,label={[label distance=-1.75pt, align=center]90:\scriptsize ground truth}] (adv.south) at +(7.3,-3.75) {\includegraphics[width=1.8cm,height=1.8cm,trim={1cm 1cm 1cm 1cm},clip]{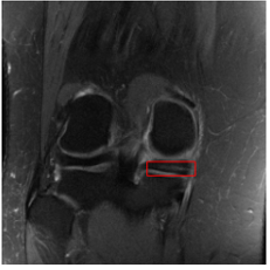}};
\end{tikzpicture}
\caption{Robustness notions considered in this work. The images are U-net reconstructions from 4x accelerated data. 
As adversarial perturbation, we select an additive error that has small $\ell_2$-norm but induces a large reconstruction error.}
\label{fig:robustness}
\end{figure*}

In this work, we study the robustness of three popular families of compressive sensing reconstruction methods for MRI:
\begin{enumerate}
    \item[i)] {\bf Neural networks trained end to end.} All leading methods in terms of image reconstruction quality of the 2019 and 2020 fastMRI competitions belong to this class of methods. We consider U-net-based recovery~\citep{ronneberger2015u}, as it is a simple baseline, and the end-to-end variational network \citep{sriram2020end}, which is a state-of-the-art method on the fastMRI dataset~\citep{zbontar2018fastmri}.
    \item[ii)] {\bf Traditional CS methods.} We consider imposing a sparsity prior and image reconstruction with $\ell_1$-norm minimization, which is the leading classical compressive sensing method~\cite{lustig2007sparse}. 
    \item[iii)] {\bf Un-trained neural networks.} We consider a variation of the Deep Image Prior~\citep{ulyanov2018deep} and Deep Decoder~\citep{heckel2019deep} for MRI introduced in~\citep{darestani2020can}, which is a neural network based method, but works without training data.
\end{enumerate}

We study the aforementioned methods with respect to the following three notions of robustness (see~Figure~\ref{fig:robustness}): 
\begin{enumerate}
\item[i)] {\bf Small adversarial perturbations.} 
Small perturbations are selected adversarially to cause a large reconstruction error for a given image and an image reconstruction method. \citet{antun2020instabilities} studied the effect of such perturbations, but only selected them for neural networks trained end-to-end.
However, for a comparison among methods, it is essential to find these perturbations separately for each method. 
While there is no explicit adversary in an MRI scanner (as an MRI scanner is a closed system), studying adversarial perturbations is important since measurement perturbations can be due to scanner-induced artifacts (coil failure, B0 magnet drift, bad gradient shimming, etc.) or patient-induced artifacts (motion, off-resonance, etc.)~\citep{bellon1986mr}.
\item[ii)] {\bf Robustness to distribution shifts.} Neural networks
are typically evaluated by
first collecting a set of images and measurements, 
second partitioning the data into training and test sets, 
and third training and evaluating the network on the train and test data. 
Thus, both train and test data are drawn from the same distribution. However, in practice train and test distributions may vary: 
for example, a network can be  trained on data from one set of patients, anatomies, and contrasts, but may be used for varying datasets at test time. 
For classification, it is well known that distribution shifts are common in practice and have a large impact on performance~\citep{recht2019imagenet}, but for image reconstruction methods the effect of distribution shifts is unknown.
\item[iii)]{\bf Robustness in recovering fine details.} Small features in an MRI image (e.g., 10-pixel sized features or even smaller) are often important for accurate diagnostics. Therefore, it is crucial for a reconstruction method to be able to recover such features in the image.
\end{enumerate}

\begin{figure}[t]
\begin{center}
\begin{tikzpicture}
\begin{groupplot}[
y tick label style={/pgf/number format/.cd,fixed,precision=4},
scaled y ticks = false, xticklabel style={
        /pgf/number format/fixed,
        /pgf/number format/precision=2
},
legend style={at={(1.75,1)} , nodes={scale=0.75}, draw={none}, fill = none, text opacity=1,
/tikz/every even column/.append style={column sep=-0.1cm}
 },
         group
         style={group size= 1 by 1, xlabels at=edge bottom, ylabels at=edge left,
         xticklabels at=edge bottom,
         horizontal sep=1.5cm, vertical sep=0.6cm,
         }, 
         width=0.36\textwidth,height=0.24\textwidth,
         ylabel={\footnotesize SSIM on Stanford},
         xlabel={\footnotesize SSIM on fastMRI},
         scaled x ticks=false,
         ymax = 0.88,
         ymin = 0.35,
         xmin = 0.58,
         xmax = 0.82,
         legend cell align=left,
         ]
\nextgroupplot[no markers]
    \addplot +[mark=none,steelblue,thick] table[x=x,y=linfit]{./fastmri_v2/dataset_shift.txt};
	\addplot +[mark=none,dashed,red!70,thick] table[x=x,y=x]{./fastmri_v2/dataset_shift.txt};
	\addplot +[scatter, only marks, steelblue!60, scatter src = explicit symbolic,mark size=1pt,scatter/classes={
            a={mark=*,mark options={solid,draw=none,fill=fgreen}, fgreen, mark size = 1.8pt},
            b={mark=*,mark options={solid,draw=none,fill=blue},blue,mark size = 1.8pt},
            c={mark=*,mark options={solid,draw=none,fill=red},red,mark size = 1.8pt},
            d={mark=*,mark options={solid,draw=none,fill=black},black,mark size = 1.8pt}
        },
              error bars/.cd, 
              error bar style={line width=0.6pt},
              y dir=both, x dir=both,
              x explicit, y explicit
              ]
       table [x=x, y=y, x error=xerr, y error=yerr, meta=class]{./fastmri_v2/dataset_shift.txt};
    \legend{best linear fit,$y=x$,$\ell_1$ group,un-trained group,U-net group,VarNet group}
\end{groupplot}
\end{tikzpicture}
\caption{\textbf{Reconstruction methods trained on one knee dataset (fastMRI) do not necessarily generalize to another knee dataset (Stanford dataset), if there are differences in how the datasets are obtained.}
All of the considered reconstruction methods lose a similar amount of accuracy when evaluated on the Stanford set (through a dataset shift). The word ``group" in the legend refers to variants of the respective method and error bars are $95\%$ confidence intervals.}
\label{fig:fastmriv2}
\end{center}
\end{figure}
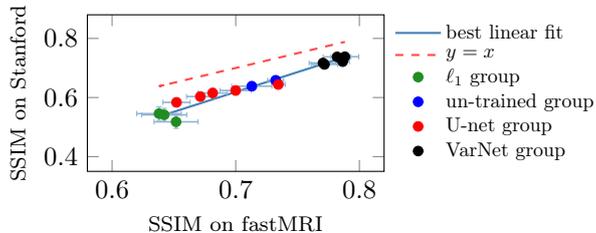

Understanding the robustness of algorithms with respect to these perturbations is important, especially in medical imaging where errors may result in a faulty diagnosis. Towards improving our understanding in this realm, the main contributions of our work and findings of our study are:
\begin{itemize}
    \item All studied methods---trained and un-trained---are sensitive to small, adversarially-selected perturbations, and the performance loss is not unique to neural networks trained end-to-end. 
    This result is in contrast to that of~\citet{antun2020instabilities}, which found that ``deep learning typically yields unstable methods,'' whereas $\ell_1$-minimization methods ``are not affected by the perturbation.''
    Those two results can co-exist because \citet{antun2020instabilities} found that an adversarial perturbation \emph{selected for a particular neural network} does not affect $\ell_1$-minimization much. In this work (in contrast to~\citet{antun2020instabilities}), we also compute adversarial perturbations for $\ell_1$-minimization and show that while those significantly impair the performance of $\ell_1$-minimization, they affect neural networks less. 
    \item We provide the first study of distribution shifts in the context of image reconstruction. 
    Even un-trained methods such as $\ell_1$-minimization are affected by distribution shifts, because their hyper-parameters are tuned on a given distribution.
    We study three notions of \emph{natural} distribution shifts for accelerated MRI reconstruction: (i) slight shift to a close domain (i.e., same anatomy but a dataset with a different acquisition technique), (ii) anatomy shift (i.e., training on brains and testing on knees), and (iii) adversarially-filtered shifts. 
    Perhaps surprisingly, we find that both un-trained and trained methods are similarly affected by all three types of distribution shifts (see Figure~\ref{fig:fastmriv2} as an example for case (i)), and typically the best performing method is also the best performing method under a distribution shift.
    For adversarially-filtered shifts, we find that challenging-to-reconstruct-images are difficult to reconstruct for all methods, from which we conclude that some images are naturally difficult to reconstruct. 
    
    \item Finally, we quantify the recovery of small features in the images i) through introducing artificial small features, and ii) through studying small, clinically relevant features in real data. From studying artificial small feature recovery, 
    we find that each reconstruction method is sensitive to specific regions in an image and faces difficulty in recovering small features in those regions.
    By studying small pathologic features in real data, we find that traditional CS methods are less robust in recovering small details compared to neural networks. This confirms the intuition that small feature recovery ability should correlate with overall reconstruction performance.
\end{itemize}

In a nutshell, the take-away of our study is that the deep learning methods that perform best based on reconstruction quality are also best under realistic distribution shifts and for small feature recovery, and we could not find them to be more sensitive to adversarial perturbations than classical methods.

\section{Problem setup: Accelerated multi-coil MRI}\label{sec:prob_statement}

We study robustness in the context of accelerated multi-coil MRI reconstruction, because this is one of the most popular applications of compressive sensing and an important medical imaging technology. 

In multi-coil MRI, $n_c$ multiple radiofrequency coils each record a measurement that is sensitive to a spatially local anatomical region. 
We consider multi-coil MRI over simpler single-coil MRI experiments since using multiple coils is common in clinical practice. 
The goal of accelerated multi-coil MRI reconstruction is to reconstruct an image $\vx^* \in \mathbb{C}^N$ from a set of measurements (often called $k$-space measurements) obtained as
\[
\vy_i = \mM \mF \mS_i \vx^* + \text{noise} \in \mathbb{C}^M, 
\quad i=1,\ldots, n_c.
\]
In the above equation, $\mS_i$ is a complex-valued position-dependent coil sensitivity map, that is applied through element-wise multiplication to the image $\vx^*$, $\mF$ implements the 2D discrete Fourier transform, and $\mM$ is a mask that implements under-sampling of $k$-space data.

For all experiments, we use the fastMRI dataset~\citep{zbontar2018fastmri}, designed for training and evaluating deep-learning-based MRI reconstruction methods. The fastMRI dataset consists of fully-sampled measurements (i.e., taken with an identity mask $\mM = \mI$) of knees taken with $n_c = 15$ coils, and of brains taken with a varying number of coils.
The dataset also contains reference images that are obtained by reconstructing the coil images from each full coil measurement as $\hat \vx_i = \inv{\mF} \vy_i$ and then combining them via the root-sum-of-squares (RSS) algorithm to a final single image:
$
    \hat{\vx} = \sqrt{\sum_{i=1}^{n_c} \abs{\hat{\vx}_i}^2}.
$
Here, $\abs{\cdot}$ and $\sqrt{\cdot}$ denote element-wise absolute value and squared root operations. Alternatively, one could use SENSE~\citep{roemer1990nmr}, a signal-to-noise ratio optimal method, to combine coil images.

To under-sample $k$-space measurements (for acceleration), we employ a standard 1D mask (random or equi-spaced vertical or horizontal lines in the frequency domain), which is the default in the fastMRI challenge. We consider 4x acceleration throughout the paper as this is one of the acceleration factors considered in the fastMRI challenge, although 2x acceleration is often used in clinical practice~\citep{deshmane2012parallel}. For evaluation, we compare to the reference images reconstructed from the full $k$-space.

\section{Image reconstruction methods}\label{sec:methodology}

We consider three families of image reconstruction methods: i) trained neural networks, ii) traditional sparsity-based CS methods, and iii) un-trained neural networks. 
In this section, we provide a brief overview of each method. 

\paragraph{Trained neural networks.} 
Convolutional neural networks are trained either to map the measurement directly to an artifact-free image, or to map from a coarse least-squares reconstruction from the under-sampled measurement to an artifact-free image. 

Let $\{(\vx_1,\vy_1),\ldots,(\vx_n,\vy_n)\}$ be a training set consisting of pairs of target image $\vx_j$ and a under-sampled measurement $\vy_j$. The measurement in our setup consists of the physical $k$-space measurements from multiple receiver coils. A network $f_\vtheta$ with parameters (weights) $\vtheta$ takes as input the measurement and generates an image. The network is typically trained by minimizing the loss
\begin{align*}
\mc L(\vtheta) = \frac{1}{n} \sum_{j=1}^n \norm[2]{ \vx_j - f(\vy_j) }^2,
\end{align*}
which yields the trained method $f_{\hat \vtheta}$. At test time, the network generates an image as $f_{\hat \vtheta}(\vy)$ based on the measurement $\vy$.

The best performing methods in the fastMRI competition are all trained networks, and yield significant improvements over classical methods~\citep{putzky2019invert,sriram2019grappanet,sriram2020end}.
We consider two methods of this type: U-net~\citep{ronneberger2015u} based reconstruction~\cite{jin2017deep},
as this is a simple baseline, and the end-to-end variational network (VarNet)~\citep{sriram2020end}, as this is the current state-of-the-art network.

U-net based reconstruction simply trains a U-net end-to-end on the training set to map the under-sampled measurement to the original image. 
VarNet-based reconstruction~\citep{sriram2020end} is more intricate, with a more complicated network including  coil sensitivity estimation, image-domain refinement, and data consistency steps, but conceptually similar as it is also trained end-to-end.

\paragraph{Traditional compressive sensing methods.} 
Traditional compressive sensing methods are sparsity based and either impose a sparse representation by minimizing the $\ell_1$-norm, or perform total-variation norm minimization. 
Traditional CS methods are popular for MRI reconstruction~\citep{chen2012compressive,block2007undersampled,lustig2007sparse}, and are used commonly in clinical practice. 
$\ell_1$-norm minimization relies on assuming sparsity in a transform domain. We impose Wavelet sparsity following common practice in MRI~\citep{chen2012compressive,lustig2007sparse}. 
Sparsity-based reconstruction recovers an image by minimizing the (convex) loss
\begin{equation}
    \mathcal{L}_1(\vx) = \sum_{i=1}^{n_c} \norm[2]{\vy_i - \mA \hat \mS \vx}^2 + \lambda \norm[1]{\mH \vx}.
    \label{eq:l1-loss}
\end{equation}
Here, $\mH$ denotes the 2D Wavelet transform and $\hat \mS$ are coil sensitivity maps estimated from the under-sampled measurement using the ESPIRiT method~\citep{uecker2014espirit}.
Sparsity-based approaches provably succeed provided that the signal is sparse~\citep{candes2006robust,lustig2007sparse}.

\paragraph{Un-trained neural networks.}
Perhaps surprisingly, convolutional neural networks can regularize inverse problems without training, as has first been demonstrated by the Deep Image Prior~\citep{ulyanov2018deep} for denoising, super-resolution, and inpainting problems. 

Such un-trained networks are also powerful for compressive sensing~\citep{van2018compressed}, 
and simple convolutional architectures such as the Deep Decoder~\citep{heckel2019deep} work well in practice. 
\citet{arora2020untrained} and \citet{darestani2020can} used variants of the deep decoder for multi-coil MRI and achieved noticeable improvements over traditional CS methods. \citet{darestani2020can} further demonstrated that un-trained networks even perform similar to the U-net---the trained approach mentioned earlier.
Un-trained neural networks provably recover smooth signals from few measurements~\cite{heckel2020compressive}.

In a nutshell, an un-trained network recovers an image by first fitting a randomly initialized, image generating network to a measurement, and then taking the network output as the recovered image. The network is not trained, and uses the structure of the network alone as a prior for the images. 

We finally note that there is a fourth class of neural network-based reconstruction methods, pioneered by~\cite{bora2017compressed}, which impose a learned prior~\cite{bora2017compressed,hand2018global,asim2020invertible,Daras_intermediate_2021}. 
At the time of writing, this class of neural network-based reconstruction methods has not been applied to MRI, and was therefore not included in our study. In the meantime, \citet{Kelkar_Bhadra_Anastasio_2020} applied this method to MRI images, and it would be interesting to include this class of methods into further robustness studies. 


\section{Small, adversarially-selected perturbations} \label{sec:adversarial-perturbation}

Adversarial perturbations are often studied to evaluate the robustness of machine learning systems. 
For classification problems, there exists a large body of literature on the sensitivity of neural networks to adversarial perturbations~\citep{goodfellow2014explaining, bruna2013intriguing,moosavi2017universal}. 
These works show that the predicted label for a test image can be altered by adding a small perturbation to that image. This observation is attributed to test images lying close to a learned decision boundary, and thus slightly altering the image allows to cross this boundary.

The study of adversarial robustness in classification is motivated by the concept of an adversary that may alter the input of a machine learning system with imperceptible perturbations. In image reconstruction problems, there is typically no such adversary, but studying adversarial robustness in image reconstruction methods is important as it enables certifying worst-case robustness.

The study of small perturbations for image reconstruction problems was initiated by~\citet{cohen2018distribution,huang2018some,antun2020instabilities}. \citet{antun2020instabilities} generated these perturbations for a given network and image through optimization and concluded that ``deep learning for inverse problems is typically unstable'' because of the impact of adversarial perturbations. 
In a recent subsequent work, \citet{genzel2020rob} unrolled a classical method, total variation minimization (TV), and generated small perturbations for this unrolled network in the same way as \citet{antun2020instabilities} did, and found ``superior robustness of the learned reconstruction schemes [a U-net] over TV.''
Thus there is a disagreement on whether neural network or classical methods are more sensitive to adversarial perturbations. 

In this section we study small adversarial perturbations, but in contrast to those previous works, (i) we include un-trained neural networks in our study,
(ii) we include the state-of-the-art method VarNet beyond the baseline trained method U-net, 
and 
(iii) we introduce a method to generate small adversarial perturbations for un-trained methods without unrolling and thus without treating un-trained methods as a neural network. 

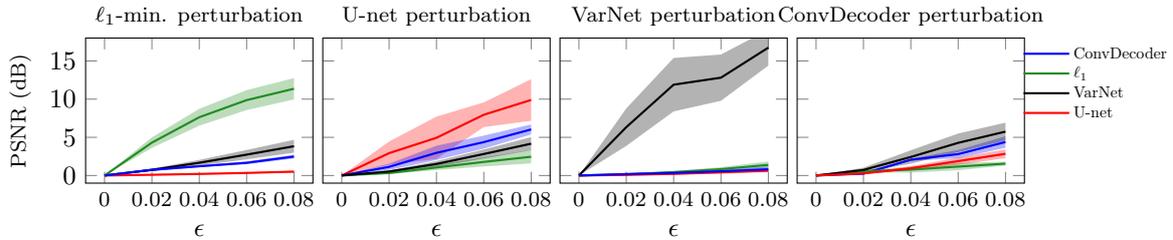
\begin{figure*}[t!]
\begin{center}
\begin{tikzpicture}

\begin{groupplot}[
y tick label style={/pgf/number format/.cd,fixed,precision=4}, every x tick label/.append style={font=\footnotesize},
scaled y ticks = false, xticklabel style={
        /pgf/number format/fixed,
        /pgf/number format/precision=2
},
legend style={at={(1.68,1)} , nodes={scale=0.6}, draw={none}, fill = none, text opacity=1,
/tikz/every even column/.append style={column sep=-0.1cm}
 },
         group
         style={group size= 4 by 1, xlabels at=edge bottom, ylabels at=edge left,
         yticklabels at=edge left,
         xticklabels at=edge bottom,
         horizontal sep=0.13cm, vertical sep=0.6cm,
         }, 
         width=0.3\textwidth,height=0.23\textwidth,
         ylabel={\footnotesize PSNR (dB)},
         xlabel = {$\epsilon$},
         scaled x ticks=false,
         ymax=18,
         ymin=-1,
         legend cell align=left,
         ]
\nextgroupplot[title={\footnotesize $\ell_1$-min. perturbation}, title style={yshift=-5pt},]
	\addplot +[mark=none,fgreen,thick] table[x expr=\thisrowno{0}/100,y index=1]{./files/adv_plot.csv};
	
	\addplot +[mark=none,black,thick] table[x expr=\thisrowno{0}/100,y index=4]{./files/adv_plot.csv};
	
	\addplot +[mark=none,red,thick] table[x expr=\thisrowno{0}/100,y index=7]{./files/adv_plot.csv};
	
	\addplot +[mark=none,blue,thick] table[x expr=\thisrowno{0}/100,y index=10]{./files/adv_plot.csv};
	
	\addplot +[name path=upper,draw=none, mark=none] table[x expr=\thisrowno{0}/100,y index = 2] {./files/adv_plot.csv};
    \addplot +[name path=lower,draw=none,mark=none] table[x expr=\thisrowno{0}/100,y index = 3] {./files/adv_plot.csv};
    \addplot +[fill=fgreen,opacity=0.3] fill between[of=upper and lower];
    
    \addplot +[name path=upper,draw=none, mark=none] table[x expr=\thisrowno{0}/100,y index = 5] {./files/adv_plot.csv};
    \addplot +[name path=lower,draw=none,mark=none] table[x expr=\thisrowno{0}/100,y index = 6] {./files/adv_plot.csv};
    \addplot +[fill=black,opacity=0.3] fill between[of=upper and lower];
    
    \addplot +[name path=upper,draw=none, mark=none] table[x expr=\thisrowno{0}/100,y index = 8] {./files/adv_plot.csv};
    \addplot +[name path=lower,draw=none,mark=none] table[x expr=\thisrowno{0}/100,y index = 9] {./files/adv_plot.csv};
    \addplot +[fill=red,opacity=0.3] fill between[of=upper and lower];
    
    \addplot +[name path=upper,draw=none, mark=none] table[x expr=\thisrowno{0}/100,y index = 11] {./files/adv_plot.csv};
    \addplot +[name path=lower,draw=none,mark=none] table[x expr=\thisrowno{0}/100,y index = 12] {./files/adv_plot.csv};
    \addplot +[fill=blue,opacity=0.3] fill between[of=upper and lower];

\nextgroupplot[title={\footnotesize U-net perturbation}, title style={yshift=-5pt}]
	\addplot +[mark=none,red,thick] table[x expr=\thisrowno{0}/100,y index=25]{./files/adv_plot.csv};
	\addplot +[mark=none,fgreen,thick] table[x expr=\thisrowno{0}/100,y index=28]{./files/adv_plot.csv};
	\addplot +[mark=none,blue,thick] table[x expr=\thisrowno{0}/100,y index=31]{./files/adv_plot.csv};
	\addplot +[mark=none,black,thick] table[x expr=\thisrowno{0}/100,y index=34]{./files/adv_plot.csv};
	
	\addplot +[name path=upper,draw=none, mark=none] table[x expr=\thisrowno{0}/100,y index = 26] {./files/adv_plot.csv};
    \addplot +[name path=lower,draw=none,mark=none] table[x expr=\thisrowno{0}/100,y index = 27] {./files/adv_plot.csv};
    \addplot +[fill=red,opacity=0.3] fill between[of=upper and lower];
    
    \addplot +[name path=upper,draw=none, mark=none] table[x expr=\thisrowno{0}/100,y index = 29] {./files/adv_plot.csv};
    \addplot +[name path=lower,draw=none,mark=none] table[x expr=\thisrowno{0}/100,y index = 30] {./files/adv_plot.csv};
    \addplot +[fill=fgreen,opacity=0.3] fill between[of=upper and lower];
    
    \addplot +[name path=upper,draw=none, mark=none] table[x expr=\thisrowno{0}/100,y index = 32] {./files/adv_plot.csv};
    \addplot +[name path=lower,draw=none,mark=none] table[x expr=\thisrowno{0}/100,y index = 33] {./files/adv_plot.csv};
    \addplot +[fill=blue,opacity=0.3] fill between[of=upper and lower];
    
    \addplot +[name path=upper,draw=none, mark=none] table[x expr=\thisrowno{0}/100,y index = 35] {./files/adv_plot.csv};
    \addplot +[name path=lower,draw=none,mark=none] table[x expr=\thisrowno{0}/100,y index = 36] {./files/adv_plot.csv};
    \addplot +[fill=black,opacity=0.3] fill between[of=upper and lower];
\nextgroupplot[title={\footnotesize VarNet perturbation}, title style={yshift=-5pt}]
	\addplot +[mark=none,black,thick] table[x expr=\thisrowno{0}/100,y index=37]{./files/adv_plot.csv};
	\addplot +[mark=none,fgreen,thick] table[x expr=\thisrowno{0}/100,y index=40]{./files/adv_plot.csv};
	\addplot +[mark=none,red,thick] table[x expr=\thisrowno{0}/100,y index=43]{./files/adv_plot.csv};
	\addplot +[mark=none,blue,thick] table[x expr=\thisrowno{0}/100,y index=46]{./files/adv_plot.csv};

	\addplot +[name path=upper,draw=none, mark=none] table[x expr=\thisrowno{0}/100,y index = 38] {./files/adv_plot.csv};
    \addplot +[name path=lower,draw=none,mark=none] table[x expr=\thisrowno{0}/100,y index = 39] {./files/adv_plot.csv};
    \addplot +[fill=black,opacity=0.3] fill between[of=upper and lower];
    
    \addplot +[name path=upper,draw=none, mark=none] table[x expr=\thisrowno{0}/100,y index = 41] {./files/adv_plot.csv};
    \addplot +[name path=lower,draw=none,mark=none] table[x expr=\thisrowno{0}/100,y index = 42] {./files/adv_plot.csv};
    \addplot +[fill=fgreen,opacity=0.3] fill between[of=upper and lower];
    
    \addplot +[name path=upper,draw=none, mark=none] table[x expr=\thisrowno{0}/100,y index = 44] {./files/adv_plot.csv};
    \addplot +[name path=lower,draw=none,mark=none] table[x expr=\thisrowno{0}/100,y index = 45] {./files/adv_plot.csv};
    \addplot +[fill=red,opacity=0.3] fill between[of=upper and lower];
    
    \addplot +[name path=upper,draw=none, mark=none] table[x expr=\thisrowno{0}/100,y index = 47] {./files/adv_plot.csv};
    \addplot +[name path=lower,draw=none,mark=none] table[x expr=\thisrowno{0}/100,y index = 48] {./files/adv_plot.csv};
    \addplot +[fill=blue,opacity=0.3] fill between[of=upper and lower];
\nextgroupplot[title={\footnotesize ConvDecoder perturbation}, title style={yshift=-5pt}]
	\addlegendentry{ConvDecoder}
	\addplot +[mark=none,blue,thick] table[x expr=\thisrowno{0}/100,y index=13]{./files/adv_plot.csv};
	\addlegendentry{$\ell_1$}
	\addplot +[mark=none,fgreen,thick] table[x expr=\thisrowno{0}/100,y index=16]{./files/adv_plot.csv};
	\addlegendentry{VarNet}
	\addplot +[mark=none,black,thick] table[x expr=\thisrowno{0}/100,y index=19]{./files/adv_plot.csv};
	\addlegendentry{U-net}
	\addplot +[mark=none,red,thick] table[x expr=\thisrowno{0}/100,y index=22]{./files/adv_plot.csv};
	
	\addplot +[name path=upper,draw=none, mark=none] table[x expr=\thisrowno{0}/100,y index = 14] {./files/adv_plot.csv};
    \addplot +[name path=lower,draw=none,mark=none] table[x expr=\thisrowno{0}/100,y index = 15] {./files/adv_plot.csv};
    \addplot +[fill=blue,opacity=0.3] fill between[of=upper and lower];
    
    \addplot +[name path=upper,draw=none, mark=none] table[x expr=\thisrowno{0}/100,y index = 17] {./files/adv_plot.csv};
    \addplot +[name path=lower,draw=none,mark=none] table[x expr=\thisrowno{0}/100,y index = 18] {./files/adv_plot.csv};
    \addplot +[fill=fgreen,opacity=0.3] fill between[of=upper and lower];
    
    \addplot +[name path=upper,draw=none, mark=none] table[x expr=\thisrowno{0}/100,y index = 20] {./files/adv_plot.csv};
    \addplot +[name path=lower,draw=none,mark=none] table[x expr=\thisrowno{0}/100,y index = 21] {./files/adv_plot.csv};
    \addplot +[fill=black,opacity=0.3] fill between[of=upper and lower];
    
    \addplot +[name path=upper,draw=none, mark=none] table[x expr=\thisrowno{0}/100,y index = 23] {./files/adv_plot.csv};
    \addplot +[name path=lower,draw=none,mark=none] table[x expr=\thisrowno{0}/100,y index = 24] {./files/adv_plot.csv};
    \addplot +[fill=red,opacity=0.3] fill between[of=upper and lower];
\end{groupplot}
\end{tikzpicture}
\caption{\textbf{Both trained and un-trained reconstruction methods are vulnerable to small adversarial perturbations.} Performance loss as a function of the perturbation strength,
$\epsilon = \frac{\norm[2]{\text{perturbation}}}{\norm[2]{k-\text{space}}}$,
for all methods. 
In each plot, the perturbations are obtained by attacking one method (specified in the plot title), and are applied to all methods. The results are averaged over 10 randomly-chosen proton density knee images from the fastMRI validation set. Shaded areas denote $95\%$ confidence intervals.
}
\label{fig:adv-pert-transfer}
\end{center}
\end{figure*}

Our experimental setup is as follows. We consider 10 randomly-chosen proton-density-weighted knee MRI images from the fastMRI validation set. 
For each image, 
we generate a small perturbation added to the measurement ($k$-space) of a given $\ell_2$ norm. 
For $\ell_1$-minimization and the ConvDecoder, we generate adversarial perturbations through a new optimization-based method detailed in Appendix~\ref{sec:find-adversarial-perturbtion}.
For U-net- and VarNet-based reconstruction, we generate perturbations 
with Projected Gradient Descent (PGD) as described in Appendix~\ref{sec:find-adversarial-perturbtion}. 
We then apply each perturbation to each image, and reconstruct with all four methods. Figure~\ref{fig:adv-pert-transfer} shows the results and the supplement contains reconstructions examples.

Our experiment shows that {\bf both trained and un-trained methods are sensitive to small adversarial perturbations.} 
Recall that \citep{antun2020instabilities,gottschling2020troublesome} found that perturbations adversarially selected for neural networks only mildly affect sparsity-based methods, and concluded that traditional CS methods are more robust than trained methods to such small perturbations. 
In contrast, our results show that while adversarially-selected perturbation for a trained network (e.g., U-net) have a relatively mild impact on $\ell_1$-norm minimization, the converse is also true: perturbations found for $\ell_1$-norm minimization have a significant effect for $\ell_1$-minimization, but only a mild effect on U-net's performance. 
Thus, both methods are vulnerable to perturbations specifically tailored to them.

We note that at first sight, the findings in Figure~\ref{fig:adv-pert-transfer} suggest that ConvDecoder is slightly more robust than both U-net and $\ell_1$-based reconstruction, which in turn is slightly more robust than VarNet-based reconstruction. 
However, it is not possible to draw such absolute comparisons, simply because in order to find adversarial perturbations, we solve a non-convex optimization problem with a numerical method (PGD), and because of the non-convexity, we are not guaranteed to find a worst-case perturbation. It could be that for $\ell_1$-minimization we find a worst-case perturbation but for U-net we do not. 
This issue is inherent to the problem setup and applies to all current methods for finding adversarial perturbations, including those from~\citep{antun2020instabilities,genzel2020rob}.

\section{Distribution shifts}\label{sec:dist-shift}

While robustness to distribution shifts has gained lots of attention over the past few years for image classification tasks~\citep{recht2019imagenet,ovadia2019can, taori2020measuring,hendrycks2020many}, there is little understanding on the effect of distribution shifts in image reconstruction problems. In particular, we are not aware of a systematic study on distribution shifts in accelerated MRI reconstruction.
Understanding the robustness to distribution shifts in image reconstruction problems is important, since given the paucity of training datasets, it may be common to train a method on one patient population, but use the trained model on another; or train on the machine of one manufacturer and test on the machine of another, etc.

In this section, we study three variants of distribution shifts: 
(i) A dataset shift from the fastMRI-knee dataset to a knee dataset from \href{mridata.org}{mridata.org} that we call the Stanford set,
(ii) an anatomy shift from brain images to knee images and vice versa, and finally
(iii) an adversarially-filtered shift to evaluate different models on a set of difficult-to-reconstruct images, inspired by adversarially-filtered shifts for classification introduced by~\citet{hendrycks2021natural}. 

We note all methods, even un-trained ones are affected by distribution shifts, because un-trained methods have hyper-parameters (such as the penalty $\lambda$ in $\ell_1$-minimization), that are tuned on a given distribution. Moreover, we consider multiple variants of each method (e.g., VarNet with different hyper-parameters) in order to obtain a larger variety of models. The details of each reconstruction method and its variants are provided in the supplement.

Our overarching finding is that the performance drop under each of the distribution shifts is similar for all considered trained and un-trained methods. 
Thus, out-of-distribution performance is strongly correlated with in-distribution performance. 
We found that surprising because un-trained methods depend only very mildly on the distribution through hyper-parameter tuning.  
As a consequence, the advantage achieved by a given method over another typically retains this advantage even under distribution shifts. 

This complements an emerging line of works starting with~\cite{recht2019imagenet} that finds a strong correlation between out-of-distribution and in-distribution generalization for a large variety of datasets and models for image classification problems~\cite{miller_linearfits_2021,taori2020measuring,Yadav_Bottou_2019} and even for question-answering models~\cite{Miller_Krauth_Recht_Schmidt_2020}. 
Our results, presented next, indicate that this relation persists even in the context of image recovery, and even when including un-trained methods.

\subsection{Dataset shift}\label{sec:fastmri-v2}


We start with studying the performance of models trained or tuned on the fastMRI knee dataset, but tested on a different knee dataset. 
Specifically, we test on the Stanford dataset retrieved by collecting all available 18 knee volumes from \href{mridata.org}{mridata.org}~\citep{epperson2013creation}. 
The Stanford set contains knees of the same size as the fastMRI images ($320\times320$), but the dataset is different in that
i) the frequency-domain representation of the Stanford set has a $320\times320$ resolution as opposed to $640\times360$ on average for fastMRI,
ii) the slice thickness is lower for the Stanford dataset (0.6mm vs 3mm), resulting in lower SNR, and iii) the Stanford set is acquired using 3D MRI (single volumetric MRI measurement) vs. 2D fastMRI (multiple slice-wise measurements), resulting in varying blurring and SNR. All those slight differences induce a clinically relevant distribution shift. 

Since all of the Stanford set samples are fat-suppressed images, when considering the shift from fastMRI to the Stanford set, we only consider fat-suppressed images from the fastMRI dataset as well.
Our main finding is that {\bf all reconstruction methods perform worse on the new MRI samples, but the absolute performance drop is similar. In-distribution and out-distribution performances are linearly correlated.} Figure~\ref{fig:fastmriv2} shows average SSIM values when training on the fastMRI dataset and evaluating on both fastMRI and Stanford datasets. Reconstruction examples are provided in the supplement. 

We finally remark that when naively applied, the state-of-the-art VarNet is particularly sensitive to this dataset shift, in that the frequency resolution changes in the Stanford set and this affects VarNet as it is based on estimating the unknown $k$-space. This is not reflected in Figure~\ref{fig:fastmriv2}, since we manually increased the resolution of Stanford set data points for VarNet to have a fair comparison among all methods with respect to only the dataset shift (and not resolution shift). For an example of VarNet reconstruction without such \emph{resolution fix}, we refer to the supplement.

\subsection{Anatomy shift}\label{sec:data-shift}


We next consider an anatomy shift where we move from a certain image type (knees) to another (brains), and vice versa.
To understand the robustness to anatomy shifts, we perform the following experiment: We train U-net and VarNet on the whole knee training set and also optimize the hyper-parameters of the ConvDecoder and $\ell_1$-norm minimization on that set. Then we test all methods on the brain validation set. We conversely train on brain MRIs and test on knee MRIs.

\begin{figure}[ht!]
\begin{center}
\begin{tikzpicture}

\begin{groupplot}[
y tick label style={/pgf/number format/.cd,fixed,precision=4},
scaled y ticks = false, xticklabel style={
        /pgf/number format/fixed,
        /pgf/number format/precision=2
},
legend style={at={(3.25,1)} , nodes={scale=0.75}, draw={none}, fill = none, text opacity=1,
/tikz/every even column/.append style={column sep=-0.1cm}
 },
         group
         style={group size= 2 by 1, xlabels at=edge bottom, ylabels at=edge left,
         horizontal sep=2cm, vertical sep=1.9cm,
         }, 
         width=0.36\textwidth,height=0.23\textwidth,
         scaled x ticks=false,
         legend cell align=left,
         ]
\nextgroupplot[ylabel style={align=center},xlabel style={align=center},xlabel={\footnotesize SSIM on knee \\ (trained on knee)}, ylabel={\footnotesize SSIM on knee \\ (trained on brain)},xmax = 0.89,ymax = 0.9,ymin = 0.6,xmin = 0.66,]
    \addplot +[mark=none,steelblue,thick] table[x=x,y=linfit]{./files/anatomy_shift_knee.txt};
	\addplot +[mark=none,dashed,red!70,thick] table[x=x,y=x]{./files/anatomy_shift_knee.txt};
	\addplot +[scatter, only marks, steelblue!60, scatter src = explicit symbolic,mark size=1pt,scatter/classes={
            a={mark=*,mark options={solid,draw=none,fill=fgreen}, fgreen, mark size = 1.8pt},
            b={mark=*,mark options={solid,draw=none,fill=blue},blue,mark size = 1.8pt},
            c={mark=*,mark options={solid,draw=none,fill=red},red,mark size = 1.8pt},
            d={mark=*,mark options={solid,draw=none,fill=black},black,mark size = 1.8pt}
        },
              error bars/.cd, 
              error bar style={line width=0.6pt},
              y dir=both, x dir=both,
              x explicit, y explicit
              ]
       table [x=x, y=y, x error=xerr, y error=yerr, meta=class]{./files/anatomy_shift_knee.txt};
    \legend{best linear fit,$y=x$,$\ell_1$ group,un-trained group,U-net group,VarNet group}
	%
\nextgroupplot[ylabel style={align=center},xlabel style={align=center},xlabel={\footnotesize SSIM on brain \\ (trained on brain)}, ylabel={\footnotesize SSIM on brain \\ (trained on knee)},xmax = 0.97,ymax = 0.97,ymin = 0.7,xmin = 0.73,no markers]
    \addplot +[mark=none,steelblue,thick] table[x=x,y =linfit]{./files/anatomy_shift_brain.txt};
	\addplot +[mark=none,dashed,red!70,thick] table[x=x, y=x]{./files/anatomy_shift_brain.txt};
	\addplot +[scatter, only marks,steelblue!60,scatter src = explicit symbolic, mark size=1pt,scatter/classes={
            a={mark=*,mark options={solid,draw=none,fill=fgreen}, fgreen, mark size = 1.6pt},
            b={mark=*,mark options={solid,draw=none,fill=blue},blue,mark size = 1.6pt},
            c={mark=*,mark options={solid,draw=none,fill=red},red,mark size = 1.6pt},
            d={mark=*,mark options={solid,draw=none,fill=black},black,mark size = 1.6pt}
        },
              error bars/.cd, 
              error bar style={line width=0.6pt},
              y dir=both, x dir=both,
              x explicit, y explicit
              ]
       table [x=x, y=y, x error=xerr, y error=yerr, meta= class]
       {./files/anatomy_shift_brain.txt};
\end{groupplot}
\end{tikzpicture}
\caption{\textbf{Both trained and un-trained methods are not robust to an anatomy distribution shift.} Validation results for a group of 100 brain images and a group of 100 knee images from the fastMRI validation set. $x$ and $y$ axes denote SSIM scores given the training domain for trained neural networks and hyper-parameter learning domain for un-trained methods.}
\label{fig:dataset-shift-unseen}
\end{center}
\end{figure}
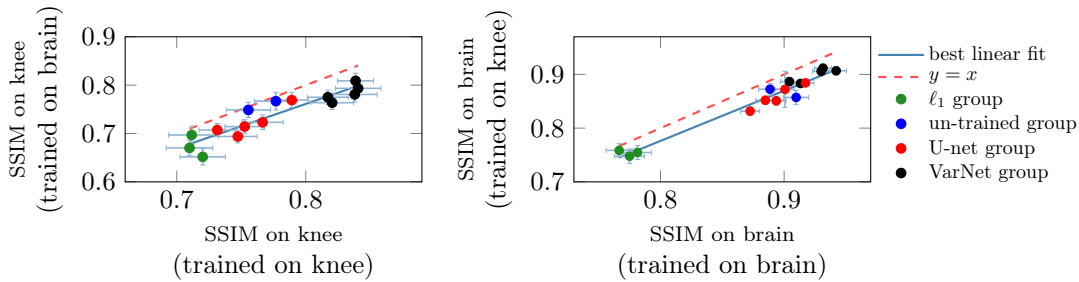

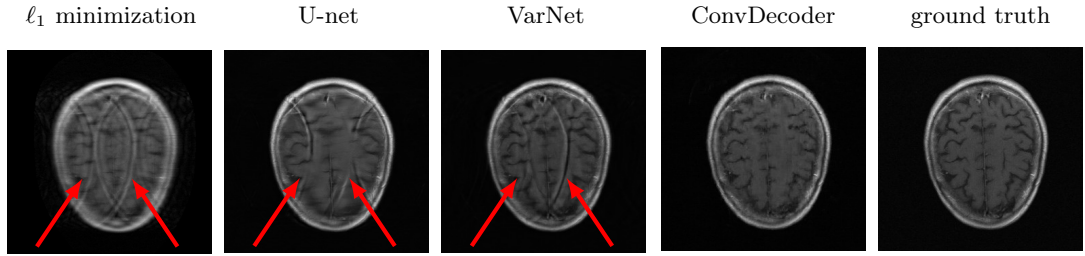
\begin{figure*}[h!]
\centering
  \begin{subfigure}[t]{0.18\textwidth}
  \caption*{$\ell_1$ minimization}
  \hspace{5pt}
  \begin{tikzpicture}
  \node[anchor=south west,inner sep=0] at (2,0) {\includegraphics[scale=0.234]{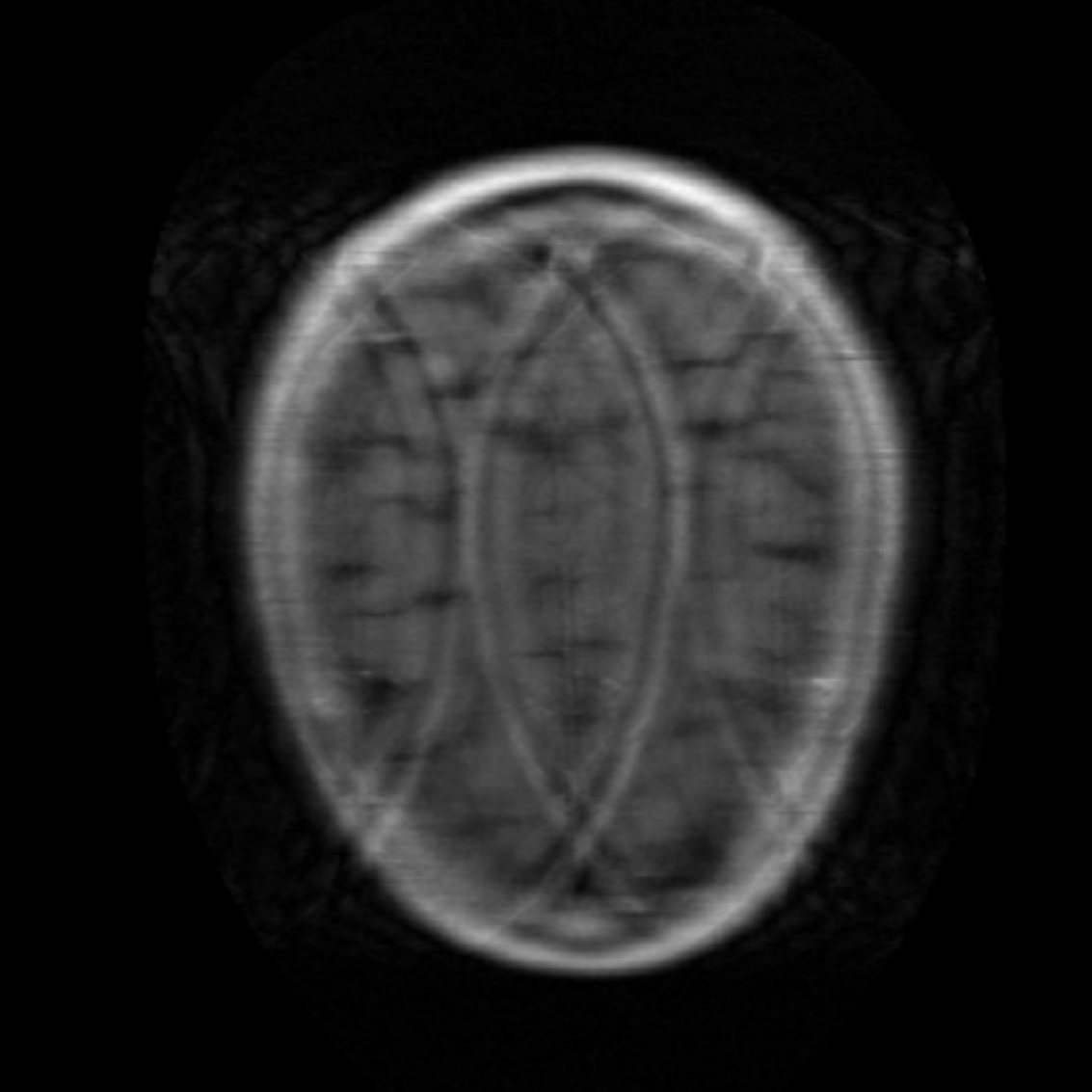}};
  \draw [-latex, ultra thick, red] (2.4,0.1) -- (3,1);
  \draw [-latex, ultra thick, red] (4.25,0.1) -- (3.65,1);
  \end{tikzpicture}
  \end{subfigure}
  \begin{subfigure}[t]{0.18\textwidth}
  \caption*{U-net}
  \hspace{5pt}
  \begin{tikzpicture}
  \node[anchor=south west,inner sep=0] at (2.5,0) {\includegraphics[scale=0.35]{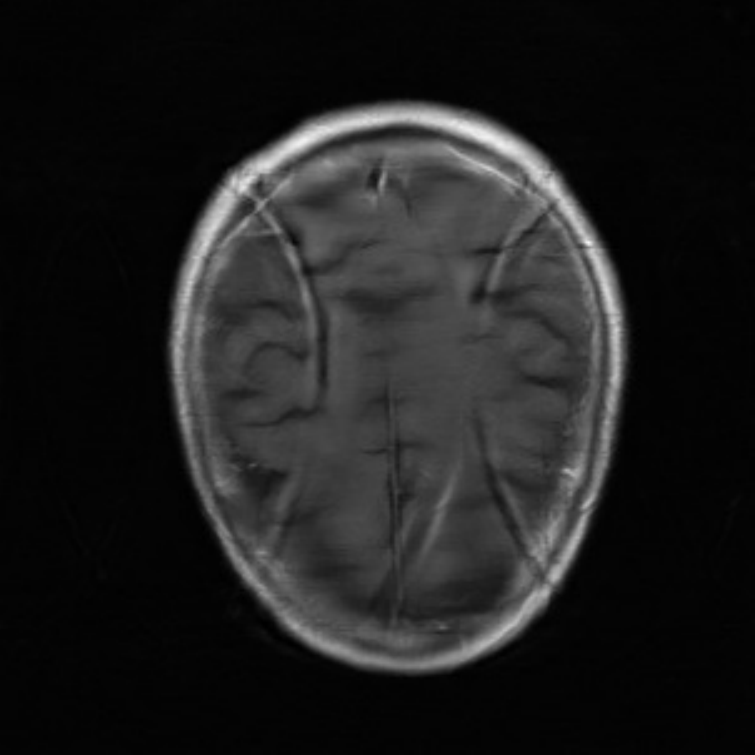}};
  \draw [-latex, ultra thick, red] (2.9,0.1) -- (3.5,1);
  \draw [-latex, ultra thick, red] (4.75,0.1) -- (4.15,1);
  \end{tikzpicture}
  \end{subfigure}
  \begin{subfigure}[t]{0.18\textwidth}
  \caption*{VarNet}
  \hspace{5pt}
  \begin{tikzpicture}
  \node[anchor=south west,inner sep=0] at (3.5,0) {\includegraphics[scale=0.35]{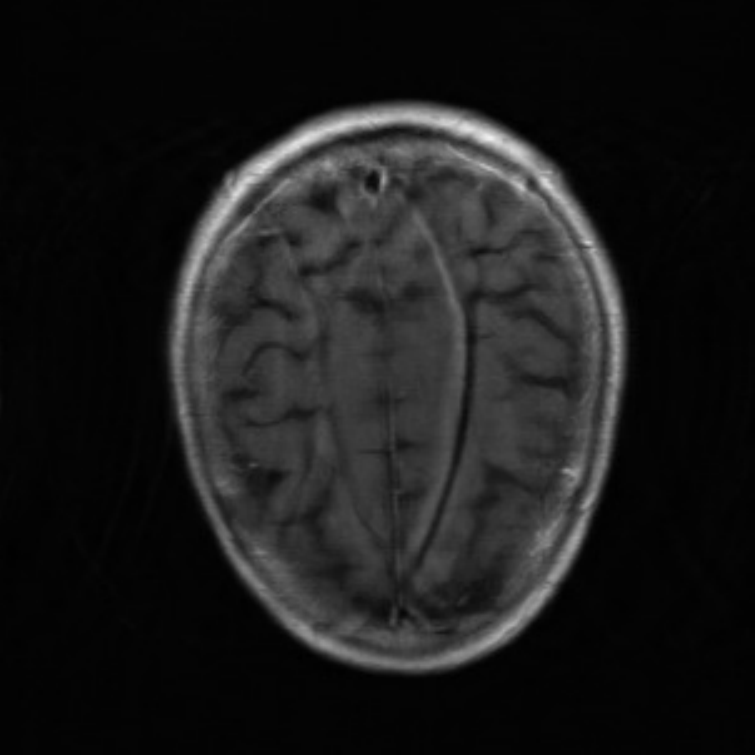}};
  \draw [-latex, ultra thick, red] (3.9,0.1) -- (4.5,1);
  \draw [-latex, ultra thick, red] (5.75,0.1) -- (5.15,1);
  \end{tikzpicture}
  \end{subfigure}
  \begin{subfigure}[t]{0.18\textwidth}
  \caption*{ConvDecoder}
  \centering\includegraphics[scale=0.35]{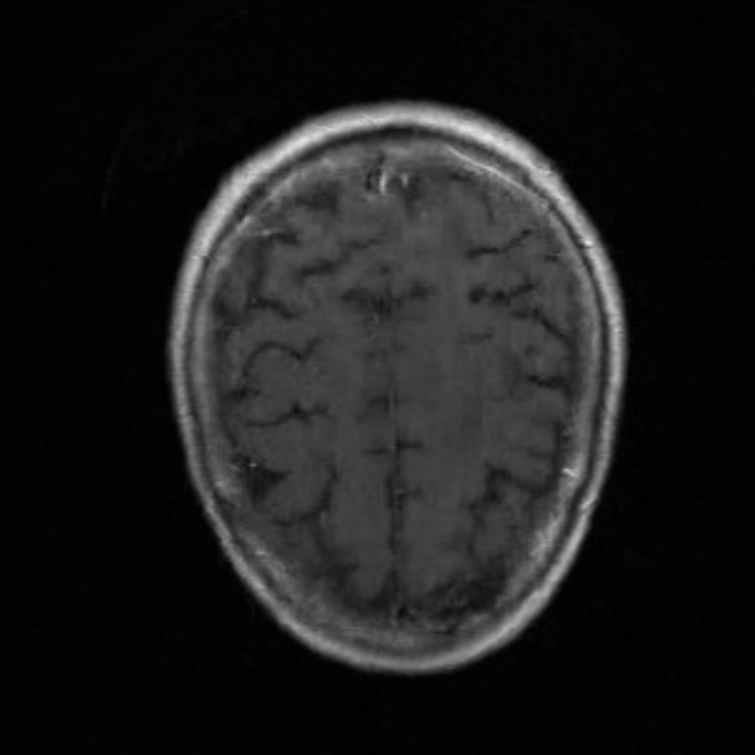}
  \end{subfigure}
  \begin{subfigure}[t]{0.18\textwidth}
  \caption*{ground truth}
  \centering\includegraphics[scale=0.35]{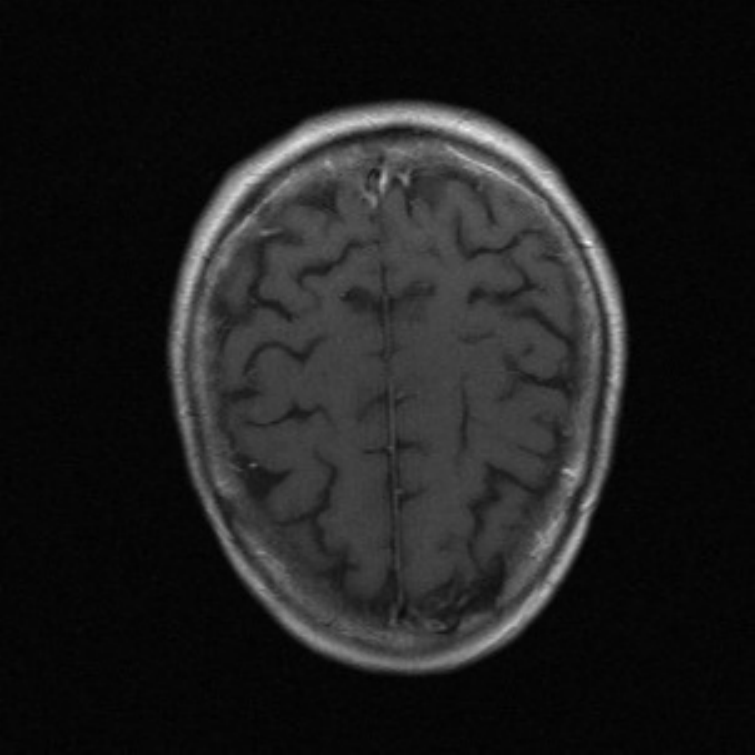}
  \end{subfigure}\par\medskip 
\caption{\textbf{Anatomy shift from knee to brain.} End-to-end variational network (VarNet) and U-net cannot remove under-sampling aliasing artifacts for some images when being trained on knee and tested on brain. $\ell_1$-norm minimization (an un-trained method), also induces aliasing artifacts (irrespective of the distribution shift) similar to most of the traditional CS methods. The ConvDecoder (an un-trained network) is more stable in this setup.}
\label{fig:dshift-sample-recs}
\end{figure*}

Our main finding from the corresponding results in Figure~\ref{fig:dataset-shift-unseen}
is that, again, {\bf all reconstruction methods perform worse on the new anatomy, but the absolute performance drop is similar}. 
Note that there is more variance in the scores for knee images since there are two considerably different knee image types. Those different types arise from different contrasts and different MRI sequences for obtaining the measurements. For the brain images, there are 5 categories of images which are close distribution-wise (i.e., similar contrasts), and most images belong to one of the categories. 


As a final remark, Figure~\ref{fig:dshift-sample-recs} shows that for some images, trained neural networks generate aliasing artifacts in their reconstruction caused by the distribution shift. 
The results of the transfer track of the fastMRI 2020 challenge~\citep[Fig.~3]{muckley2020state} which considers a shift from one scanner to another, also showed that such shifts can induce visible artifacts in reconstructions. 
Thus, the forward reconstruction model ($k$-space to image space) for trained neural networks is conditioned on the learned distribution. Consequently, such trained networks may not learn a scan-agnostic reconstruction model and that their generalizability to overcome aliasing artifacts relies explicitly on the distribution they are trained on (knees in the case of Figure~\ref{fig:dshift-sample-recs}). 

\subsection{Adversarially-filtered shift}\label{sec:adv-filt-shift}

We finally study the performance on images that are particularly difficult to reconstruct as measured by their reconstruction error, in order to understand whether any of the considered methods degrade or shine on such naturally difficult examples. This experiment is inspired by the study of ``adversarially-filtered'' data in image classification. Such points refer to a set of \emph{challenging} samples that cause a significant performance loss to most of the classifiers, and was introduced by \citet{hendrycks2021natural} as ImageNet-A.
ImageNet-A consists of all ImageNet~\citep{deng2009imagenet} images that ResNet-50 misclassifies. 

Here, we create fastMRI-A, a subset of the fastMRI dataset 
that contains the most \emph{challenging} to reconstruct samples, and test all methods on this set of difficult images.
We construct the fastMRI-A (A for adversarial) dataset as follows. We take 5 mid-slice images from each of the 199 knee validation volumes of the fastMRI dataset. This results in a set of 995 images. From this set, we select the images that result in the 100 lowest SSIM scores (bottom 10\%) when reconstructing them via the i-RIM architecture~\citep{putzky2019invert}. We use the i-RIM architecture, a fifth reconstruction method, because this network is one of the best-performing methods in the fastMRI competition, and the winner of the single-coil challenge track. 
Note that for our experiment, it is important not to choose the difficult examples with any of the methods we study (i.e., VarNet, U-net, ConvDecoder, and $\ell_1$-norm minimization), because the goal is to understand whether challenging samples for one method are also challenging for other methods.
We include a few examples from fastMRI-A and their corresponding 4x-accelerated reconstructions in the supplementary materials.

\begin{figure}[ht!]
\begin{center}
\begin{tikzpicture}

\begin{groupplot}[
y tick label style={/pgf/number format/.cd,fixed,precision=4},
scaled y ticks = false, xticklabel style={
        /pgf/number format/fixed,
        /pgf/number format/precision=2
},
legend style={at={(1.75,1)} , nodes={scale=0.75}, draw={none}, fill = none, text opacity=1,
/tikz/every even column/.append style={column sep=-0.1cm}
 },
         group
         style={group size= 1 by 1, xlabels at=edge bottom, ylabels at=edge left,
         xticklabels at=edge bottom,
         horizontal sep=1.5cm, vertical sep=0.6cm,
         }, 
         width=0.36\textwidth,height=0.24\textwidth,
         ylabel={\footnotesize SSIM on fastMRI-A},
         xlabel={\footnotesize SSIM on fastMRI},
         scaled x ticks=false,
         ymax = 0.92,
         ymin = 0.48,
         xmin = 0.63,
         xmax = 0.87,
         legend cell align=left,
         ]
\nextgroupplot[no markers]
	\addplot +[mark=none,steelblue,thick] table[x=x,y=linfit]{./files/adv_filter.txt};
	\addplot +[mark=none,dashed,red!70,thick] table[x=x,y=x]{./files/adv_filter.txt};
	\addplot +[scatter, only marks, steelblue!60, scatter src = explicit symbolic,mark size=1pt,scatter/classes={
            a={mark=*,mark options={solid,draw=none,fill=fgreen}, fgreen, mark size = 1.8pt},
            b={mark=*,mark options={solid,draw=none,fill=blue},blue,mark size = 1.8pt},
            c={mark=*,mark options={solid,draw=none,fill=red},red,mark size = 1.8pt},
            d={mark=*,mark options={solid,draw=none,fill=black},black,mark size = 1.8pt}
        },
              error bars/.cd, 
              error bar style={line width=0.6pt},
              y dir=both, x dir=both,
              x explicit, y explicit
              ]
       table [x=x, y=y, x error=xerr, y error=yerr, meta=class]{./files/adv_filter.txt};
    \legend{best linear fit,$y=x$,$\ell_1$ group,un-trained group,U-net group,VarNet group}
	
\end{groupplot}
\end{tikzpicture}
\caption{\textbf{Challenging data points are \emph{naturally challenging} and lower scores for these data points are not due to learning.} Trained and un-trained methods perform equally poorly on these samples as there is a constant gap between $y=x$ and the best linear fit.}
\label{fig:adv-filter}
\end{center}
\end{figure}
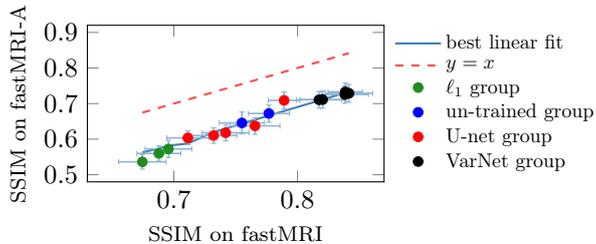

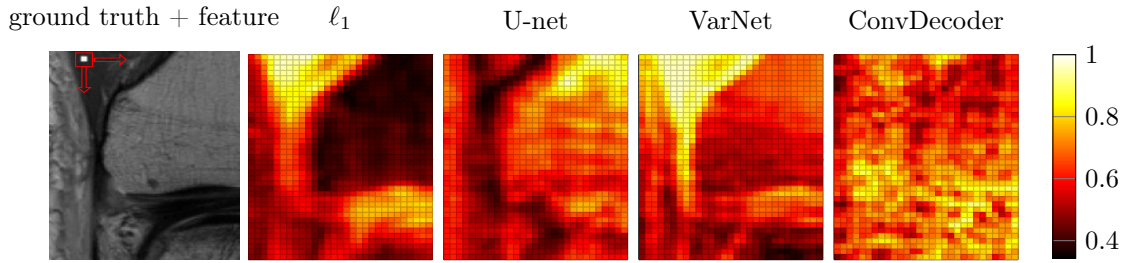
\begin{figure*}[h!]
\begin{center}
\begin{tikzpicture}
\begin{groupplot}[
    group style={group size=5 by 1,
                group name = heat_plots,
                 xlabels at=edge bottom,
                 ylabels at=edge left,
                 yticklabels at=edge left,
                 xticklabels at= edge bottom,
                 horizontal sep=0.15cm, vertical sep=0.4cm,
                },
    width=4cm,
    height=4.3cm,
    colormap/hot2, 
    view={0}{270}]
\node[above left,label={[label distance=1pt]90:ground truth + feature}] (feature) at (0,-0.12) {\includegraphics[width=2.5cm,height=2.75cm]{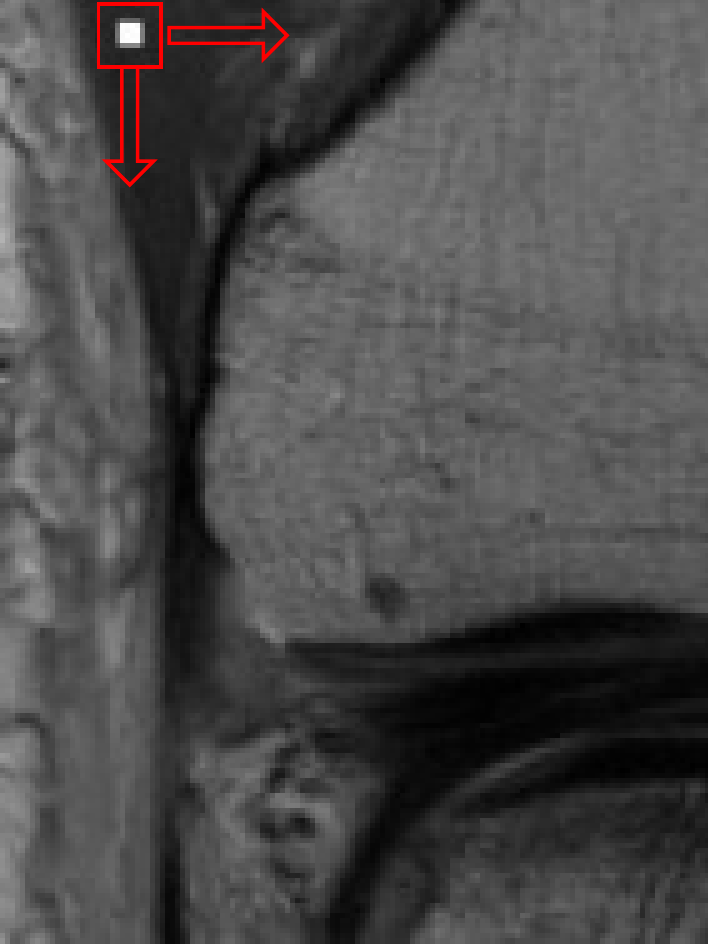}};

\nextgroupplot[colorbar, colorbar style={at={(4.5,1)},anchor=north east, width=0.3cm, height=2.7cm},yticklabels={,,},xticklabels={,,},tickwidth=0,title = $\ell_1$]
    \addplot3[surf] table {./files/heatl.txt};

\nextgroupplot[yticklabels={,,},xticklabels={,,},tickwidth=0,title = U-net]
    \addplot3[surf] table {./files/heatu.txt};
    
\nextgroupplot[yticklabels={,,},xticklabels={,,},tickwidth=0, title = VarNet]
    \addplot3[surf] table {./files/heatv.txt};

\nextgroupplot[yticklabels={,,},xticklabels={,,},tickwidth=0,title = ConvDecoder]
    \addplot3[surf] table {./files/heatc_gd.txt};

\end{groupplot}
\end{tikzpicture}
\captionsetup{skip=3pt}
\caption{Reconstruction heat maps for the end-to-end variational network (VarNet), U-net, ConvDecoder, and $\ell_1$-norm minimization. Each heat map shows the information loss at different locations of an image (i.e., the MSE for each $3\times3$ region) across the image. The left-most image on the top row is the ground-truth image on which we slide a window of fixed information. All heat maps are normalized to the interval $[0,1]$. 
}
\label{fig:small_features}
\end{center}
\end{figure*}

Figure~\ref{fig:adv-filter} shows the performance of the four methods we consider on the fastMRI-A dataset. We plot the performance on those images as a function of the performance on the fastMRI dataset, in order to compare the relative performance change caused by the distribution shift. 
For testing these methods on the fastMRI-A dataset, we used the same set of validation images that we chose in Section~\ref{sec:data-shift}. 

There is a constant gap between the linear interpolation shown in Figure~\ref{fig:adv-filter} and $y = x$ which demonstrates that all four reconstruction methods are equally sensitive to a shift to naturally challenging samples. 
Thus, we again find (see Figure~
\ref{fig:adv-filter}) {\bf that the performance of all methods drops by a similar amount under this distribution shift---meaning that neither un-trained nor trained methods degrade or shine on the difficult-to-reconstruct images.} 
This establishes that there are difficult images to reconstruct, on which all methods perform worse than on an average image. 

Moreover we find that {\bf challenging images are \emph{naturally} difficult to reconstruct,} since both trained and un-trained methods are equally prone to this shift.
Our hypothesis for this natural difficulty is that fastMRI-A samples contain more high-frequency information compared to an average fastMRI example, and thus are harder to reconstruction because less high-frequency information is available in the measurements. The experiments in the supplement confirm this hypothesis.

\section{Recovering small features in an image}\label{sec:small-features}

For medical applications, it is important to recover small details of an image, because such details can be critical for a diagnosis. 
Figure 2b in \cite{knoll2020advancing} shows that for a given image in the fastMRI challenge, all participating trained neural networks fail to recover a small but clinically relevant detail (a tear of the meniscus, which may require surgery for the patients). 

In this section, we study the ability of the four methods 
to recover such small details.
We first propose a simple framework based on artificial features in order to determine whether there is any location dependency for a given reconstruction method when recovering the small feature.
Furthermore, we evaluate the ability to recover small features on a set of 22 annotated fastMRI knee images which contain real-world pathologies. 

\paragraph{Artificial feature recovery.}
Our proposed framework for small feature recovery is as follows. We take a small window consisting of $3\times3$ pixels, fill it with the maximum value of a pixel in the image and slide it through the image to have this spot in different locations. 
We then generate a measurement from this perturbed image and perform the reconstruction using the four mentioned methods. We then measure the error in reconstructing this feature only (i.e., we compute the Mean-Squared Error (MSE) for the $3\times3$ region). We performed this experiment for every location in a test image to understand whether there is a location dependency for different methods. 


The results, displayed in Figure~\ref{fig:small_features}, show that
{\bf different reconstruction methods are sensitive to errors at different regions in the image.} In Figure~\ref{fig:small_features}, we see that $\ell_1$-minimization and VarNet perform worse in dark regions of the image (note that the feature itself is bright).
Note that this location dependency is not due to learning, in that both trained and un-trained methods have location dependent recovery performance. 

\begin{figure}[ht!]
\begin{center}
\begin{tikzpicture}

\begin{groupplot}[
y tick label style={/pgf/number format/.cd,fixed,precision=4},
scaled y ticks = false, xticklabel style={
        /pgf/number format/fixed,
        /pgf/number format/precision=2
},
legend style={at={(1,1)} , nodes={scale=0.6}, draw={none}, fill = none, text opacity=1,
/tikz/every even column/.append style={column sep=-0.1cm}
 },
         group
         style={group size= 1 by 1, xlabels at=edge bottom, ylabels at=edge left,
         yticklabels at=edge left,
         xticklabels at=edge bottom,
         horizontal sep=0.5cm, vertical sep=1.5cm,
         }, 
         width=0.37\textwidth,height=0.23\textwidth,
         ylabel style={align=center},
         ylabel={pathology recovery \\ error (NMSE)},
         scaled x ticks=false,
         legend cell align=left,
         ]
\nextgroupplot[title=Real-world pathologies, xlabel={reconstruction error (NMSE)}, no markers,ymax=0.02,xmin=0.007,xmax=0.025]
	\addplot +[scatter, only marks,steelblue,mark size=1pt,
              error bars/.cd, 
              error bar style={line width=0.6pt},
              y dir=both, x dir=both,
              x explicit, y explicit]
       table [x index=2, y index=0, x error index=3, y error index=1]
       {./real_paths/real_abs_nmse.csv};
	\node[text width=5pt] at (0.008,0.013) {\tiny \steelb{VarNet}};
	\node[text width=5pt] at (0.011,0.0145) {\tiny \steelb{$\text{U-net}$}};
	\node[text width=5pt] at (0.013,0.016) {\tiny \steelb{ConvDecoder}};
	\node[text width=5pt] at (0.022,0.0175) {\tiny \steelb{$\ell_1$}};

\end{groupplot}
\end{tikzpicture}
\captionsetup{skip=3pt}
\caption{\textbf{Reconstruction accuracy is correlated with robustness.} Small feature recovery error based on reconstruction accuracy (SSIM) computed on a set of 22 fastMRI images which contain natural small features. The state-of-the-art VarNet performs best both in terms of reconstruction quality and in terms of small feature recovery error.}
\label{fig:small-features-real}
\end{center}
\end{figure}
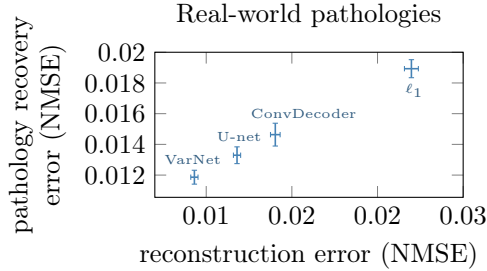

\paragraph{Real-world feature recovery.}
Next, we study natural features by performing an evaluation on a set of 22 annotated images~\citep{cheng2020addressing} from the fastMRI knee dataset which contain real-world pathologies.
We study the performance of those methods in recovering the small features by measuring the MSE only for the region in which the feature is located. 
The results, depicted in Figure~\ref{fig:small-features-real}, show that the ranking of the methods in terms of small-feature-recovery performance is VarNet > U-net > ConvDecoder > $\ell_1$-norm minimization, which coincides with the ranking in terms of overall reconstruction quality. 
The figure also shows that the reconstruction quality is perfectly linearly correlated with accurate recovery of fine details, which is intuitively expected.

With regards to prior works on small feature recovery, \citet{cheng2020addressing} made the first step toward this direction by proposing an optimization framework to synthesize small features for trained networks, by generating different random perturbations of small spatial extent and selecting the worst one for a given method. However, we found it difficult to build a comparison among different methods based on such framework, mainly because of the reasons we discussed for optimization-based perturbations in Section~\ref{sec:adversarial-perturbation} (i.e., (i) worst-case features are not guaranteed, and (ii) hyper-parameters are different across methods). 

\section{Conclusion}\label{sec:conclusion}

Deep-learning-based image reconstruction methods yield high-quality reconstructions from under-sampled data. However, recent works raised the concern that the improvements in reconstruction quality come at a loss in robustness. 
In this paper, we have studied the robustness 
(i) against small adversarial perturbations,
(ii)  to distribution shifts, and
(iii) in recovering fine details, for 
three families of MRI reconstruction methods: trained deep networks, un-trained deep networks, and classical sparsity-based approaches.

We find that both deep-learning-based as well as classical sparsity-based image reconstruction methods are sensitive to small, adversarially-selected perturbations. 

Moreover, the reconstruction quality is correlated with small feature recovery, and hence improving overall reconstruction performance also improves performance for recovering fine details of an image. 

Finally, we find that the performance drop under each of three different realistic distribution shifts is similar for all considered trained and un-trained methods.
The out-of-distribution accuracy is linearly correlated with the in-distribution accuracy, and the performance ranking of the methods typically remains accurate even under distribution shifts. 
This is perhaps surprising, because un-trained methods only depend on the training distribution through hyper-parameter tuning. 

To improve performance in practice, it is important overcome the performance drop due to distribution shifts: Distribution shifts occur in practice and incur a significant loss in performance. 
For un-trained methods, very little data for hyper-parameter tuning is needed and thus the performance gap may be closed via access to only few images of the new domain, and it might even be possible to do hyper-parameter tuning on a single under-sampled measurement~\cite{darestani2020can}. 

For trained methods, this problem might be addressed through larger and more diverse datasets, or through data-augmentation~\cite{fabian_data_2021}. However, distribution shifts are difficult to overcome: 
\citet{taori2020measuring} finds for image classification that robustness enhancing methods---apart from training on large and more diverse datasets---help little for natural distribution shifts, and conclude that ``distribution shifts arising in real data are currently an open research problem.''

In a nutshell, the take-away of our study is that the deep learning methods that perform best based on reconstruction quality are also best under realistic distribution shifts and for small feature recovery, and we could not find them to be more sensitive to adversarial perturbations. 

\section*{Reproducibility}

The code to reproduce all results in this paper is available at 
\href{https://github.com/MLI-lab/Robustness-CS}
{https://github.com/MLI-lab/Robustness-CS}.

\section*{Acknowledgment}\label{sec:ack}

M. Zalbagi Darestani and R. Heckel are (partially) supported by NSF award IIS-1816986, and R. Heckel acknowledges support of the IAS at TUM and the DFG. R. Heckel also thanks Ludwig Schmidt for discussions on distribution shifts, and particular on linear fit phenomena. 
A. Chaudhari is supported by National Institutes of Health (NIH) grants NIH R01 AR063643, R01 EB002524, K24 AR062068, P41 EB015891, received research support from GE Healthcare and Philips, provided consulting services to Skope MR Inc, Subtle Medical, Image Analysis Group, Edge Analytics, Culvert Engineering, and Chondrometrics GmBH, and is a share-holder of Subtle Medical, LVIS Corporation, and Brain Key.



\printbibliography

@string{iclr = {International Conference on Learning Representations (ICLR)}}

@string{icml = {International {Conference} on {Machine} {Learning} (ICML)}}

@string{nips = {Advances in Neural Information Processing Systems (NeurIPS)}}

@string{cvpr = {Conference on Computer Vision and Pattern Recognition (CVPR)}}

@inproceedings{Yadav_Bottou_2019, title={Cold case: The lost {MNIST} digits}, booktitle=nips, author={Yadav, Chhavi and Bottou, Leon}, year={2019} }

@inproceedings{fabian_data_2021,
  title = {Data augmentation for deep learning based accelerated {MRI} reconstruction with limited data},
  booktitle = icml,
  author = {Zalan Fabian and Reinhard Heckel and Mahdi Soltanolkotabi},
  year = {2021},
}

@inproceedings{miller_linearfits_2021,
  title = {Accuracy on the line: On the strong correlation between out-of-distribution and in-distribution generalization},
  booktitle = icml,
  author = {Miller, Joh and Taori, Rohan and Raghunathan, Aditi and Sagawa, Shiori and  Wei Koh, Pang and  Shankar, Vaishaal and Liang, Percy and Carmon, Yair and Schmidt, Ludwig},
  year = {2021},
}

@inproceedings{Miller_Krauth_Recht_Schmidt_2020, title={The effect of natural distribution shift on question answering models}, booktitle=icml,
 author={Miller, John and Krauth, Karl and Recht, Benjamin and Schmidt, Ludwig}, year={2020} }

@inproceedings{deng2009imagenet,
  title={{I}mage{N}et: A large-scale hierarchical image database},
  author={J. Deng and W. Dong and R. Socher and L. Li and K. Li and L. Fei-Fei},
  booktitle={Conference on Computer Vision and Pattern Recognition (CVPR)},
  year={2009}
}

@inproceedings{Daras_intermediate_2021, title={Intermediate layer optimization for inverse problems using deep generative models},  author={Daras, Giannis and Dean, Joseph and Jalal, Ajil and Dimakis, Alexandros G},
booktitle={International Conference on Machine Learning (ICML)},
year={2021},
}

@inproceedings{Kelkar_Bhadra_Anastasio_2020, 
title={Compressible latent-space invertible networks for generative model-constrained image reconstruction}, 
booktitle={IEEE Transactions on Computational Imaging}, 
author={Kelkar, Varun A. and Bhadra, Sayantan and Anastasio, Mark A.}, 
pages={209--223}, 
year={2021}}

@inproceedings{heckel2020denoising,
  title={Denoising and regularization via exploiting the structural bias of convolutional generators},
  author={R. Heckel and M. Soltanolkotabi},
  booktitle={International Conference on Learning Representations (ICLR)},
  year={2020}
}

@inproceedings{zbontar2018fastmri,
  title={fast{MRI}: An open dataset and benchmarks for accelerated {MRI}},
  author={J. Zbontar and F. Knoll and A. Sriram and M. J. Muckley and M. Bruno and A. Defazio and M. Parente and K. J. Geras and J. Katsnelson and H. Chandarana and others},
  booktitle={arXiv: 1811.08839 [cs.CV]},
  year={2018}
}

@inproceedings{wang2019pyramid,
  title={Pyramid convolutional {RNN} for {MRI} reconstruction},
  author={P. Wang and E. Z. Chen and T. Chen and V. M. Patel and S. Sun},
  booktitle={arXiv: 1912.00543 [eess.IV]},
  year={2019}
}

@inproceedings{ronneberger2015u,
  title={U-net: Convolutional networks for biomedical image segmentation},
  author={O. Ronneberger and P. Fischer and T. Brox},
  booktitle={International Conference on Medical Image Computing and Computer-Assisted Intervention},
  pages={234--241},
  year={2015},
  organization={}
}

@inproceedings{schlemper2017deep,
  title={A deep cascade of convolutional neural networks for {MR} image reconstruction},
  author={J. Schlemper and J. Caballero and J. V. Hajnal and A. Price and D. Rueckert},
  booktitle={International Conference on Information Processing in Medical Imaging},
  pages={647--658},
  year={2017},
  organization={}
}

@inproceedings{putzky2019invert,
  title={Invert to learn to invert},
  author={P. Putzky and M. Welling},
  booktitle={Advances in Neural Information Processing Systems (NeurIPS)},
  pages={444--454},
  year={2019}
}

@inproceedings{ulyanov2018deep,
  title={Deep image prior},
  author={D. Ulyanov and A. Vedaldi and V. Lempitsky},
  booktitle={IEEE Conference on Computer Vision and Pattern Recognition (CVPR)},
  pages={9446--9454},
  year={2018}
}

@inproceedings{heckel2019deep,
  title={Deep Decoder: Concise image representations from untrained non-convolutional networks},
  author={R. Heckel and P. Hand},
  booktitle={International Conference on Learning Representations (ICLR)},
  year={2019}
}

@inproceedings{van2018compressed,
  title={Compressed sensing with deep image prior and learned regularization},
  author={D. Van Veen and A. Jalal and M. Soltanolkotabi and E. Price and S. Vishwanath and A. G. Dimakis},
  booktitle={arXiv: 1806.06438 [stat.ML]},
  year={2018}
}

@inproceedings{kingma2014adam,
  title={Adam: A method for stochastic optimization},
  author={D. P. Kingma and J. Ba},
  booktitle={ International Conference on Learning Representations (ICLR)},
  year={2015}
}

@inproceedings{block2007undersampled,
  title={Undersampled radial {MRI} with multiple coils. {I}terative image reconstruction using a total variation constraint},
  author={K. T. Block and M. Uecker and J. Frahm},
  booktitle={Magnetic Resonance in Medicine: An Official Journal of the International Society for Magnetic Resonance in Medicine},
  volume={},
  number={},
  pages={1086--1098},
  year={2007},
  publisher={}
}

@inproceedings{heckel2020compressive,
  title={Compressive sensing with un-trained neural networks: Gradient descent finds the smoothest approximation},
  author={Heckel, Reinhard and Soltanolkotabi, Mahdi},
  booktitle={International Conference on Machine Learning (ICML)},
  year={2020}
}

@inproceedings{sriram2020end,
  title={End-to-end variational networks for accelerated {MRI} reconstruction},
  author={A. Sriram and J. Zbontar and T. Murrell and A. Defazio and C. L. Zitnick and N. Yakubova and F. Knoll and P. Johnson},
  booktitle={International Conference on Medical Image Computing and Computer-Assisted Intervention},
  pages={64--73},
  year={2020},
}

@inproceedings{sriram2019grappanet,
  title={GrappaNet: Combining parallel imaging with deep learning for multi-coil {MRI} reconstruction},
  author={S. Sriram and J. Zbontar and T. Murrell and C. L. Zitnick and A. Defazio and D. K. Sodickson},
  booktitle={IEEE/CVF Conference on Computer Vision and Pattern Recognition (CVPR)},
  pages={14315--14322},
  year={2020}
}

@inproceedings{arora2020untrained,
  title={Untrained modified deep decoder for joint denoising parallel imaging reconstruction},
  author={S. Arora and V. Roeloffs and M. Lustig},
  booktitle={International Society for Magnetic Resonance in Medicine Annual Meeting},
  volume={},
  number={},
  pages={},
  year={2020},
  publisher={}
}

@inproceedings{antun2020instabilities,
  title={On instabilities of deep learning in image reconstruction and the potential costs of {AI}},
  author={V. Antun and F. Renna and C. Poon and B. Adcock and A. C. Hansen},
  booktitle={Proceedings of the National Academy of Sciences (PNAS)},
  year={2020},
}

@inproceedings{gottschling2020troublesome,
  title={The troublesome kernel: Why deep learning for inverse problems is typically unstable},
  author={N. M. Gottschling and V. Antun and B. Adcock and A. C. Hansen},
  booktitle={arXiv: 2001.01258 [cs.LG]},
  year={2020}
}

@inproceedings{cheng2020addressing,
  title={Addressing the false negative problem of deep learning {MRI} reconstruction models by adversarial attacks and robust training},
  author={K. Cheng and F. Caliv{\'a} and R. Shah and M. Han and S. Majumdar and V. Pedoia},
  booktitle={Proceedings of Machine Learning Research},
  year={2020}
}

@article{darestani2020can,
  title={Accelerated {MRI} with un-trained neural networks},
  author={Zalbagi Darestani, Mohammad and Heckel, Reinhard},
  journal={arXiv: 2007.02471 [eess.IV]},
  year={2020}
}

@inproceedings{moosavi2017universal,
  title={Universal adversarial perturbations},
  author={SM. Moosavi-Dezfooli and A. Fawzi and O. Fawzi and P. Frossard},
  booktitle={Proceedings of the IEEE Conference on Computer Vision and Pattern Recognition (CVPR)},
  pages={1765--1773},
  year={2017}
}

@inproceedings{chen2012compressive,
  title={Compressive sensing {MRI} with wavelet tree sparsity},
  author={C. Chen and J. Huang},
  booktitle={Advances in Neural Information Processing Systems (NeurIPS)},
  pages={1115--1123},
  year={2012}
}

@inproceedings{uecker2014espirit,
  title={{ESPIRiT}—an eigenvalue approach to autocalibrating parallel {MRI}: Where {SENSE} meets {GRAPPA}},
  author={M. Uecker and P. Lai and M. J. Murphy and P. Virtue and M. Elad and J. M. Pauly and S. S. Vasanawala and M. Lustig},
  booktitle={Magnetic Resonance in Medicine},
  pages={990--1001},
  year={2014},
}

@inproceedings{genzel2020rob,
  title={Solving inverse problems with deep neural networks -- robustness included?},
  author={M. Genzel and J. Macdonald and M. März},
  booktitle={arXiv :2011.04268  [cs.LG]},
  year={2020}
}

@inproceedings{taori2020measuring,
  title={Measuring robustness to natural distribution shifts in image classification},
  author={R. Taori and A. Dave and V. Shankar and N. Carlini and B. Recht and L. Schmidt},
  booktitle={Advances in Neural Information Processing Systems (NeurIPS)},
  volume={},
  year={2020}
}

@inproceedings{ovadia2019can,
  title={Can you trust your model's uncertainty? {E}valuating predictive uncertainty under dataset shift},
  author={Y. Ovadia and others},
  booktitle={Advances in Neural Information Processing Systems (NeurIPS)},
  pages={13991--14002},
  year={2019}
}

@inproceedings{hendrycks2020many,
  title={The many faces of robustness: A critical analysis of out-of-distribution generalization},
  author={D. Hendrycks and others},
  booktitle={arXiv: 2006.16241 [cs.CV]},
  year={2020}
}

@inproceedings{hendrycks2021natural,
  title={Natural adversarial examples},
  author={D. Hendrycks and K. Zhao and S. Basart and J. Steinhardt and D. Song},
  booktitle={Conference on Computer Vision and Pattern Recognition (CVPR)},
  year={2021}
}

@inproceedings{huang2018some,
  title={Some investigations on robustness of deep learning in limited angle tomography},
  author={Y. Huang and T. W{\"u}rfl and K. Breininger and L. Liu and G. Lauritsch and A. Maier},
  booktitle={International Conference on Medical Image Computing and Computer-Assisted Intervention},
  pages={145--153},
  year={2018},
  organization={}
}

@inproceedings{knoll2020advancing,
  title={Advancing machine learning for {MR} image reconstruction with an open competition: Overview of the 2019 fast{MRI} challenge},
  author={F. Knoll and T. Murrell and A. Sriram and N. Yakubova and J. Zbontar and M. Rabbat and A. Defazio and M. J. Muckley and D. K. Sodickson and C. L. Zitnick and others},
  booktitle={Magnetic Resonance in Medicine},
  year={2020},
  publisher={}
}

@inproceedings{recht2019imagenet,
  title={Do {I}mage{N}et classifiers generalize to {I}mage{N}et?},
  author={B. Recht and R. Roelofs and L. Schmidt and V. Shankar},
  booktitle={International Conference on Machine Learning (ICML)},
  pages={5389--5400},
  year={2019}
}

@inproceedings{lustig2007sparse,
  title={Sparse {MRI}: The application of compressed sensing for rapid {MR} imaging},
  author={M. Lustig and D. Donoho and J. M. Pauly},
  booktitle={Magnetic Resonance in Medicine: An Official Journal of the International Society for Magnetic Resonance in Medicine},
  pages={1182--1195},
  year={2007},
}

@inproceedings{goodfellow2014explaining,
  title={Explaining and harnessing adversarial examples},
  author={I. J. Goodfellow and J. Shlens and C. Szegedy},
  booktitle=iclr,
  year={2015}
}

@inproceedings{bruna2013intriguing,
  title={Intriguing properties of neural networks},
  author={C. Szegedy and W. Zaremba and I. Sutskever and J. Bruna and D. Erhan and I. Goodfellow and R. Fergus},
  booktitle={International Conference on Learning Representations (ICLR)},
  year={2014}
}

@inproceedings{epperson2013creation,
  title={Creation of fully sampled {MR} data repository for compressed sensing of the knee},
  author={K. Epperson and AM. Sawyer and M. Lustig and M. Alley and M. Uecker},
  booktitle={Proceedings of the 22nd Annual Meeting for Section for Magnetic Resonance Technologists},
  year={2013}
}

@inproceedings{muckley2020state,
  title={State-of-the-art machine learning {MRI} reconstruction in 2020: Results of the second fast{MRI} challenge},
  author={M. J. Muckley and B. Riemenschneider and A. Radmanesh and S. Kim and G. Jeong and J. Ko and Y. Jun and H. Shin and D. Hwang and M. Mostapha and others},
  booktitle={IEEE Transactions on Medical Imaging},
  year={2021}
}

@inproceedings{hammernik2018learning,
  title={Learning a variational network for reconstruction of accelerated {MRI} data},
  author={K. Hammernik and T. Klatzer and E. Kobler and M. P. Recht and D. K. Sodickson and T. Pock and F. Knoll},
  booktitle={Magnetic Resonance in Medicine},
  pages={3055--3071},
  year={2018},
}

@inproceedings{candes2006robust,
  title={Robust uncertainty principles: Exact signal reconstruction from highly incomplete frequency information},
  author={E. J. Cand{\`e}s and J. Romberg and T. Tao},
  booktitle={IEEE Transactions on Information Theory},
  pages={489--509},
  year={2006},
}

@inproceedings{roemer1990nmr,
  title={The {NMR} phased array},
  author={P. B. Roemer and W. A. Edelstein and C. E. Hayes and S. P. Souza and O. M. Mueller},
  booktitle={Magnetic Resonance in Medicine},
  pages={192--225},
  year={1990},
}

@inproceedings{bellon1986mr,
  title={{MR} artifacts: A review},
  author={E. M. Bellon and E. M. Haacke and P. E. Coleman and D. C. Sacco and D. A. Steiger and R. E. Gangarosa},
  booktitle={American Journal of Roentgenology},
  pages={1271--1281},
  year={1986},
}

@inproceedings{deshmane2012parallel,
  title={Parallel {MR} imaging},
  author={A. Deshmane and V. Gulani and M. A. Griswold and N. Seiberlich},
  booktitle={Journal of Magnetic Resonance Imaging},
  pages={55--72},
  year={2012},
}

@inproceedings{bora2017compressed,
  title={Compressed sensing using generative models},
  author={A. Bora and A. Jalal and E. Price and A. G. Dimakis},
  booktitle={International Conference on Machine Learning (ICML)},
  pages={537--546},
  year={2017},
}

@inproceedings{asim2020invertible,
  title={Invertible generative models for inverse problems: Mitigating representation error and dataset bias},
  author={M. Asim and M. Daniels and O. Leong and A. Ahmed and P. Hand},
  booktitle={International Conference on Machine Learning (ICML)},
  pages={399--409},
  year={2020},
}

@inproceedings{hand2018global,
  title={Global guarantees for enforcing deep generative priors by empirical risk},
  author={P. Hand and V. Voroninski},
  booktitle={Conference On Learning Theory},
  pages={970--978},
  year={2018},
}

@inproceedings{cohen2018distribution,
  title={Distribution matching losses can hallucinate features in medical image translation},
  author={Cohen, Joseph Paul and Luck, Margaux and Honari, Sina},
  booktitle={International Conference on Medical Image Computing and Computer-Assisted Intervention},
  pages={529--536},
  year={2018},
}

@inproceedings{jin2017deep,
  title={Deep convolutional neural network for inverse problems in imaging},
  author={Jin, Kyong Hwan and McCann, Michael T and Froustey, Emmanuel and Unser, Michael},
  booktitle={IEEE Transactions on Image Processing},
  pages={4509--4522},
  year={2017},
}

\newpage
\onecolumn

\noindent{\huge APPENDIX\par}
\appendix

\section{Small, adversarially-selected perturbations}

Here we explain how we find adversarially-selected perturbations, and give additional details on the effect of those perturbations.

\subsection{Finding adversarial perturbations through optimization}\label{sec:find-adversarial-perturbtion}

Adversarial perturbation are designed to be small in the measurement domain, but to induce significant reconstruction artifacts in the image recovered by the under-sampled version of the perturbed ground-truth data. An adversarial perturbation depends on the image, the forward model, and the reconstruction method.

\paragraph{Finding adversarial perturbations for trained neural networks:}
Let $\mA$ be the linear forward model mapping an image $\vx \in \reals^N$ to a measurement $\vy \in \reals^M$. In our context of MRI, the forward model consists of taking the Fourier transform and applying a mask. In the multi-coil setup the forward model takes the measurements for all coils, and the measurement $\vy$ contains the measurements of all coils. 

Let $\Psi \colon \reals^M \to \reals^N$ be a trained neural network $\Psi$ mapping a measurement to an image. In order to find an adversarial perturbation $\vz \in \reals^M$ that is additive in the measurement domain for the neural network $\Psi$ and for a given image $\vx^*$, we apply projected gradient descent to the following optimization problem:
\begin{align}
\min_{ \vz \colon \norm[2]{\vz} \leq \epsilon \norm[2]{\mA \vx^\ast} } 
    \; -\frac{1}{2} 
    \norm[2]{\Psi(\mA\vx^*) - \Psi(\mA\vx^* + \vz)}^2,
    \label{eq:perturbation-loss}
\end{align}
In words, this optimization problem seeks a perturbation $\vz$ which causes maximal discrepancy between the reconstruction from a clean measurment $\mA \vx^\ast$ and a perturbed measurement $\mA\vx^* + \vz$.

In order to run projected gradient descent, we need to be able to compute gradients of the loss in equation~\eqref{eq:perturbation-loss} with respect to the perturbation $\vz$, thus the reconstruction operator $\Psi$ needs to be differentiable. This is the case for the trained neural network based reconstruction methods in this paper by using standard auto-differentiate packages like PyTorch. However, this method is not directly applicable for reconstruction with the ConvDecoder and $\ell_1$-minimization, as discussed more below. 

Also note that the optimization problem is non-convex, thus projected gradient descent might converge to a sub-optimal perturbation, i.e., a perturbation that does not induce the maximal reconstruction error. This problem is inherent to finding adversarial perturbations for most machine learning methods.

We finally note that in order to find an adversarial perturbation, 
\citet{antun2020instabilities} proposed to minimize
\begin{align*}
     \mathcal{L}(\vz) = \; -\frac{1}{2} 
    \norm[2]{\Psi(\mA\vx^*) - \Psi(\mA\vx^* + \mM\vz)}^2 + \frac{\lambda}{2}\norm[2]{\vz}^2.   
\end{align*}
Here, the parameter $\lambda$ controls the magnitude of the perturbation. We found it difficult to tune this parameter for finding a perturbation with a given norm controlled by $\epsilon$, and therefore work with projected gradient descent, which is guaranteed to give a perturbation of norm $\epsilon$. 

\paragraph{Finding adversarial perturbations for un-trained neural networks and $\ell_1$-minimization:}

In principle, we could find adversarial perturbation for any reconstruction method $\Psi$ by solving the optimization problem in equation~\eqref{eq:perturbation-loss}. However, for un-trained methods and for $\ell_1$-minimization it is not clear how to solve this optimization problem because gradients of $\Psi$ cannot easily be computed, because the reconstruction map is an optimization problem itself. 

There are at least two approaches to overcome the difficulty of computing gradients. The first is to numerically compute a gradient. For example, the numerical gradient of the first term in the loss function w.r.t $\vz$, assuming $g(\vz) = -\frac{1}{2} \norm[2]{\Psi(\mA\vx^*) - \Psi(\mA\vx^* + \vz)}^2$, can be computed as:
\begin{align*}
    \left[ \frac{\partial g(\vz)}{\partial \vz} 
    \right]_{i}
    &\approx \frac{g(\vz+\ve_i h) - g(\vz)}{h},  
\end{align*}
where $\ve_i$ is the $i$-th unit vector and $h$ a very small constant. 
However, estimating the gradient like this is computationally extremely expensive as we need to evaluate the reconstruction method $\Psi$ for each estimate of the gradient, and the perturbation $\vz$ is of high dimension in practice, so we have to evaluate the reconstruction method many times.

A second method, pursued by~\citet{genzel2020rob}, is to realize that many optimization methods including $\ell_1$-minimization are solved numerically with an iterative algorithm, and that the corresponding iterations can be written as a neural network. For this un-rolled network we can compute gradients with standard software packages for automatic differentiation. This is in principle even possible for the un-trained neural networks that we consider.

As a perhaps simpler alternative, we propose a two-step procedure for constructing adversarial perturbations for un-trained methods through optimization. 
This approach is inspired by optimizing a loss function similar to equation~\eqref{eq:perturbation-loss}. 

\paragraph{Step 1.} 
We illustrate our approach with $\ell_1$-minimization. For $\ell_1$-minimization (recall~equation~\eqref{eq:l1-loss}), given a measurement $\vy$, we minimize the loss function $\mc L_1(\vx,\vy) = \frac{1}{2}\norm[2]{\mA \vx - \vy}^2 + \lambda \norm[1]{\mH \vx}$ with respect to $\vx$, where $\mA$ is the forward model and $\mH$ is the sparsifying basis. 
The idea is to combine  the optimization problems
of (i) reconstructing the image and (ii) finding a perturbation into one optimization problem.

In the first step of our method, we minimize the loss function:
\begin{align}
    \mathcal{L}(\vz,\vx) 
    =  \mathcal{L}_1(\vx, \mA \vx^\ast + \vz) \; -\beta 
    \norm[2]{\vx - \vx^\ast}^2
    \label{eq:perturbation-loss-untrained}
\end{align}
by performing gradient descent steps with respect to the reconstructed image $\vx$ and projected gradient descent steps with respect to the perturbation $\vz$, where, as before, we project onto the $\ell_2$-ball $\{\vz \colon \norm[2]{\vz} \leq \epsilon \norm[2]{\mA \vx^\ast} \}$. 
Note that the first term reconstructs an image consistent with the measurement, and the second term penalizes solutions that are close to the original solution $\vx^\ast$. Larger values of the hyper-parameter $\beta$ give larger perturbations $\vz$. 
Thus, while $\ell_1$-norm minimization is reconstructing an image, the adversarial perturbation $\vz$ is reconstructed simultaneously.
By optimizing the loss in equation~\eqref{eq:perturbation-loss-untrained} in a few gradient iterations of Projected Gradient Descent (PGD), we obtain a candidate perturbation $\hat{\vz}$ with $\ell_2$ norm equal to $\epsilon$.

\paragraph{Step 2.} As the second step of our method, we run $\ell_1$-norm minimization on $\vy = \mA\vx^* + \vz$ to see what effect this adversarial perturbation causes to the reconstruction process. This step is necessary because the image obtained from the first step is irrelevant, in that the optimization problem solved in the first step is meant for obtaining the perturbation (and therefore the output image of the first step is just a noisy image).

We emphasize that this algorithm is applicable to both $\ell_1$-minimization, the ConvDecoder, and other, similar, optimization-based un-trained methods. 
Specifically, for the ConvDecoder, we simply apply the method to the loss function of the ConvDecoder $\mathcal{L}(\mC,\vy)$ defined in equation~\eqref{eq:untrained-loss}, in order to find adversarial perturbations for ConvDecoder:
\begin{equation}
	\mathcal{L}(\mC,\vy) = \frac{1}{2}
	\sum_{i=1}^{n_c} 
	\norm[2]{ \vy_i - \mA G_i(\mC) }^2.
	\label{eq:untrained-loss}
\end{equation}
Here, $G_i(\mC)$ is the reconstructed image for coil $i$.

\subsection{Reconstruction examples for adversarial perturbations}\label{sec:sample-recs-adv}
As discussed in Section~\ref{sec:adversarial-perturbation}, both trained and un-trained methods are vulnerable to small, adversarial perturbations. In Figure~\ref{fig:adv-sample-recs}, we provide sample reconstruction for all of the considered methods ($\ell_1$-norm minimization, ConvDecoder, U-net, and the end-to-end VarNet) for perturbations with $\epsilon = \frac{\norm[2]{\text{perturbation}}}{\norm[2]{k\text{-space}}} = 0.08$.

\begin{table*}[ht!]
\setlength{\tabcolsep}{1pt}
\centering
\begin{tabular}{cccccc}
  & $\ell_1$-minimization & U-net & VarNet & ConvDecoder & ground truth \\
  \rule{0pt}{8ex}
  \begin{tabular}{@{}c@{}}4x clean\\ reconstruction\end{tabular} & \scalebox{1}[-1]{\includegraphics[width=0.16\textwidth,valign=m]{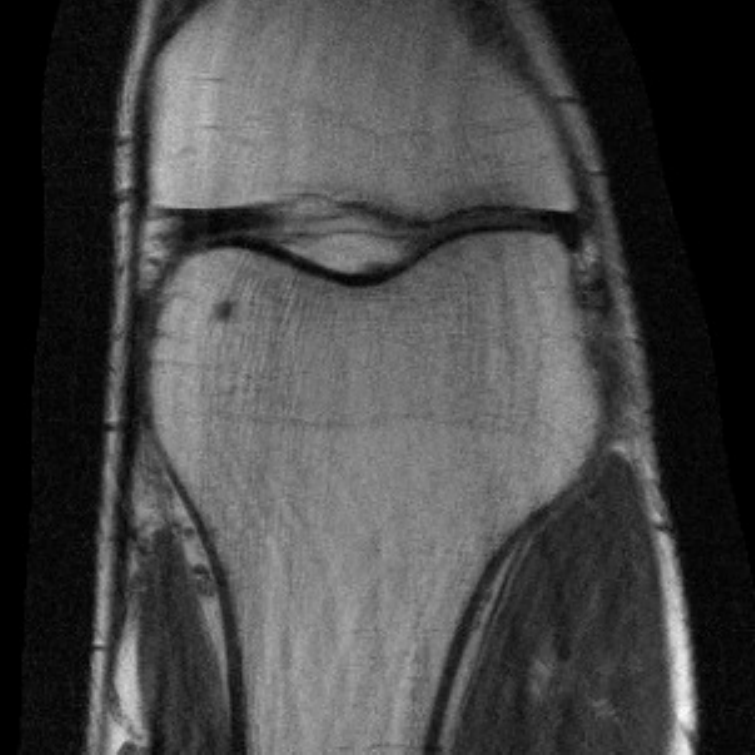}}&
  \scalebox{1}[-1]{\includegraphics[width=0.16\textwidth,valign=m]{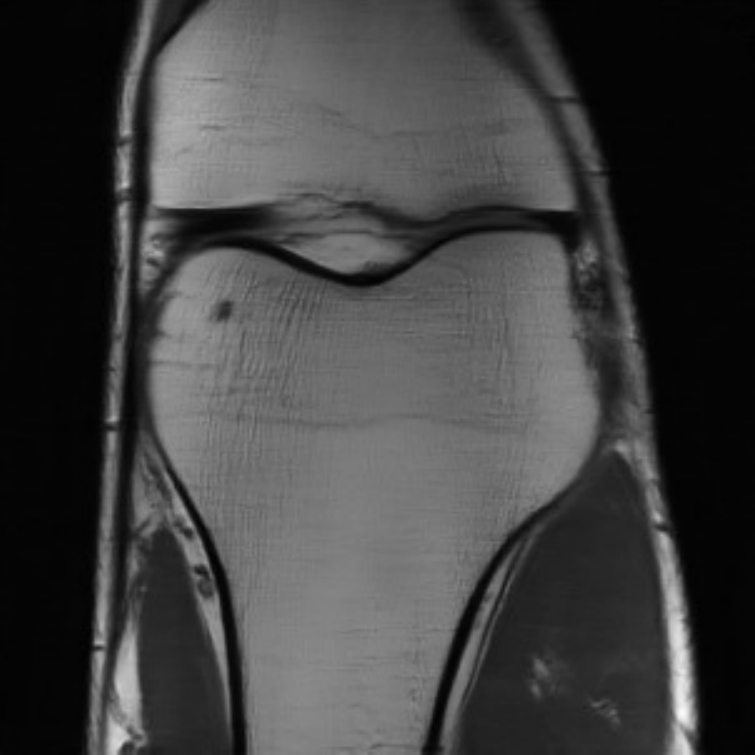}}&
  \scalebox{1}[-1]{\includegraphics[width=0.16\textwidth,valign=m]{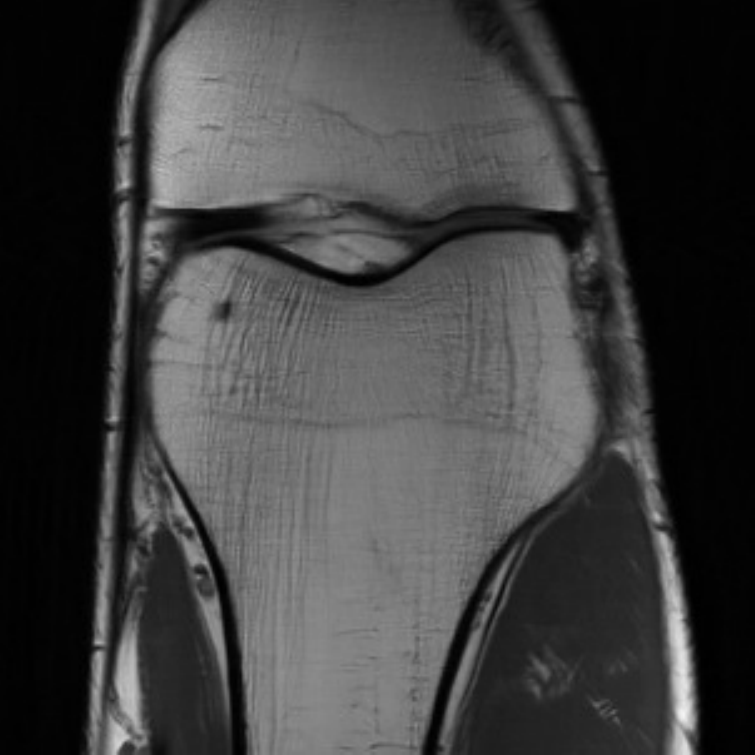}}&
  \scalebox{1}[-1]{\includegraphics[width=0.16\textwidth,valign=m]{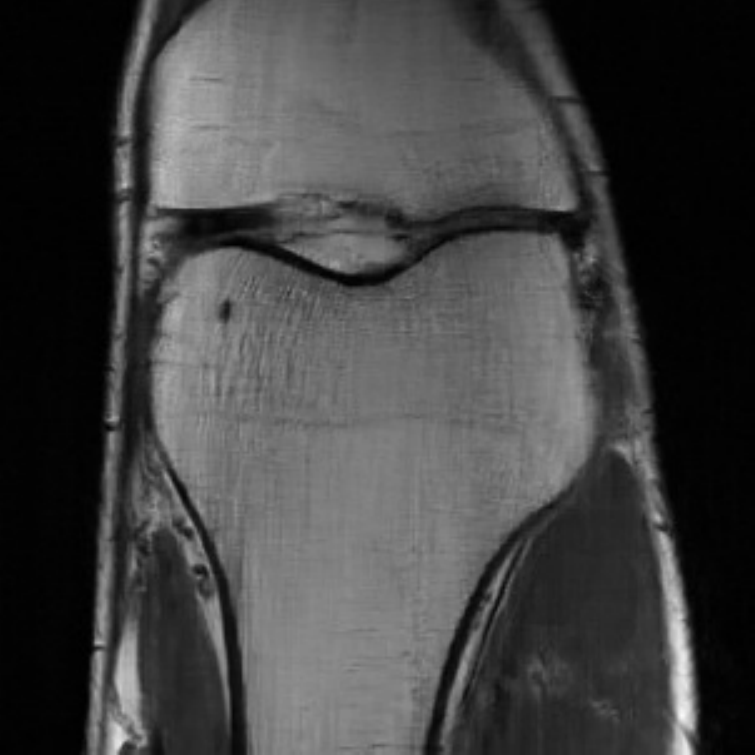}}&
  \scalebox{1}[-1]{\includegraphics[width=0.16\textwidth,valign=m]{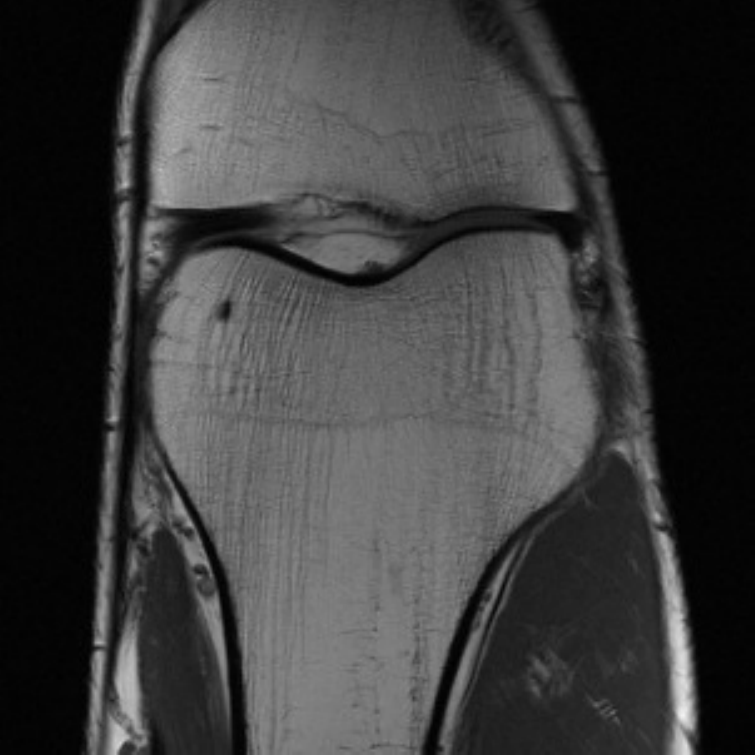}}\\
  \rule{0pt}{8ex}
  \begin{tabular}{@{}c@{}}perturbed\\ ground truth\end{tabular}& \scalebox{1}[-1]{\includegraphics[width=0.16\textwidth,valign=m]{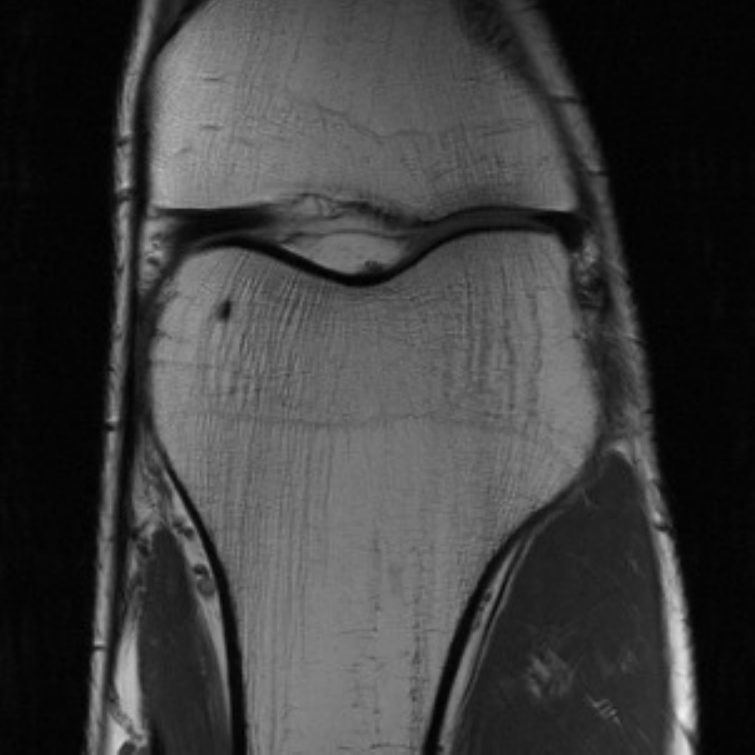}}&
  \scalebox{1}[-1]{\includegraphics[width=0.16\textwidth,valign=m]{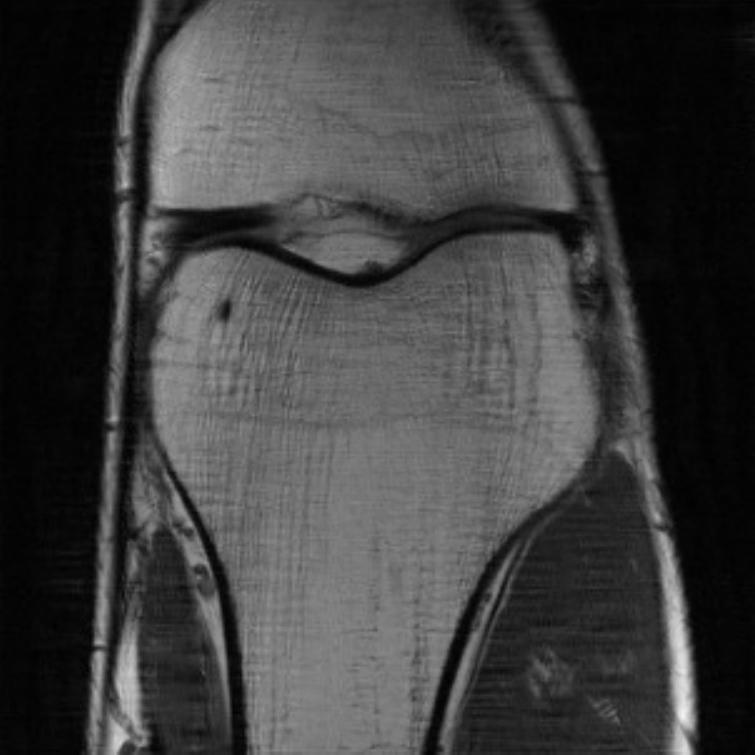}}&
  \scalebox{1}[-1]{\includegraphics[width=0.16\textwidth,valign=m]{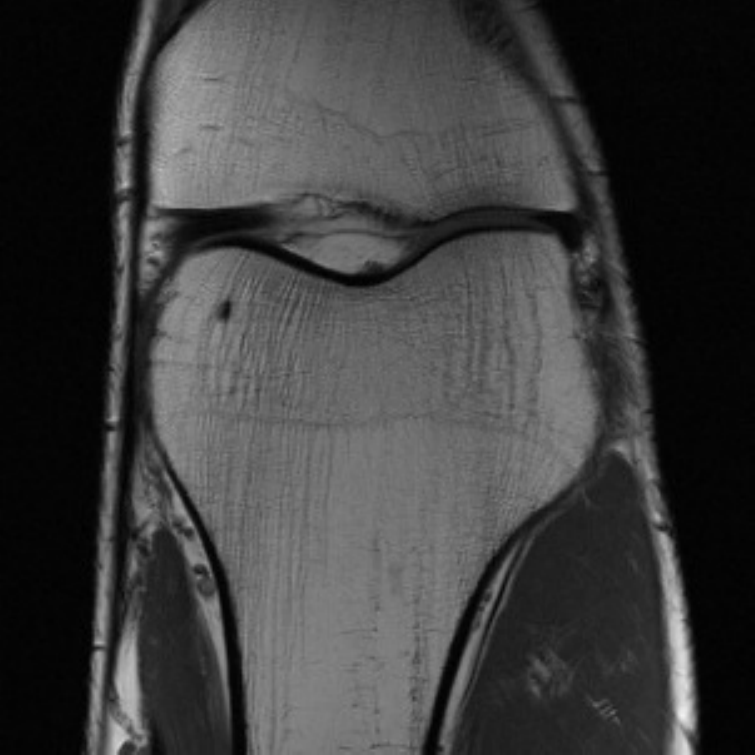}}&
  \scalebox{1}[-1]{\includegraphics[width=0.16\textwidth,valign=m]{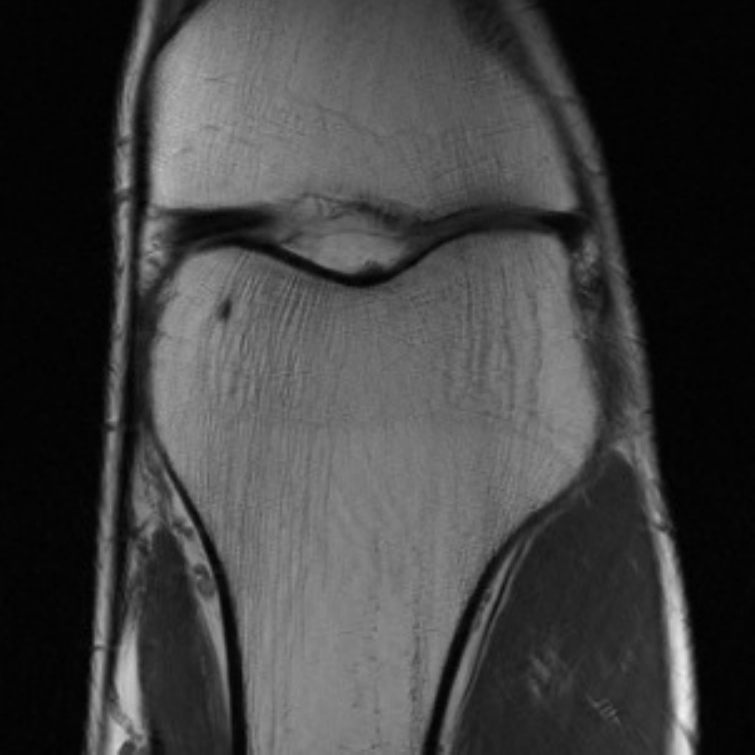}}&\\ 
  \rule{0pt}{8ex}
  \begin{tabular}{@{}c@{}}4x perturbed\\ reconstruction\end{tabular} & \scalebox{1}[-1]{\includegraphics[width=0.16\textwidth,valign=m]{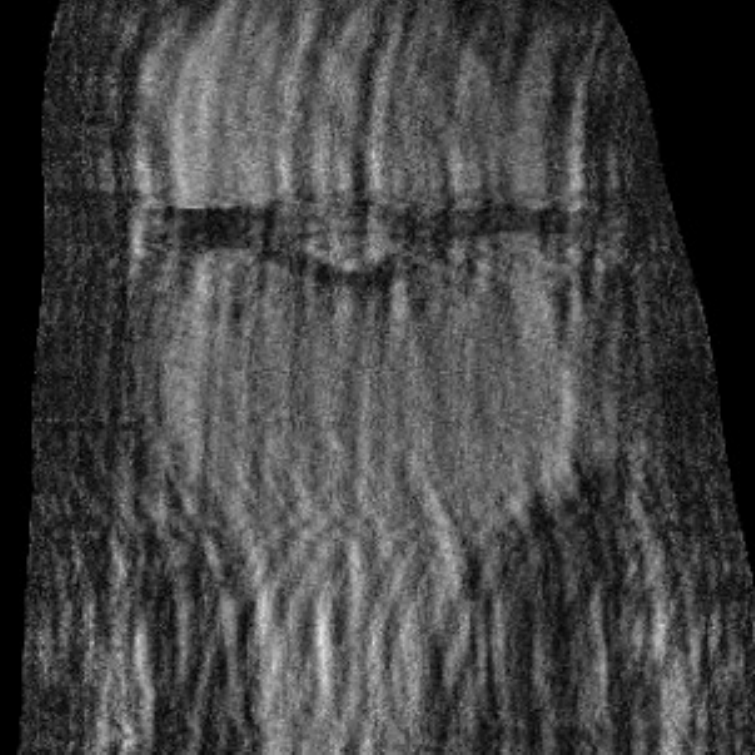}}&
  \scalebox{1}[-1]{\includegraphics[width=0.16\textwidth,valign=m]{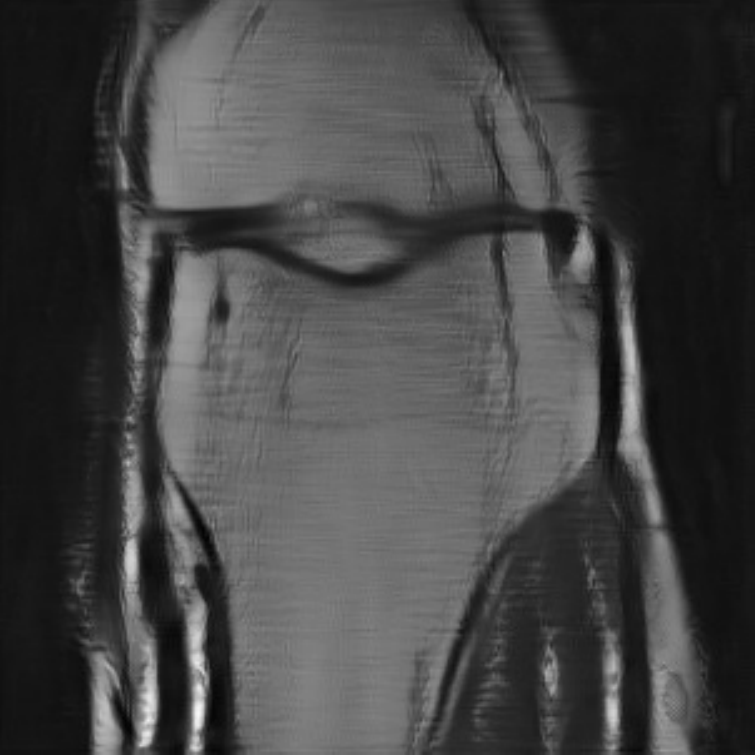}}&
  \scalebox{1}[-1]{\includegraphics[width=0.16\textwidth,valign=m]{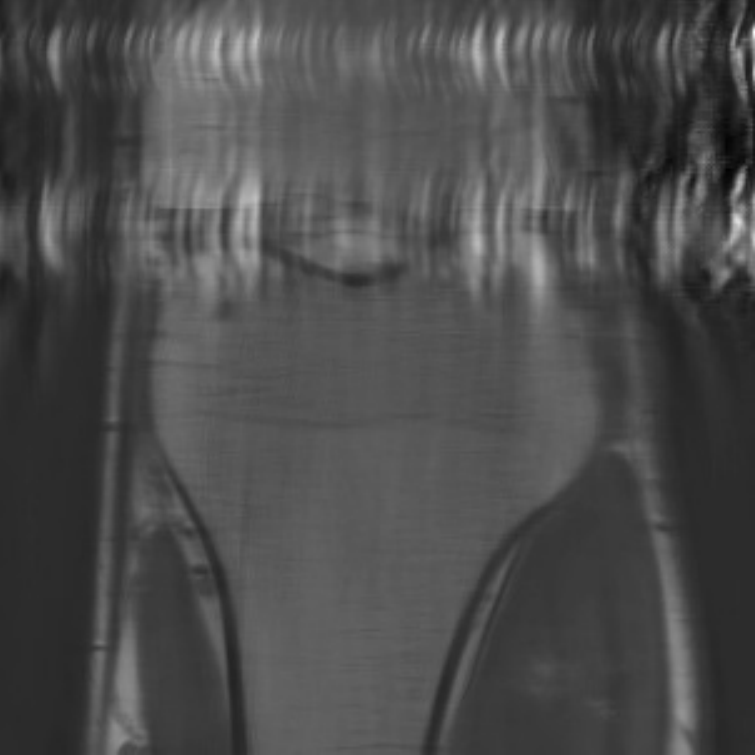}}&
  \scalebox{1}[-1]{\includegraphics[width=0.16\textwidth,valign=m]{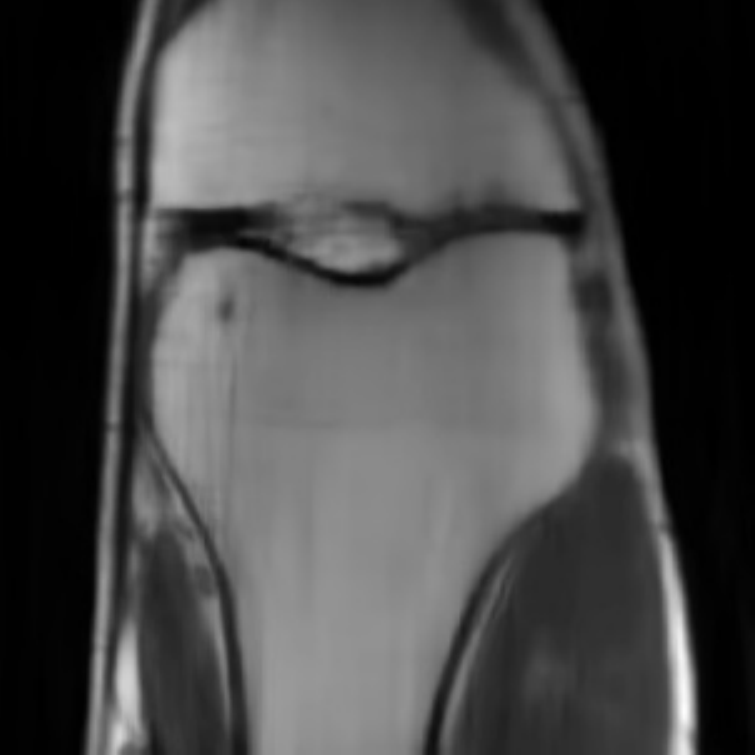}}&
\end{tabular}
\captionsetup{skip=10pt}
\captionof{figure}{\textbf{End-to-end variational network (VarNet), U-net, ConvDecoder, and $\ell_1$-norm minimization are all vulnerable to small adversarially-selected perturbations. First row:} The ground-truth image and reconstructions from the clean 4x under-sampled measurements for a validation image from the multi-coil fastMRI dataset. \textbf{Second row:} Perturbed ground-truth images after finding one perturbation specifically for each method ($\epsilon = 0.08$ which is the maximal perturbation considered in this study). \textbf{Third row:} Reconstructions from the perturbed 4x under-sampled measurements.
}
\label{fig:adv-sample-recs}
\end{table*}


\subsection{The choice of sparsifying  basis is irrelevant for small, adversarially-selected perturbations}\label{sec:basis}

In Section~\ref{sec:adversarial-perturbation}, we demonstrated that both trained and un-trained methods suffer from small perturbations in the input. 
As a sparsifying basis for sparsity based reconstruction, we chose the  Wavelet-basis as it is one of the most popular ones used in MRI. However, the particular choice of basis is not critical for our results, and our findings continue to hold for other popular sparsifying bases: Figure~\ref{fig:adv-basis} shows sample attacks for Wavelet, Fourier, and Discrete Cosine Transform (DCT) sparsifying bases, and it can be seen that for all those choices, the perturbations have a similar effect.

\begin{table*}[ht!]
\setlength{\tabcolsep}{1pt}
\centering
\begin{tabular}{ccccc}
  & Wavelet basis & Fourier basis & DCT basis & ground truth \\
  \rule{0pt}{8ex}
  \begin{tabular}{@{}c@{}}4x clean\\ reconstruction\end{tabular} & \scalebox{1}[-1]{\includegraphics[width=0.16\textwidth,valign=m]{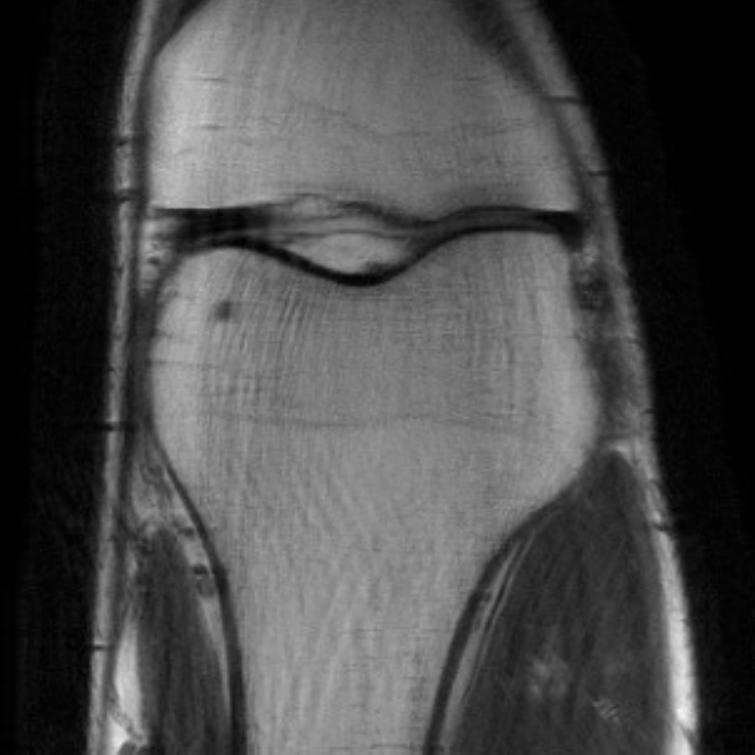}}&
  \scalebox{1}[-1]{\includegraphics[width=0.16\textwidth,valign=m]{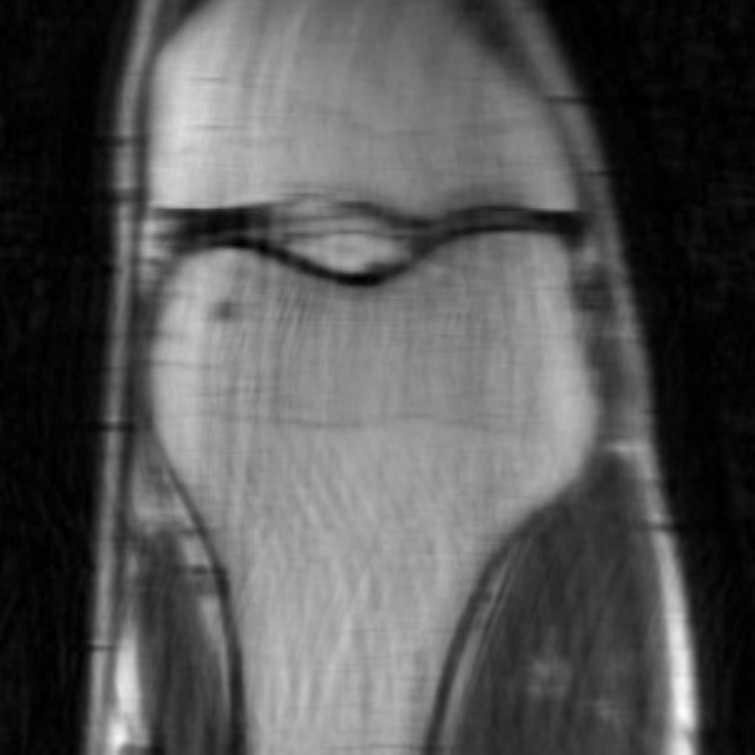}}&
  \scalebox{1}[-1]{\includegraphics[width=0.16\textwidth,valign=m]{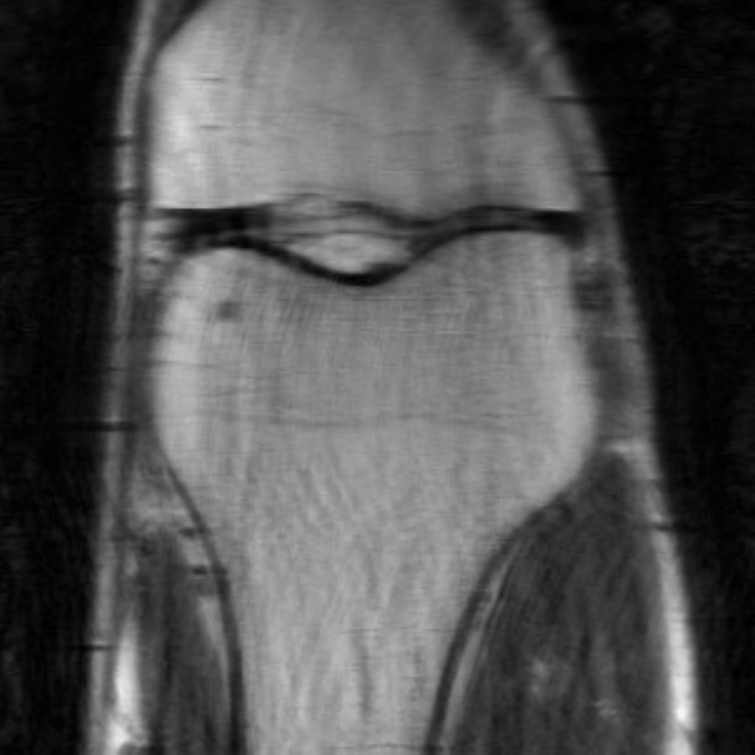}}&
  \scalebox{1}[-1]{\includegraphics[width=0.16\textwidth,valign=m]{./adv_imgs/orig-eps-converted-to}}\\
  \rule{0pt}{8ex}
  \begin{tabular}{@{}c@{}}perturbed\\ ground truth\end{tabular}& \scalebox{1}[-1]{\includegraphics[width=0.16\textwidth,valign=m]{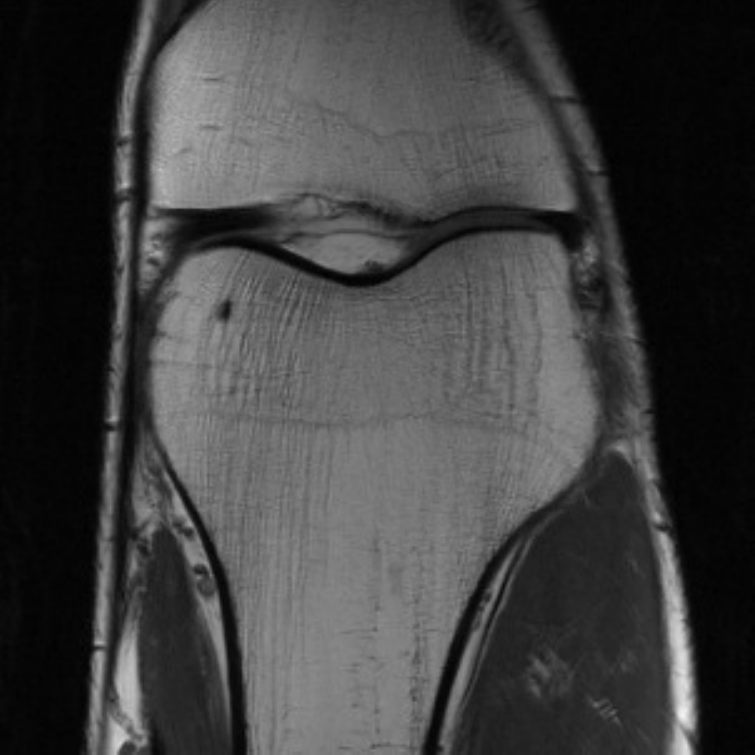}}&
  \scalebox{1}[-1]{\includegraphics[width=0.16\textwidth,valign=m]{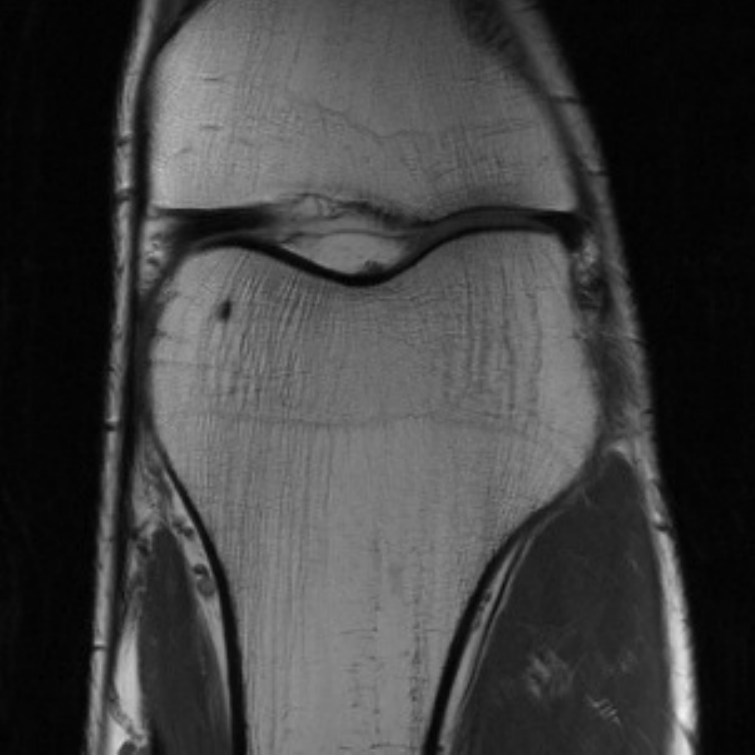}}&
  \scalebox{1}[-1]{\includegraphics[width=0.16\textwidth,valign=m]{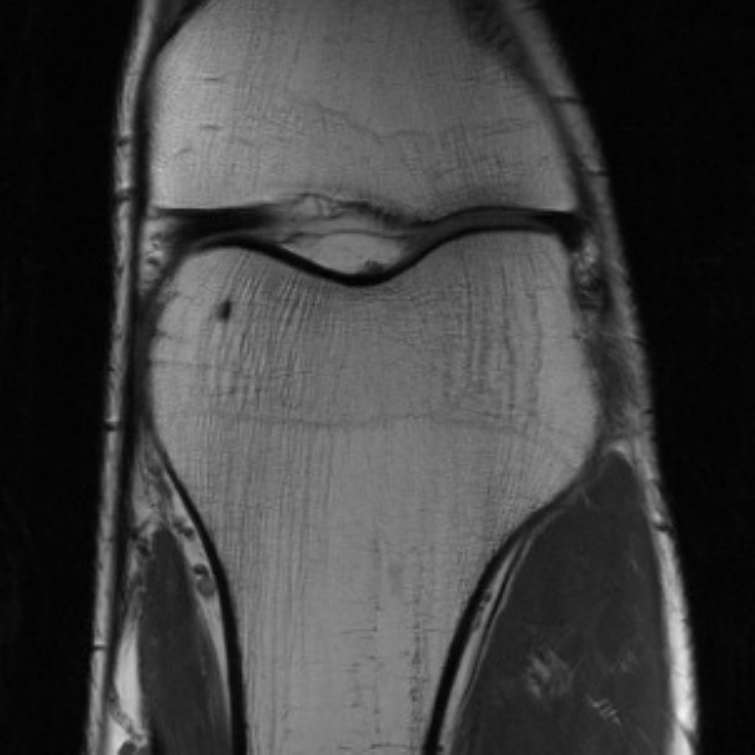}}& \\ 
  \rule{0pt}{8ex}
  \begin{tabular}{@{}c@{}}4x perturbed\\ reconstruction\end{tabular} & \scalebox{1}[-1]{\includegraphics[width=0.16\textwidth,valign=m]{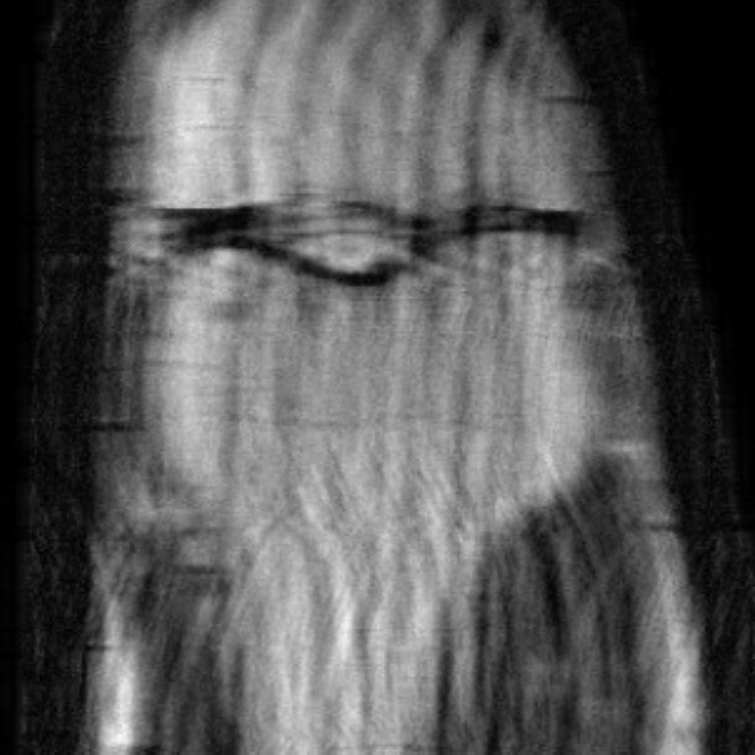}}&
  \scalebox{1}[-1]{\includegraphics[width=0.16\textwidth,valign=m]{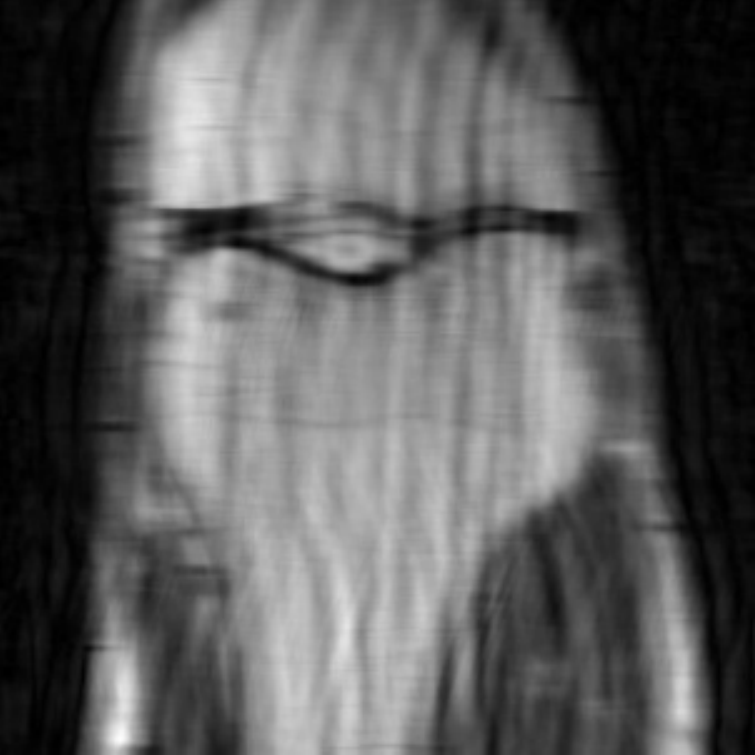}}&
  \scalebox{1}[-1]{\includegraphics[width=0.16\textwidth,valign=m]{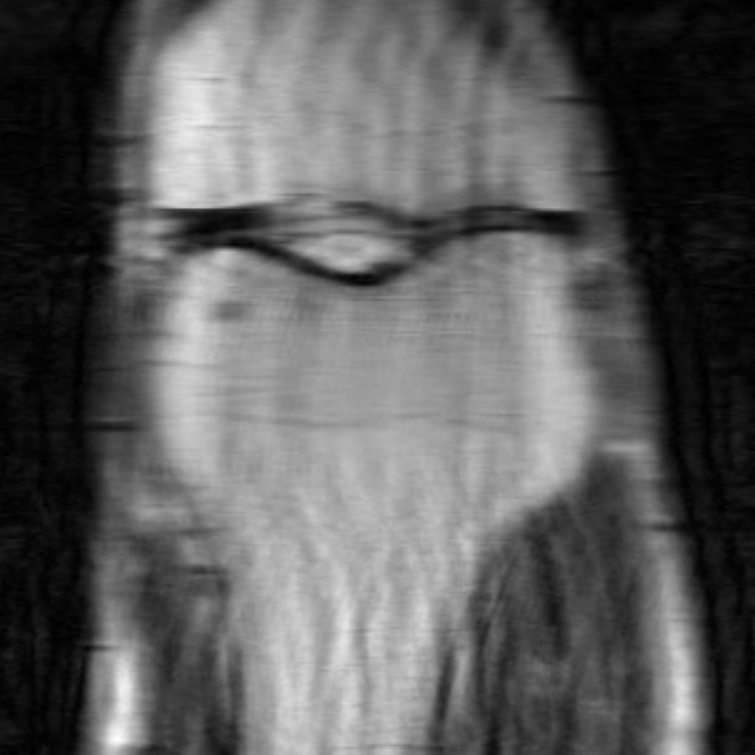}}&
\end{tabular}
\captionsetup{skip=10pt}
\captionof{figure}{
Perturbations have a similar effect for different common choices of sparsifying bases. The experiment shows that an adversarial perturbation with $\epsilon=0.08$ specific for each basis can be found.
}
\label{fig:adv-basis}
\end{table*}

\section{Distribution shifts}

In this section we give additional details on distribution shifts, in particular we provide details on the adversarially-filtered shifts.

\subsection{Reconstruction examples for the dataset shift}

\begin{figure*}[ht!]
\centering
\begin{subfigure}[t]{0.19\textwidth}
\caption*{$\ell_1$}
\centering\scalebox{1}[-1]{\includegraphics[scale=0.37]{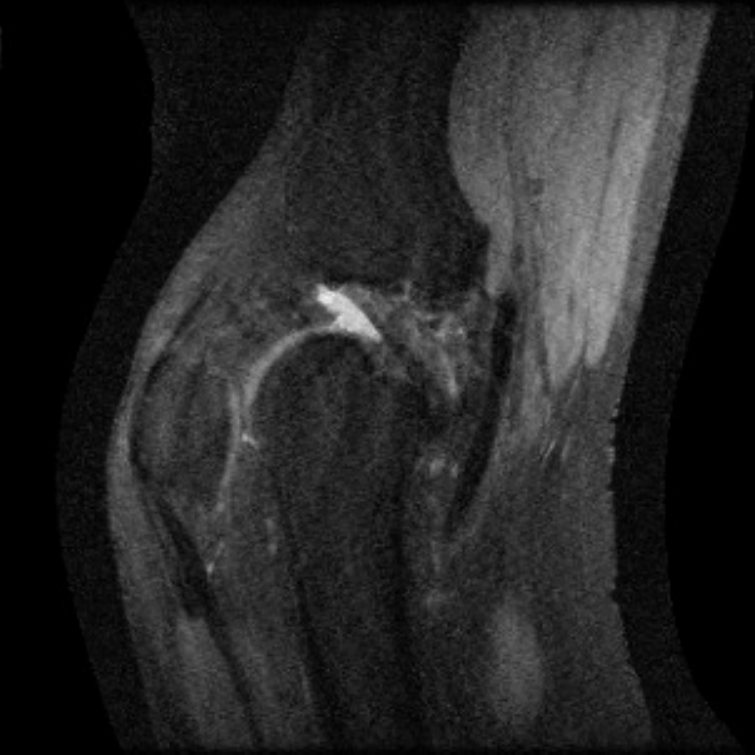}}
\end{subfigure}
\begin{subfigure}[t]{0.19\textwidth}
\caption*{U-net}
\centering\scalebox{1}[-1]{\includegraphics[scale=0.37]{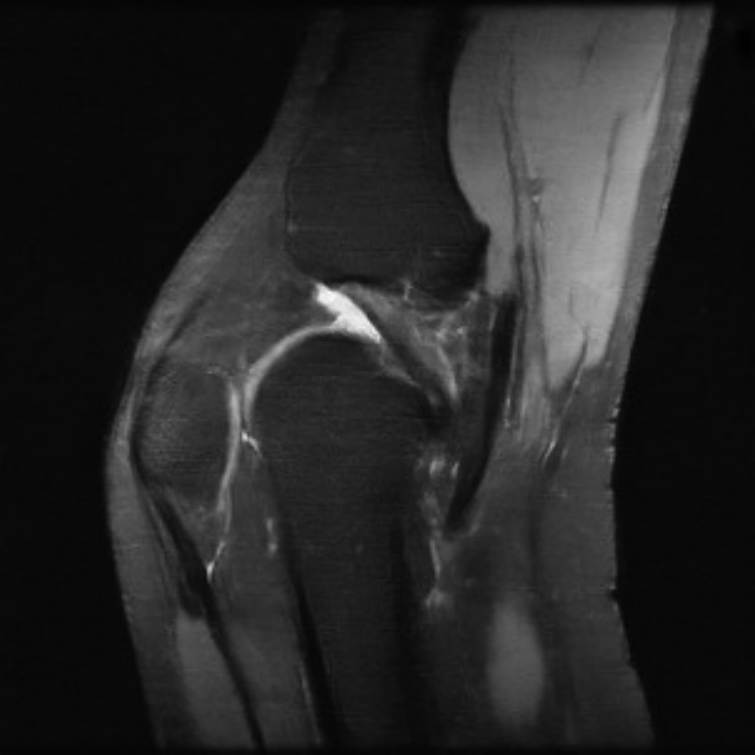}}
\end{subfigure}
\begin{subfigure}[t]{0.19\textwidth}
\caption*{VarNet}
\centering\scalebox{1}[1]{\includegraphics[scale=0.37]{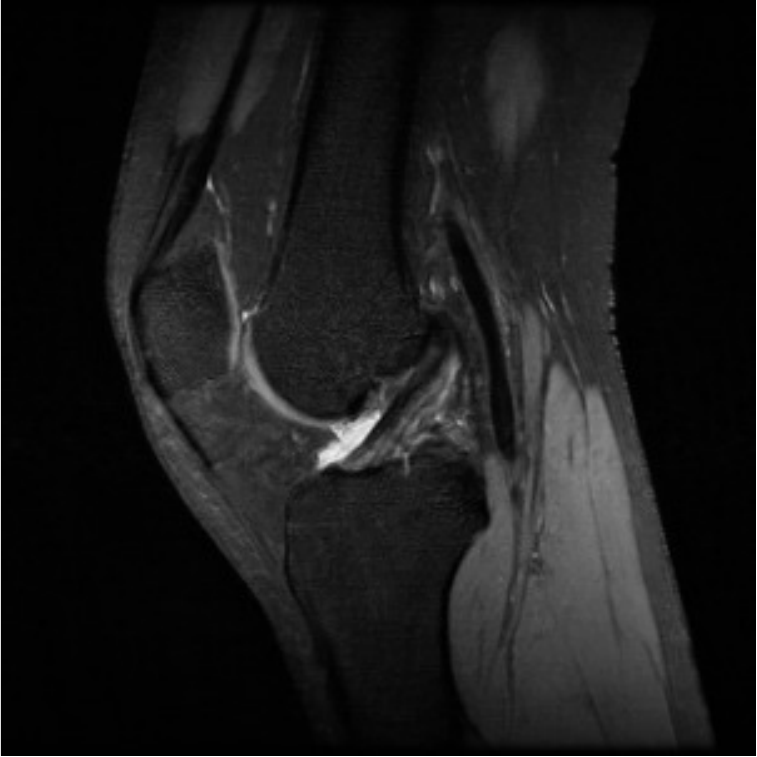}}
\end{subfigure}
\begin{subfigure}[t]{0.19\textwidth}
\caption*{ConvDecoder}
\centering\scalebox{1}[-1]{\includegraphics[scale=0.37]{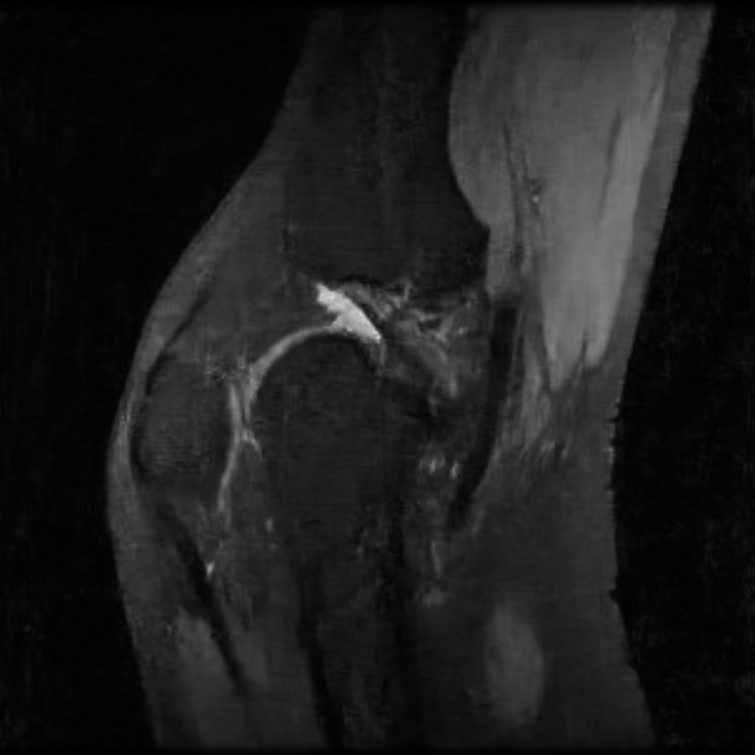}}
\end{subfigure}
\begin{subfigure}[t]{0.19\textwidth}
\caption*{ground truth}
\centering\scalebox{1}[-1]{\includegraphics[scale=0.37]{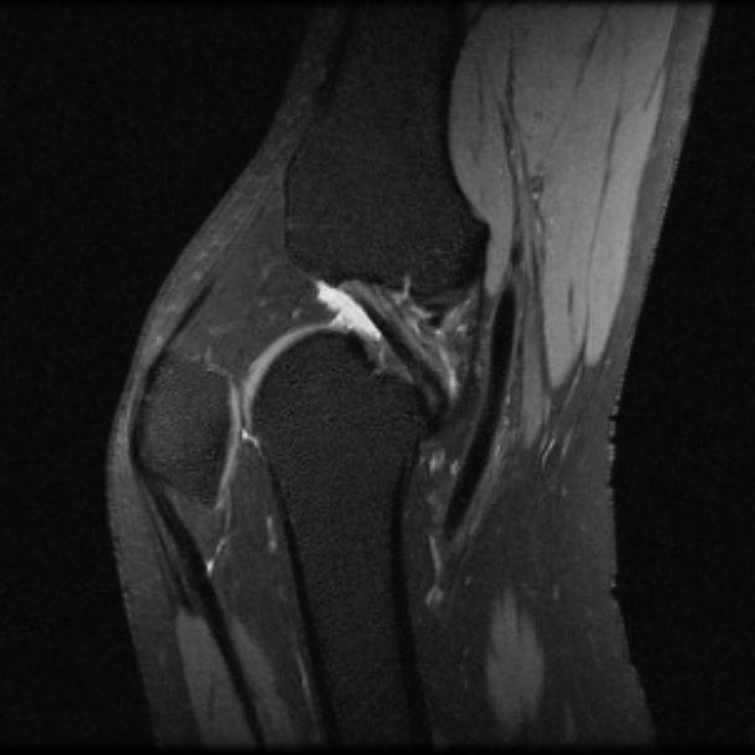}}
\end{subfigure}
\caption{\textbf{fastMRI reconstruction methods do not generalize well to the Stanford set.} All of the considered reconstruction methods linearly lose performance when being evaluated on the Stanford set. Reconstructions (for 4x acceleration) show that each method either comes with reconstruction artifacts or fails to recover all the details.
}
\label{fig:fastmriv2-recs}
\end{figure*}

As shown in Section~\ref{sec:fastmri-v2}, all of the models that have been trained or tuned on the fastMRI domain perform worse when evaluated on the Stanford set. In Figure~\ref{fig:fastmriv2-recs}, we provide reconstruction examples for such a distribution shift, showing that a distribution shift induces visible performance degradation. 




\subsection{Reconstruction examples for anatomy shift}

\begin{figure*}[tb]
\centering
  \begin{subfigure}[t]{0.19\textwidth}
  \caption*{$\ell_1$ minimization}
  \centering\scalebox{1}[-1]{\includegraphics[scale=0.37]{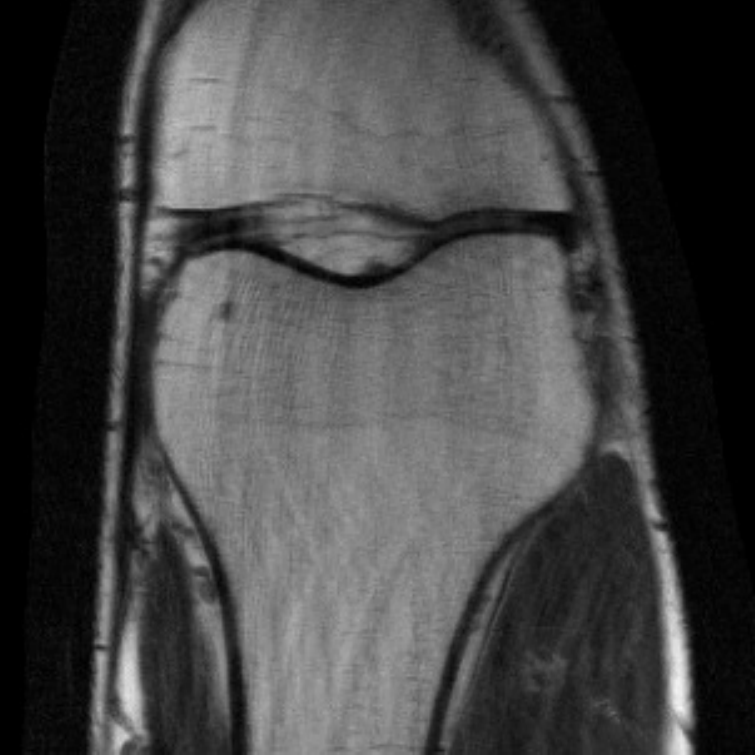}}
  \end{subfigure}
  \begin{subfigure}[t]{0.19\textwidth}
  \caption*{U-net}
  \centering\scalebox{1}[-1]{\includegraphics[scale=0.37]{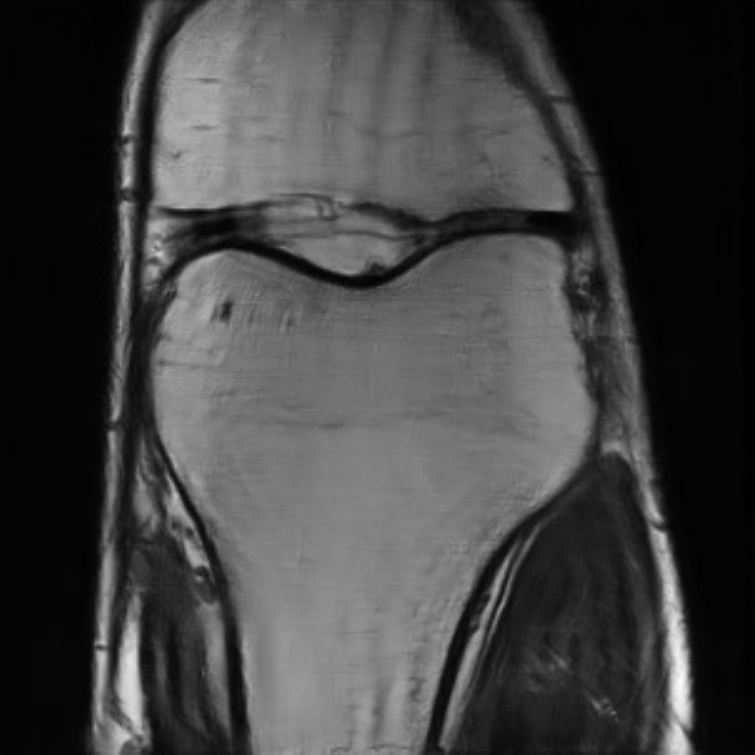}}
  \end{subfigure}
  \begin{subfigure}[t]{0.19\textwidth}
  \caption*{VarNet}
  \centering\scalebox{1}[-1]{\includegraphics[scale=0.37]{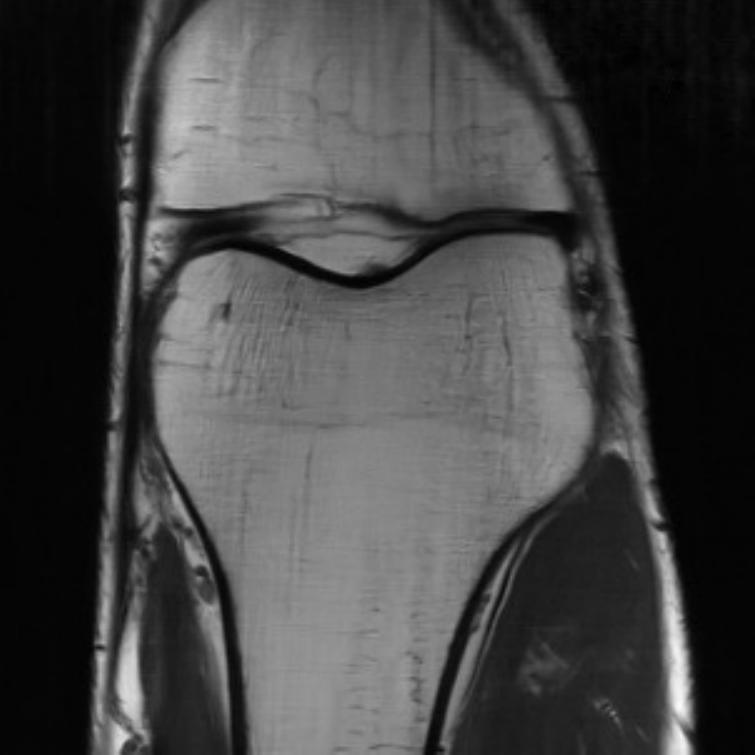}}
  \end{subfigure}
  \begin{subfigure}[t]{0.19\textwidth}
  \caption*{ConvDecoder}
  \centering\scalebox{1}[-1]{\includegraphics[scale=0.247]{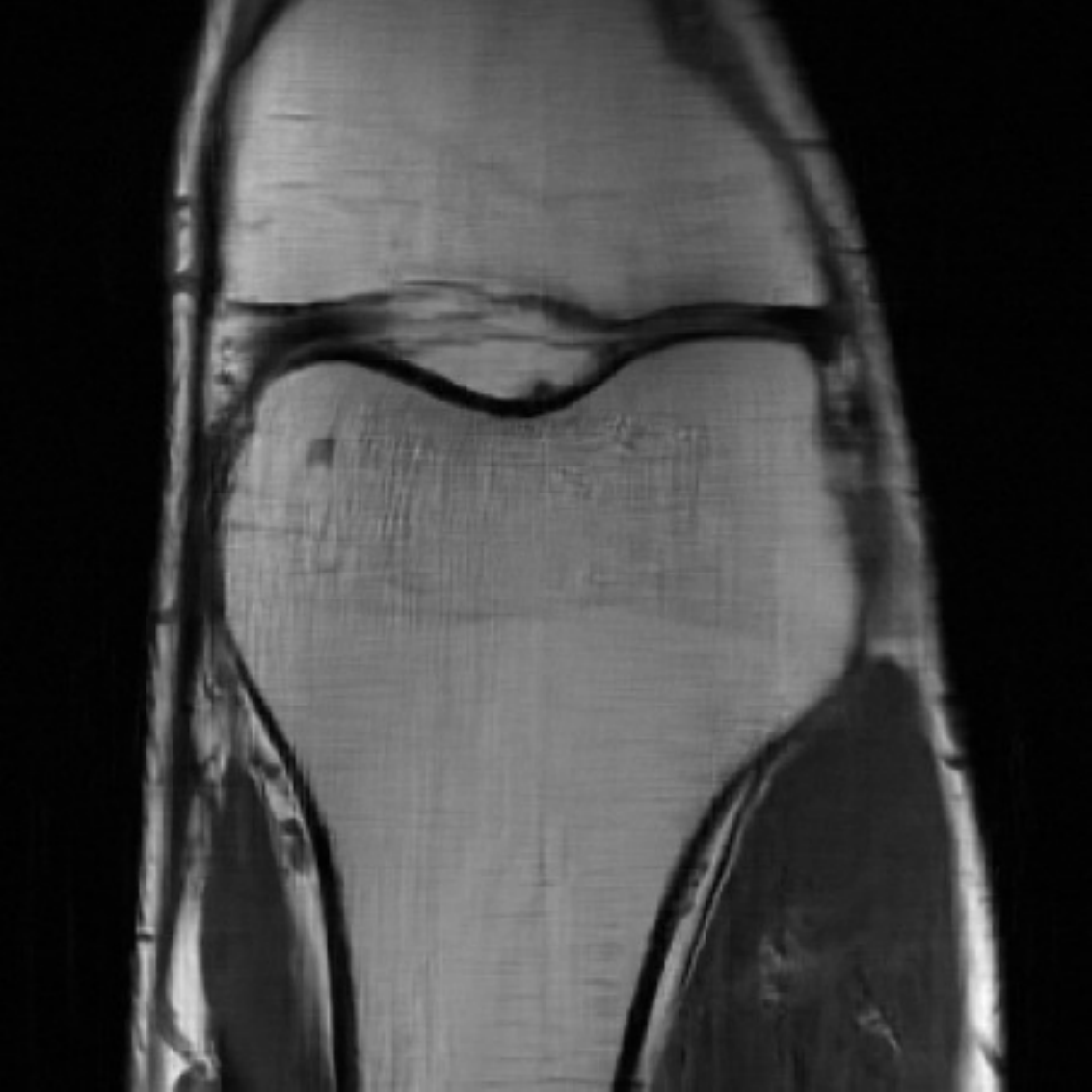}}
  \end{subfigure}
  \begin{subfigure}[t]{0.19\textwidth}
  \caption*{ground truth}
  \centering\scalebox{1}[-1]{\includegraphics[scale=0.247]{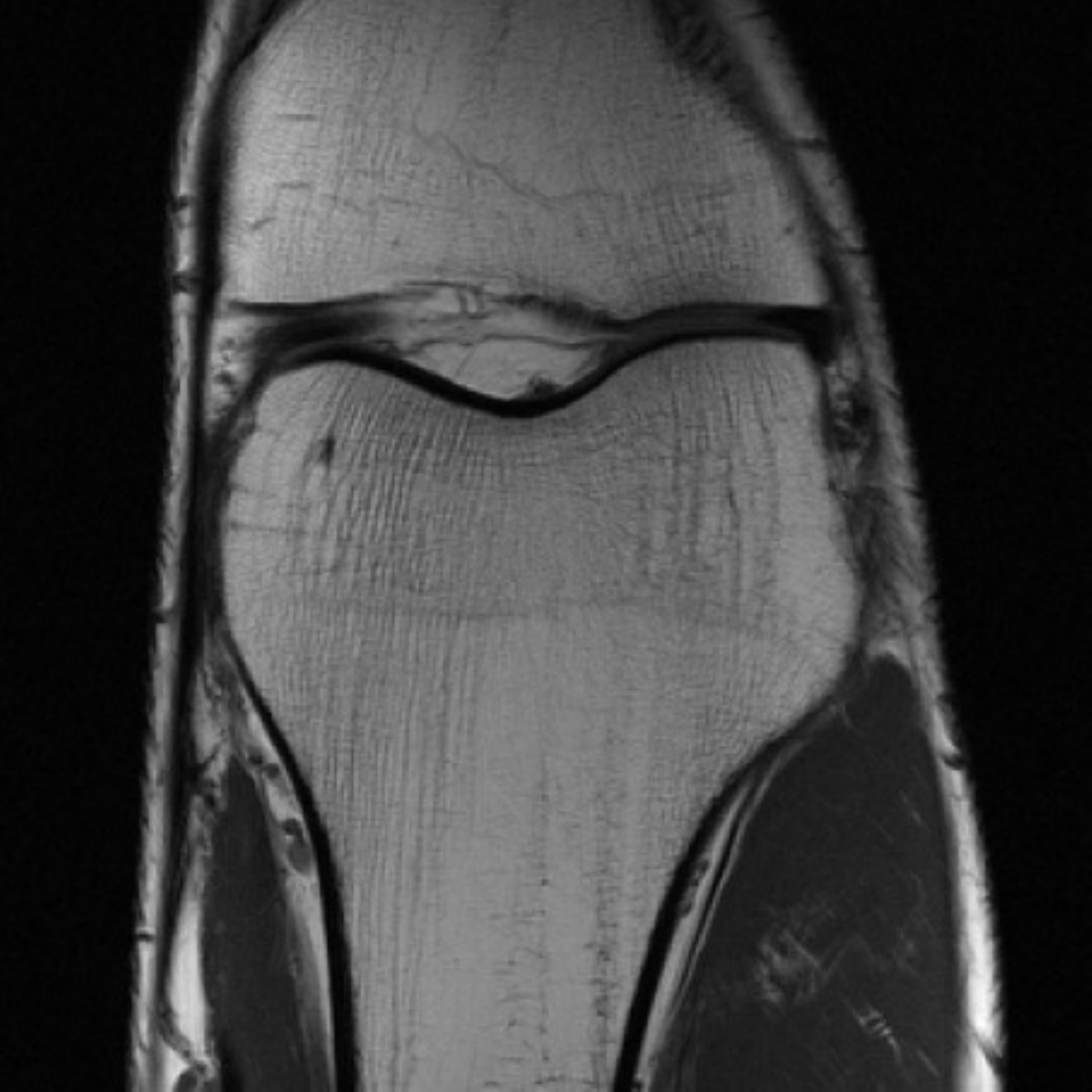}}
  \end{subfigure}
\caption{
\textbf{Anatomy shift from brain to knee.} End-to-end variational network (VarNet) and U-net generate vertical reconstruction artifacts in the lower side of the image when being trained on brain and tested on knee. Reconstructions are acquired from 4x under-sampled measurements.
}
\label{fig:dshift-sample-recs-knee}
\end{figure*}

We saw in Section~\ref{sec:data-shift} that a dataset shift (from brain to knee images and vice versa) causes a performance loss for all reconstruction methods. 
In Figure~\ref{fig:dshift-sample-recs} in the main body, we provide sample reconstructions when shifting the domain from knee to brain images which show that trained neural networks tend to generated artifacts with such a shift. 
In Figure~\ref{fig:dshift-sample-recs-knee}, we provide additional sample reconstruction when shifting the domain from brain to knee images. As shown, trained neural networks generate artifacts in this case as well.

Figure~\ref{fig:dataset-shift-unseen} in Section~\ref{sec:data-shift} shows that reconstruction quality degrades with an anatomy shift. 
To demonstrate this, we plot in panel 1 of Figure~\ref{fig:dataset-shift-unseen} the SSIM reconstruction scores of all methods when trained (or tuned) on knee images and tested on brain images, compared to the case that they are trained and tested on brain images. 

\begin{figure}[ht!]
\begin{center}
\begin{tikzpicture}

\begin{groupplot}[
y tick label style={/pgf/number format/.cd,fixed,precision=4},
scaled y ticks = false, xticklabel style={
        /pgf/number format/fixed,
        /pgf/number format/precision=2
},
legend style={at={(1.75,1)} , nodes={scale=0.75}, draw={none}, fill = none, text opacity=1,
/tikz/every even column/.append style={column sep=-0.1cm}
 },
         group
         style={group size= 2 by 1, xlabels at=edge bottom, ylabels at=edge left,
         horizontal sep=2cm, vertical sep=1.9cm,
         }, 
         width=0.37\textwidth,height=0.24\textwidth,
         scaled x ticks=false,
         legend cell align=left,
         ]
\nextgroupplot[ylabel style={align=center},xlabel style={align=center},xlabel={\footnotesize SSIM on knee \\ (trained on knee)}, ylabel={\footnotesize SSIM on brain \\ (trained on knee)},xmax = 0.88,ymax = 0.95,ymin = 0.67,xmin = 0.65,]
    \addplot +[mark=none,steelblue,thick] table[x=x,y=linfit]{./files/anatomy_shift_train_knee.txt};
	\addplot +[mark=none,dashed,red!70,thick] table[x=x,y=x]{./files/anatomy_shift_train_knee.txt};
	\addplot +[scatter, only marks, steelblue!60, scatter src = explicit symbolic,mark size=1pt,scatter/classes={
            a={mark=*,mark options={solid,draw=none,fill=fgreen}, fgreen, mark size = 1.8pt},
            b={mark=*,mark options={solid,draw=none,fill=blue},blue,mark size = 1.8pt},
            c={mark=*,mark options={solid,draw=none,fill=red},red,mark size = 1.8pt},
            d={mark=*,mark options={solid,draw=none,fill=black},black,mark size = 1.8pt}
        },
              error bars/.cd, 
              error bar style={line width=0.6pt},
              y dir=both, x dir=both,
              x explicit, y explicit
              ]
       table [x=x, y=y, x error=xerr, y error=yerr, meta=class]{./files/anatomy_shift_train_knee.txt};
	%
\nextgroupplot[ylabel style={align=center},xlabel style={align=center},xlabel={\footnotesize SSIM on brain \\ (trained on brain)}, ylabel={\footnotesize SSIM on knee \\ (trained on brain)},xmax = 0.97,ymax = 0.97,ymin = 0.6,xmin = 0.73,no markers]
    \addplot +[mark=none,steelblue,thick] table[x=x,y =linfit]{./files/anatomy_shift_train_brain.txt};
	\addplot +[mark=none,dashed,red!70,thick] table[x=x, y=x]{./files/anatomy_shift_train_brain.txt};
	\addplot +[scatter, only marks,steelblue!60,scatter src = explicit symbolic, mark size=1pt,scatter/classes={
            a={mark=*,mark options={solid,draw=none,fill=fgreen}, fgreen, mark size = 1.6pt},
            b={mark=*,mark options={solid,draw=none,fill=blue},blue,mark size = 1.6pt},
            c={mark=*,mark options={solid,draw=none,fill=red},red,mark size = 1.6pt},
            d={mark=*,mark options={solid,draw=none,fill=black},black,mark size = 1.6pt}
        },
              error bars/.cd, 
              error bar style={line width=0.6pt},
              y dir=both, x dir=both,
              x explicit, y explicit
              ]
       table [x=x, y=y, x error=xerr, y error=yerr, meta= class]
       {./files/anatomy_shift_train_brain.txt};
       \legend{best linear fit,$y=x$,$\ell_1$ group,un-trained group,U-net group,VarNet group}
\end{groupplot}
\end{tikzpicture}
\captionsetup{belowskip=0pt}
\caption{\textbf{Anatomy shift evaluation.}
Another visualization of the results in Fig.~\ref{fig:dataset-shift-unseen}: in each plot, we consider the same model trained on the same dataset, but evaluate on different test sets (knee and brain).
Validation results for a group of 100 brain images and a group of 100 knee images from the fastMRI validation set show that trained an un-trained models achieve higher scores for brain images.
}
\label{fig:dataset-shift-fixed-train}
\end{center}
\end{figure}
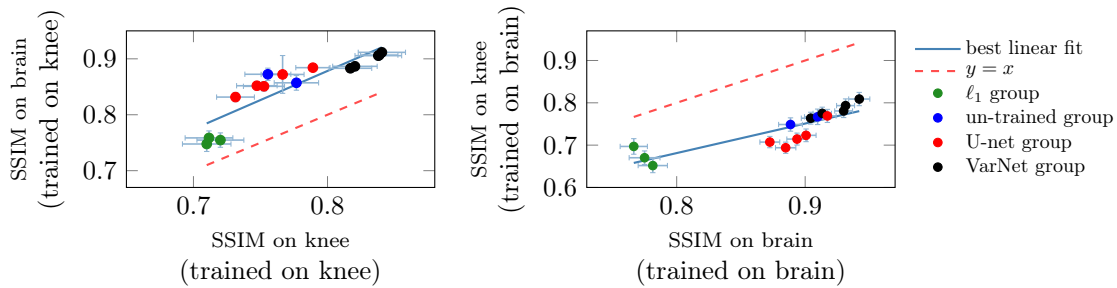

Another way to visualize the exact same data is to plot the SSIM reconstruction scores of a model trained on knee images evaluated on brains over the reconstruction scores of the same model evaluated on knees. This alternative evaluation can be found in Figure~\ref{fig:dataset-shift-fixed-train}. 
The Figure shows, in addition to our previous findings, that brain images are naturally easier to reconstruct than knee images as all models achieve higher scores for them.


\subsection{
Details on the adversarially-filtered shift
}
\label{sec:fastmriA-samples}

In Section~\ref{sec:adv-filt-shift}, we showed how a group of images can be naturally difficult to reconstruct for all image reconstruction methods. We further hypothesized that these samples are challenging due to less concentration of energy in the center region of their $k$-space, which corresponds to low-frequency components of the image. In this section, we  confirm this hypothesis and then depict a number of samples from the fastMRI-A dataset.

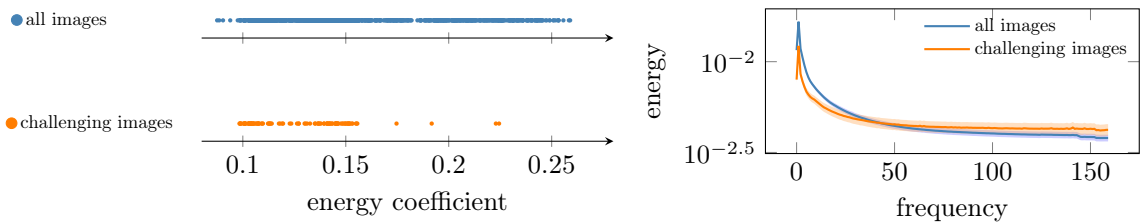
\begin{figure}[th!]
\begin{center}
\begin{tikzpicture}
\begin{groupplot}[
legend style={draw=none, nodes={scale=0.7}, 
/tikz/every even column/.append style={column sep=-0.1cm}
 },
    group style={group size=1 by 2,
                 group name=coefs,
                 xlabels at=edge bottom,
                 ylabels at=edge left,
                 yticklabels at=edge left,
                 xticklabels at= edge bottom,
                 horizontal sep=0.15cm, vertical sep=0.9cm,
                },
    xmax=0.28, xmin=0.08,
    width=7cm,
    height=2.05cm,]
\nextgroupplot[axis y line=none,axis x line=bottom,legend style={at={(-0.2,1)}}]
    \addlegendimage{steelblue,mark=*,only marks}
	\addlegendentry{all images}
    \addplot +[mark=*,draw=none,mark options={solid,fill=steelblue,scale=0.2},steelblue,thick] table[x index=0,y index=1]{./energy/coef_all.csv};
    
\nextgroupplot[axis y line=none,axis x line=bottom,xlabel={energy coefficient},legend style={at={(-0.04,1)}}]
    \addlegendimage{amber,mark=*,only marks}
	\addlegendentry{challenging images}
    \addplot +[mark=*,draw=none,mark options={solid,fill=amber,scale=0.2},amber,thick] table[x index=0,y index=1]{./energy/coef_challenging.csv};
\end{groupplot}
\begin{groupplot}[
legend style={draw=none, nodes={scale=0.7},fill=none, 
/tikz/every even column/.append style={column sep=-0.1cm}
 },
    group style={group size=1 by 1,
                 xlabels at=edge bottom,
                 ylabels at=edge left,
                 yticklabels at=edge left,
                 xticklabels at= edge bottom,
                 horizontal sep=0.15cm, vertical sep=0.4cm,
                },
    width=6.5cm,
    height=3.5cm,]
\nextgroupplot[anchor=south west, at={($(coefs c1r1.north east) + (2cm,-2cm)$)},legend style={at={(1,1)}},legend cell align=left,xlabel={frequency},ylabel={energy},ymode=log, scaled y ticks=false,]
    \addplot +[mark=none,steelblue,thick] table[x index=0,y index=1]{./energy/energy_all.csv};
    \addlegendentry{all images}
    
    \addplot +[mark=none,amber,thick] table[x index=0,y index=1]{./energy/energy_challenging.csv};
    \addlegendentry{challenging images}
    
    \addplot +[name path=upper,draw=none, mark=none] table[x index=0,y index = 2] {./energy/energy_all.csv};
    \addplot +[name path=lower,draw=none,mark=none] table[x index=0,y index = 3] {./energy/energy_all.csv};
    \addplot +[fill=blue!50,opacity=0.4] fill between[of=upper and lower];
    
    \addplot +[name path=upper,draw=none, mark=none] table[x index=0,y index = 2] {./energy/energy_challenging.csv};
    \addplot +[name path=lower,draw=none,mark=none] table[x index=0,y index = 3] {./energy/energy_challenging.csv};
    \addplot +[fill=amber!60,opacity=0.4] fill between[of=upper and lower];
\end{groupplot}
\end{tikzpicture}
\caption{
Difficult-to-reconstruct images contain more energy on the high frequencies as the average fastMRI image. 
\textbf{Left:} 
Scatter plot of the fraction of signal energy concentrated on the low, fully-sampled, center frequencies over the energy of the whole $k$-space. 
Difficult-to-reconstruct images have smaller such coefficient, indicating that more energy is on the high frequencies, which are only sub-sampled.
\textbf{Right:} 
Amount of energy as a function of the frequency along a vertical line in the $k$-space (in order to be consistent with the sampling trajectory). The cross between the two lines at frequency = 45 shows the difficult-to-reconstruct images have more energy on the high frequencies, relative to the average fastMRI image.
}
\label{fig:spectrum}
\end{center}
\end{figure}

In order to demonstrate that challenging samples are difficult to reconstruct because of the energy distribution in their frequency-domain representation, we performed the following experiment. We first define the low-frequency proportion of an image as 
\[\text{low-frequency proportion}= \frac{\text{energy of the center frequencies}}{\text{energy of the whole }k\text{-space}}.
\]
An image with a high low-frequency proportion has most energy in the low-frequency component and is therefore relatively smooth, has less detail, and is relatively easy to reconstruct. Figure~\ref{fig:spectrum}, left panel, shows that the challenging images indeed have a low low-frequency proportion, i.e., most challenging images have much of their energy concentrated in high frequency components, which are sub-sampled.

Second, we calculate the energy in the frequency domain as a function of the frequency along a vertical line in the $k$-space. The figure shows again that challenging images have much of it's energy concentrated on high frequencies.

From this we can conclude that difficult-to-reconstruct samples are difficult to reconstruction because relative to other samples, (i) they contain less energy concentrated on their low (center) frequencies, and (ii) they contain more high-frequency information which increases the probability of not being recorded during the scanning process.


\begin{figure*}[th!]
\centering
  \begin{subfigure}[t]{0.19\textwidth}
  \caption*{$\ell_1$ minimization}
  \centering\scalebox{1}[-1]{\includegraphics[scale=0.36]{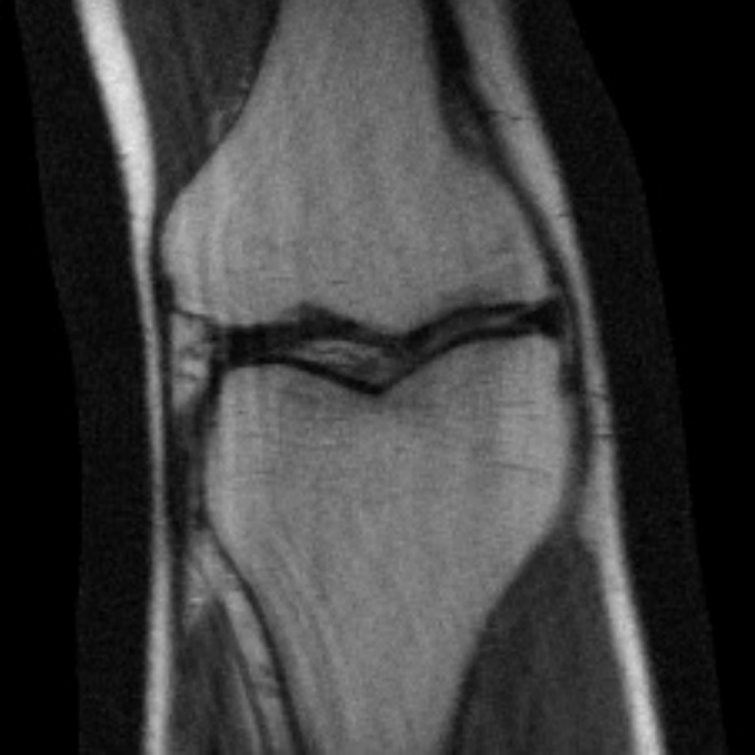}}
  \end{subfigure}
  \begin{subfigure}[t]{0.19\textwidth}
  \caption*{U-net}
  \centering\scalebox{1}[-1]{\includegraphics[scale=0.36]{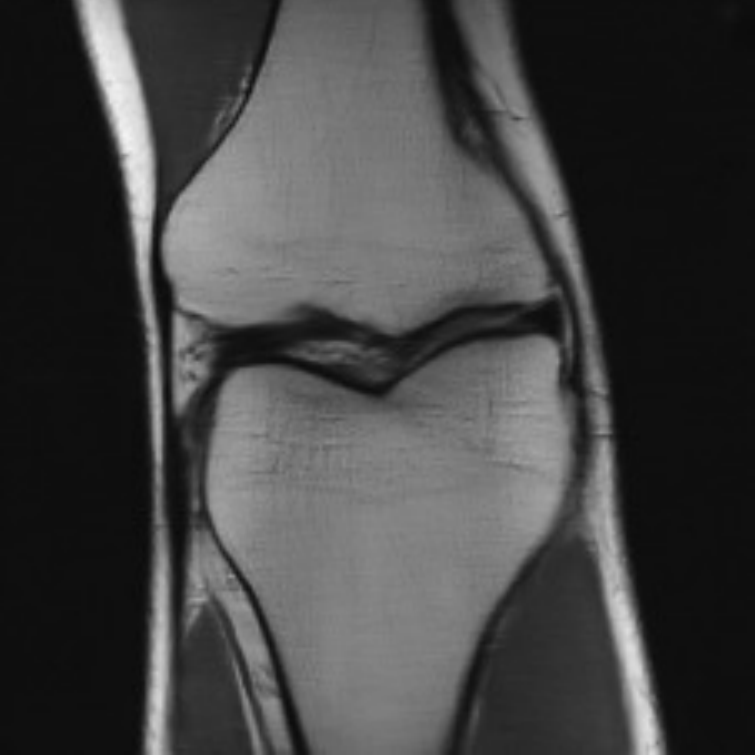}}
  \end{subfigure}
  \begin{subfigure}[t]{0.19\textwidth}
  \caption*{VarNet}
  \centering\scalebox{1}[-1]{\includegraphics[scale=0.36]{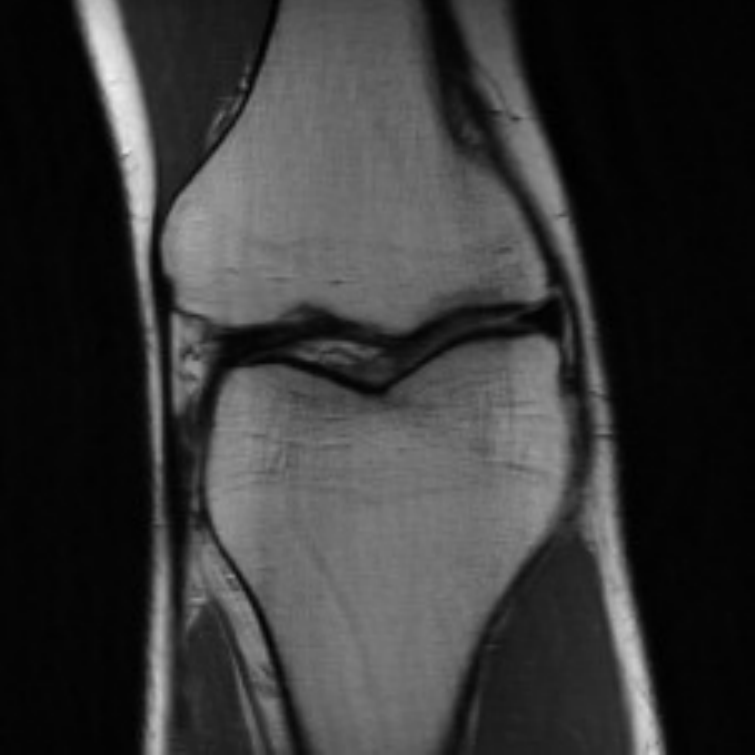}}
  \end{subfigure}
  \begin{subfigure}[t]{0.19\textwidth}
  \caption*{ConvDecoder}
  \centering\scalebox{1}[-1]{\includegraphics[scale=0.36]{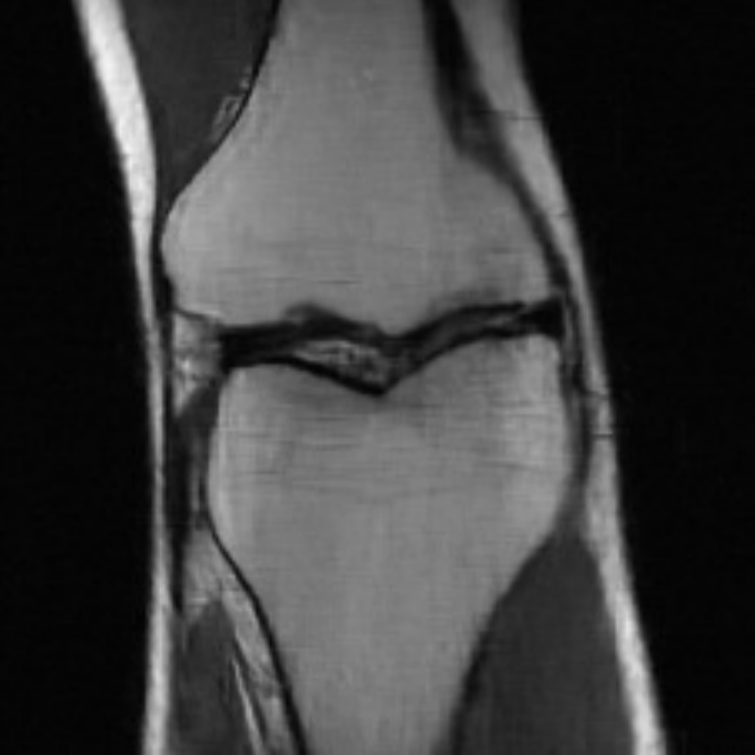}}
  \end{subfigure}
  \begin{subfigure}[t]{0.19\textwidth}
  \caption*{ground truth}
  \centering\scalebox{1}[-1]{\includegraphics[scale=0.36]{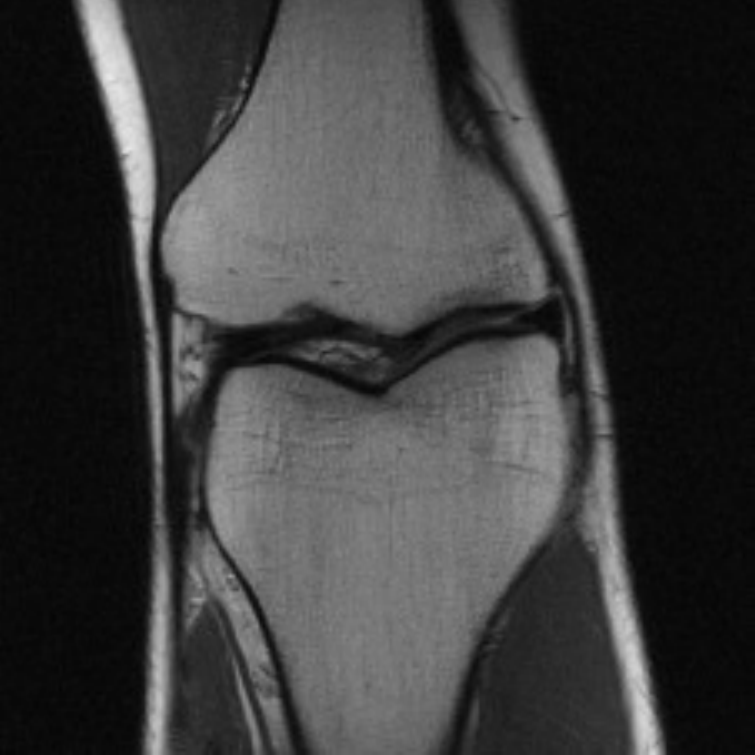}}
  \end{subfigure}\par\medskip
  \begin{subfigure}[t]{0.19\textwidth}
  \centering\scalebox{1}[-1]{\includegraphics[scale=0.36]{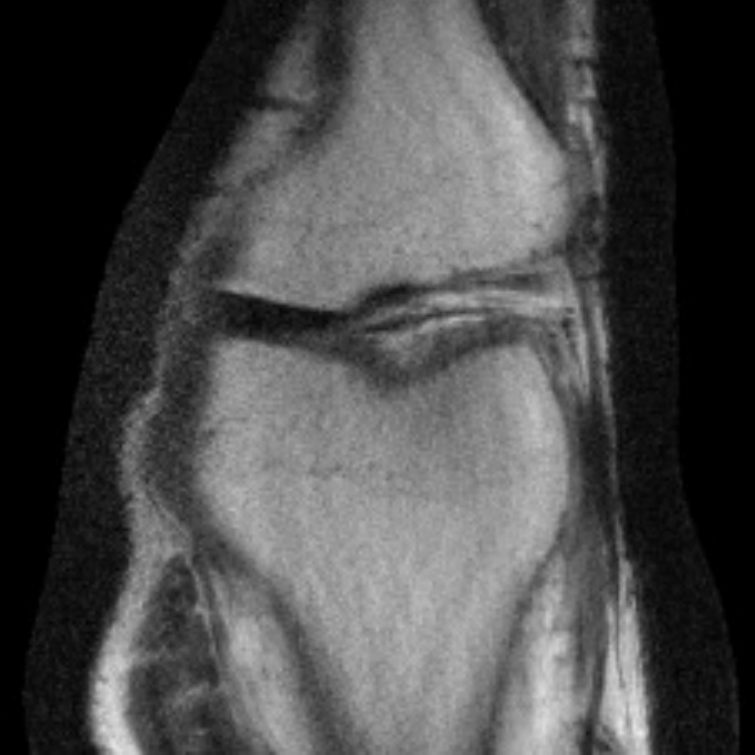}}
  \end{subfigure}
  \begin{subfigure}[t]{0.19\textwidth}
  \centering\scalebox{1}[-1]{\includegraphics[scale=0.36]{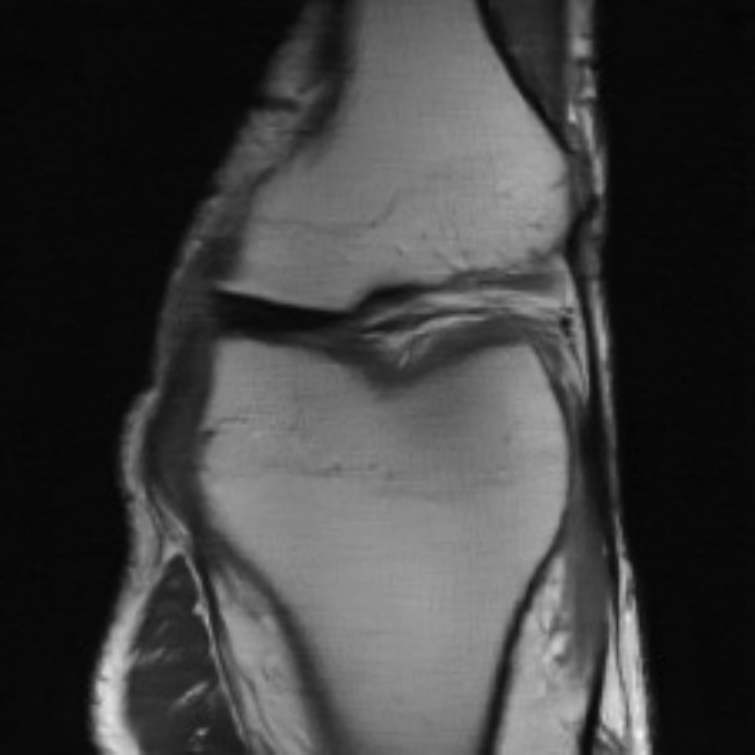}}
  \end{subfigure}
  \begin{subfigure}[t]{0.19\textwidth}
  \centering\scalebox{1}[-1]{\includegraphics[scale=0.36]{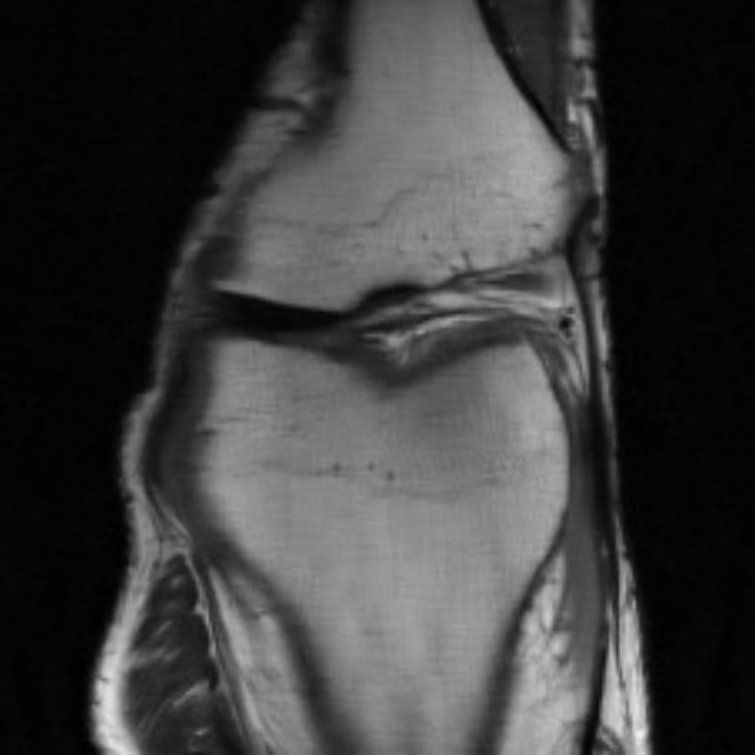}}
  \end{subfigure}
  \begin{subfigure}[t]{0.19\textwidth}
  \centering\scalebox{1}[-1]{\includegraphics[scale=0.36]{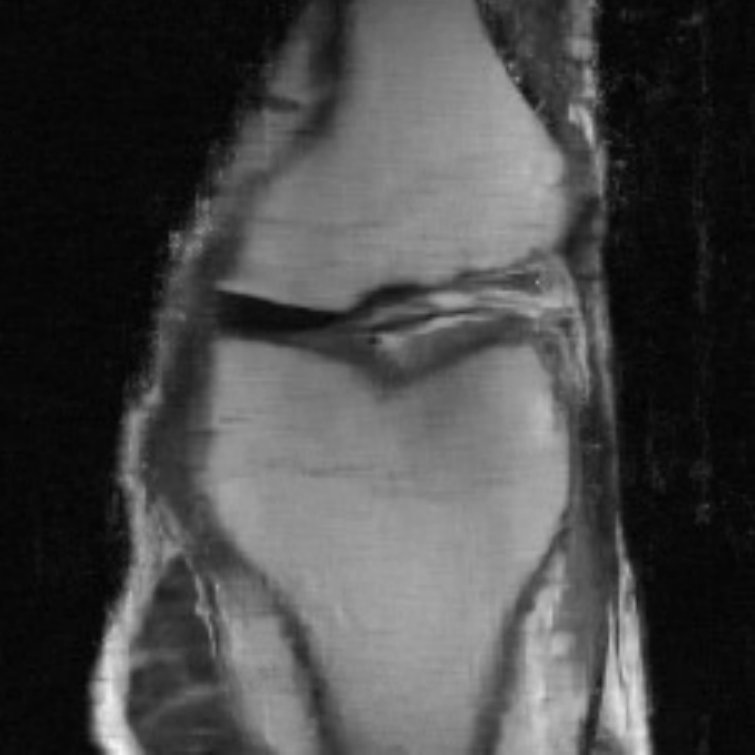}}
  \end{subfigure}
  \begin{subfigure}[t]{0.19\textwidth}
  \centering\scalebox{1}[-1]{\includegraphics[scale=0.36]{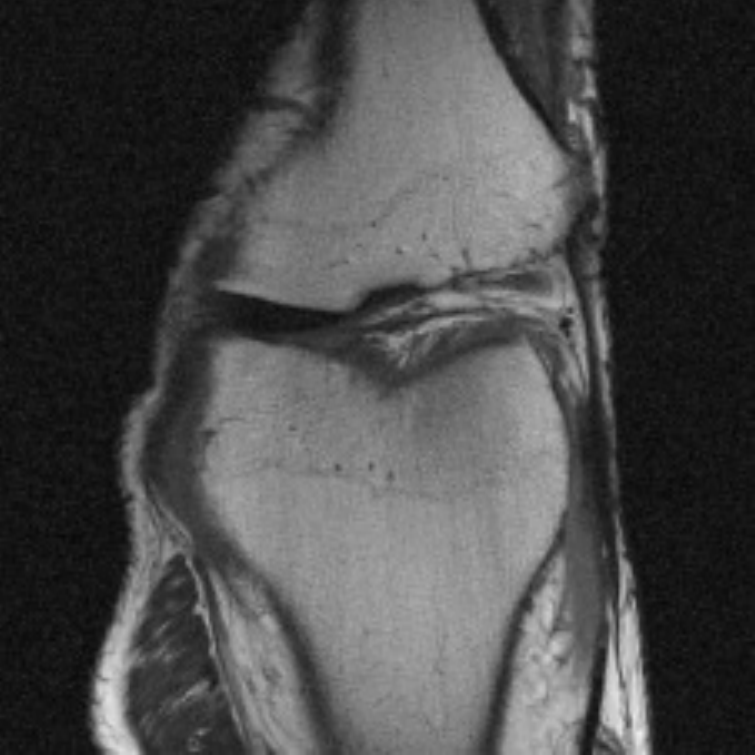}}
  \end{subfigure}\par\medskip 
  \begin{subfigure}[t]{0.19\textwidth}
  \centering\scalebox{1}[-1]{\includegraphics[scale=0.36]{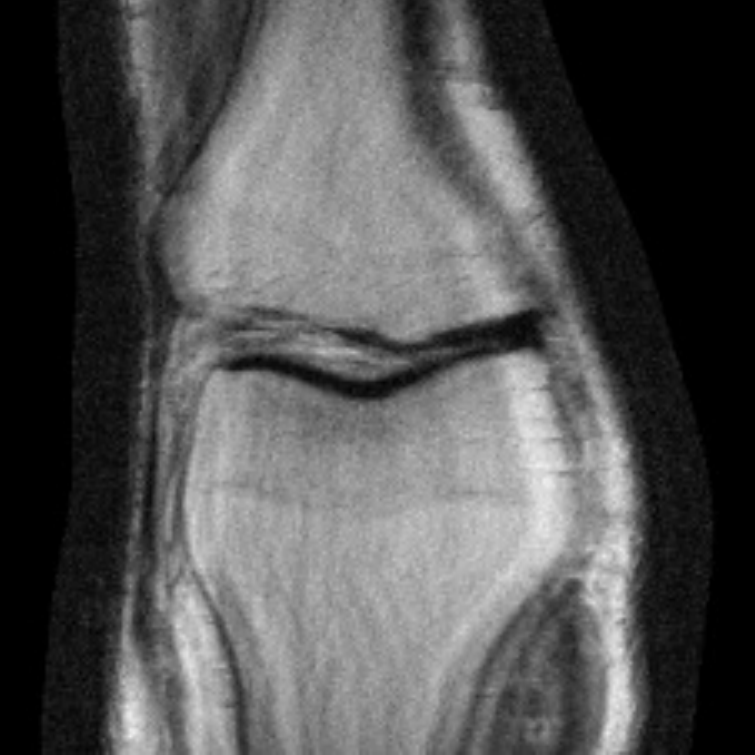}}
  \end{subfigure}
  \begin{subfigure}[t]{0.19\textwidth}
  \centering\scalebox{1}[-1]{\includegraphics[scale=0.36]{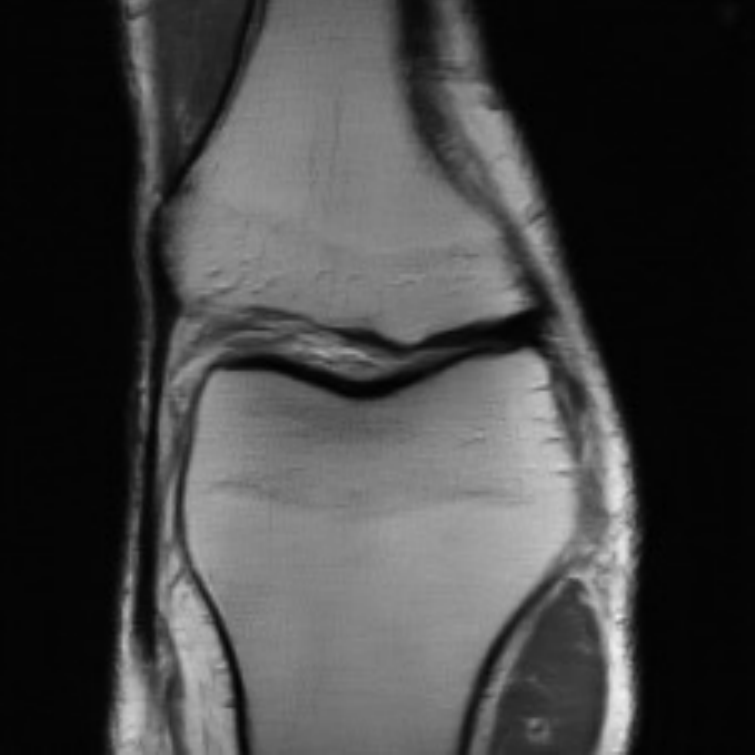}}
  \end{subfigure}
  \begin{subfigure}[t]{0.19\textwidth}
  \centering\scalebox{1}[-1]{\includegraphics[scale=0.36]{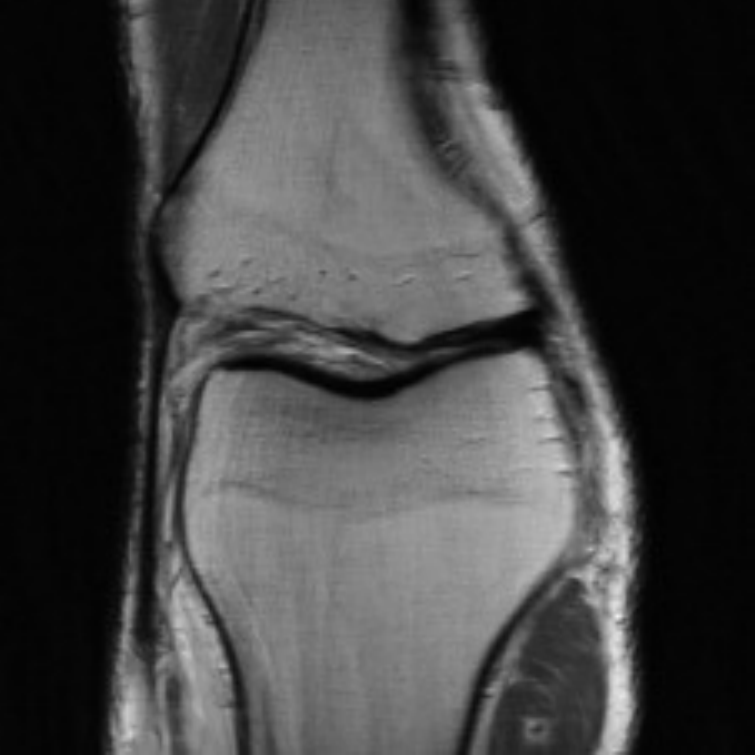}}
  \end{subfigure}
  \begin{subfigure}[t]{0.19\textwidth}
  \centering\scalebox{1}[-1]{\includegraphics[scale=0.36]{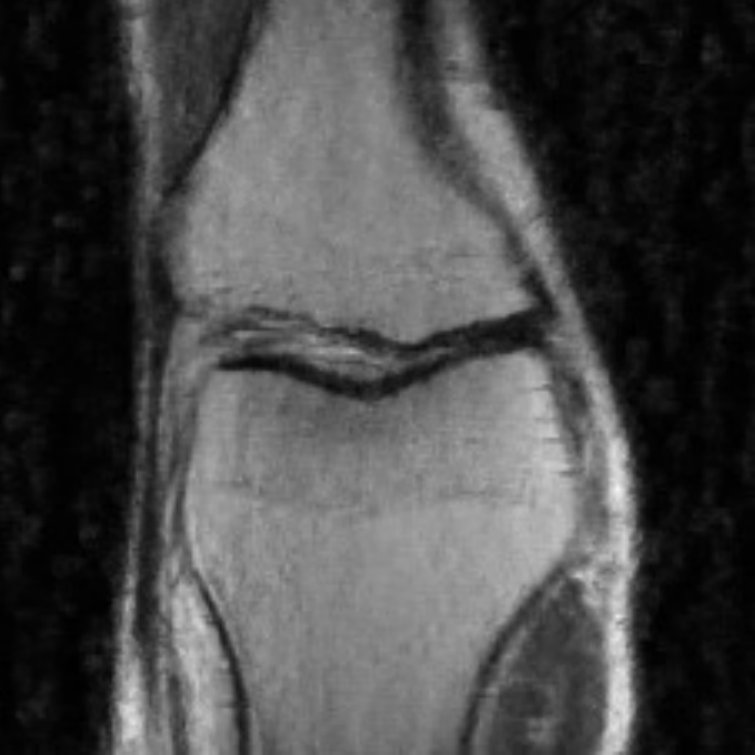}}
  \end{subfigure}
  \begin{subfigure}[t]{0.19\textwidth}
  \centering\scalebox{1}[-1]{\includegraphics[scale=0.36]{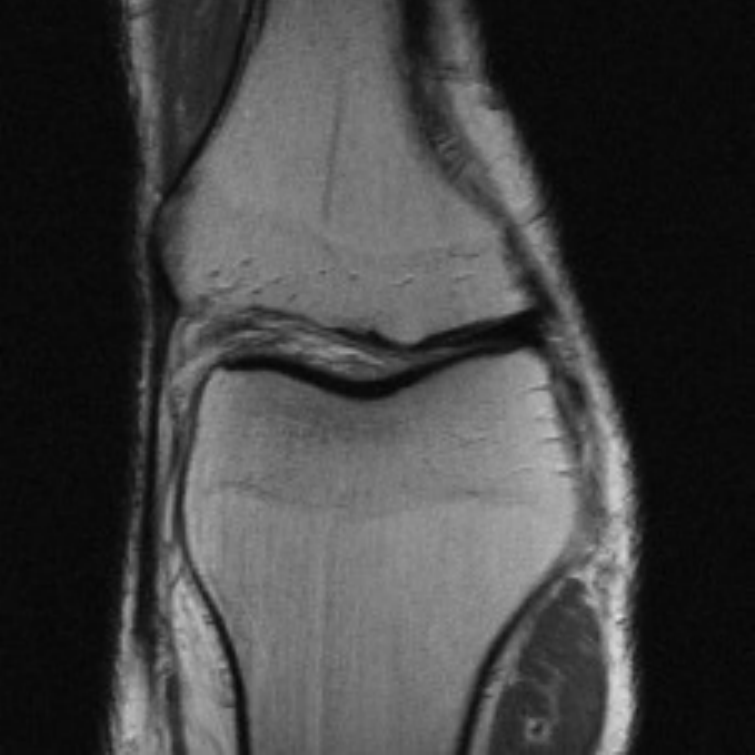}}
  \end{subfigure}\par\medskip
  \begin{subfigure}[t]{0.19\textwidth}
  \centering\scalebox{1}[-1]{\includegraphics[scale=0.36]{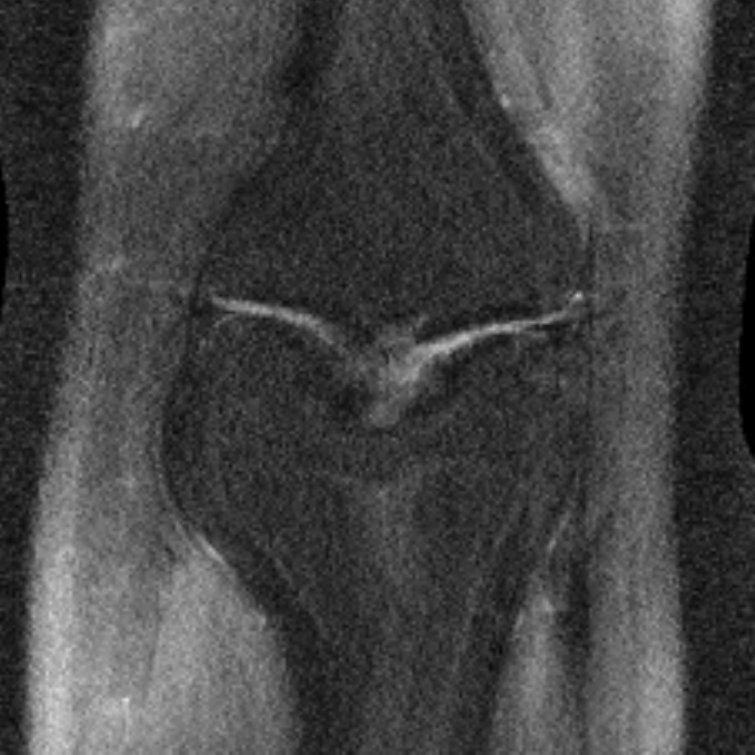}}
  \end{subfigure}
  \begin{subfigure}[t]{0.19\textwidth}
  \centering\scalebox{1}[-1]{\includegraphics[scale=0.36]{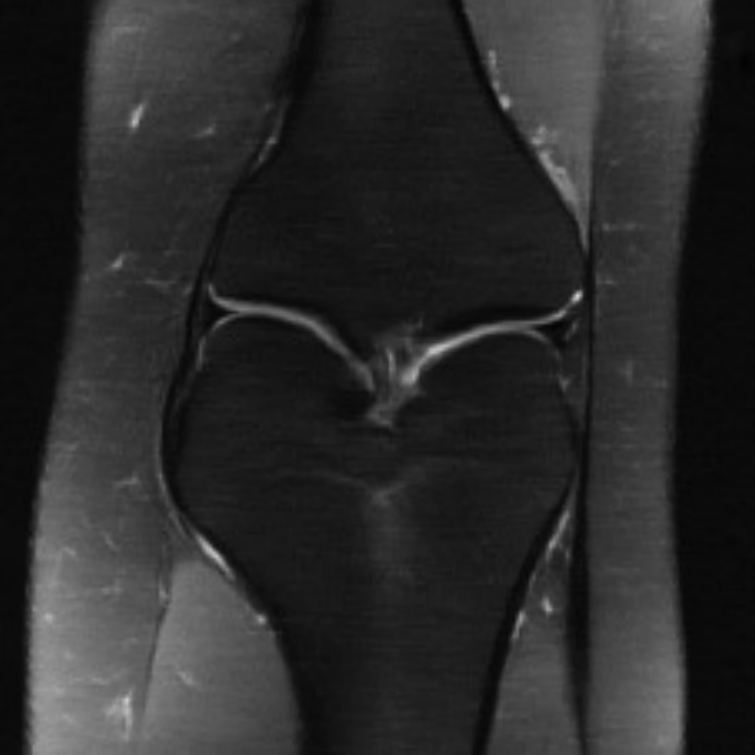}}
  \end{subfigure}
  \begin{subfigure}[t]{0.19\textwidth}
  \centering\scalebox{1}[-1]{\includegraphics[scale=0.36]{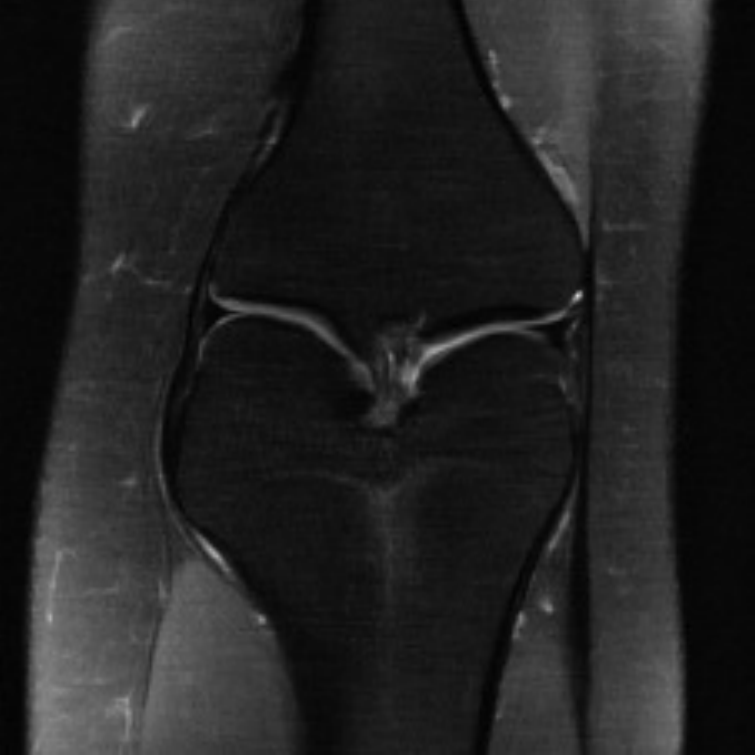}}
  \end{subfigure}
  \begin{subfigure}[t]{0.19\textwidth}
  \centering\scalebox{1}[-1]{\includegraphics[scale=0.36]{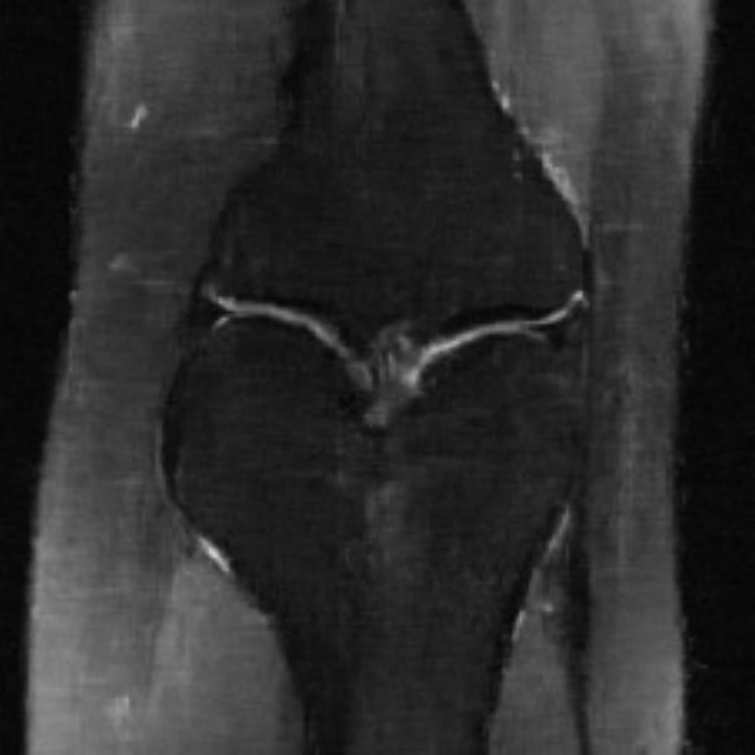}}
  \end{subfigure}
  \begin{subfigure}[t]{0.19\textwidth}
  \centering\scalebox{1}[-1]{\includegraphics[scale=0.36]{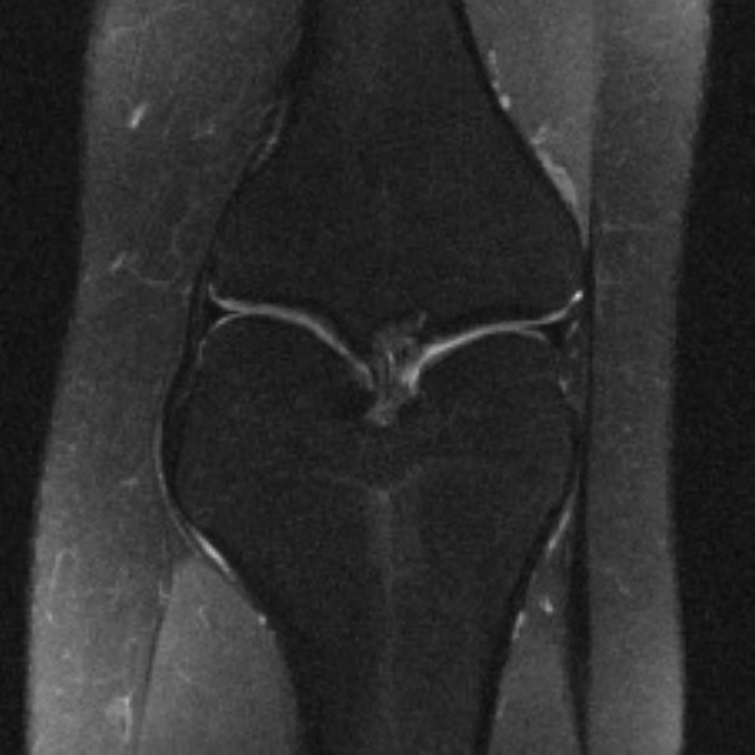}}
  \end{subfigure}\par\medskip
  \begin{subfigure}[t]{0.19\textwidth}
  \centering\scalebox{1}[-1]{\includegraphics[scale=0.36]{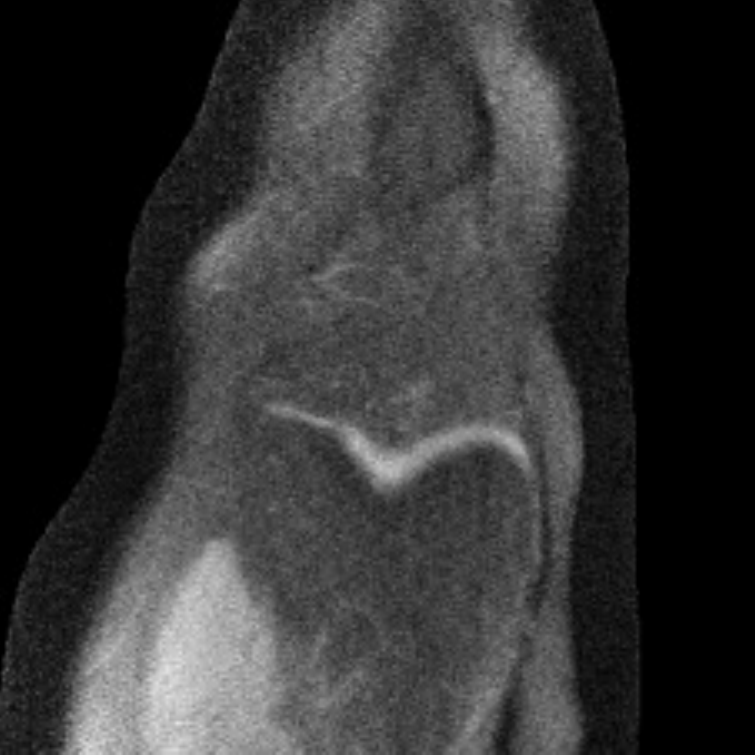}}
  \end{subfigure}
  \begin{subfigure}[t]{0.19\textwidth}
  \centering\scalebox{1}[-1]{\includegraphics[scale=0.36]{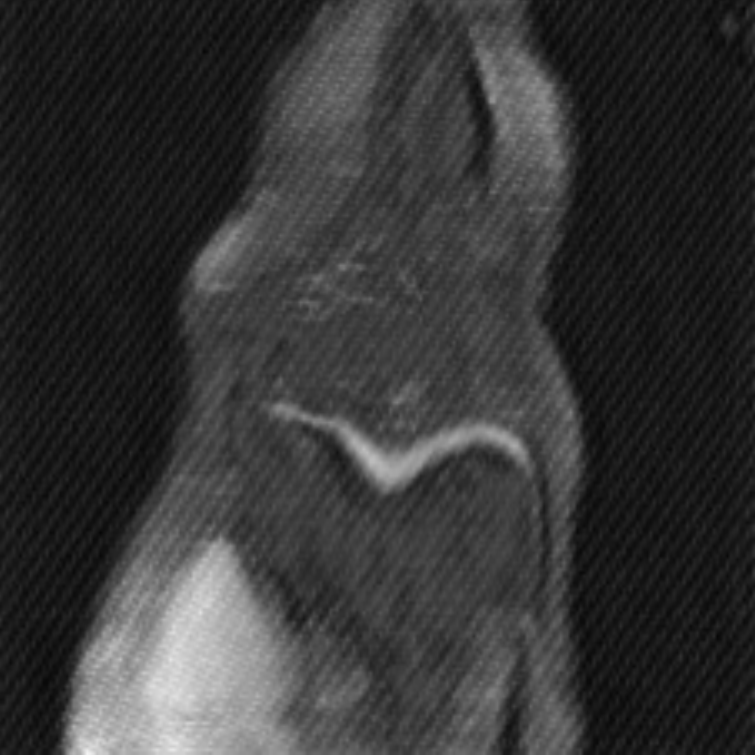}}
  \end{subfigure}
  \begin{subfigure}[t]{0.19\textwidth}
  \centering\scalebox{1}[-1]{\includegraphics[scale=0.36]{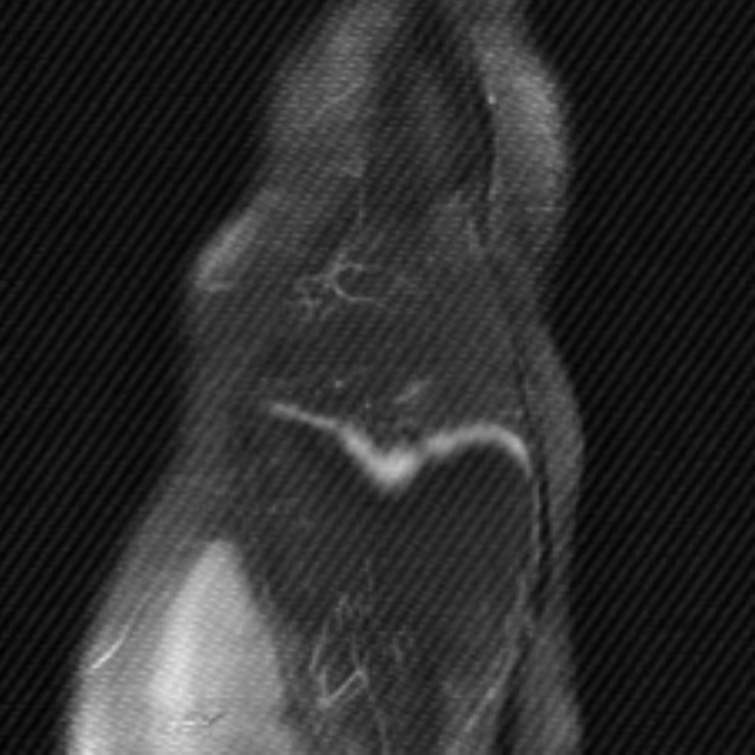}}
  \end{subfigure}
  \begin{subfigure}[t]{0.19\textwidth}
  \centering\scalebox{1}[-1]{\includegraphics[scale=0.36]{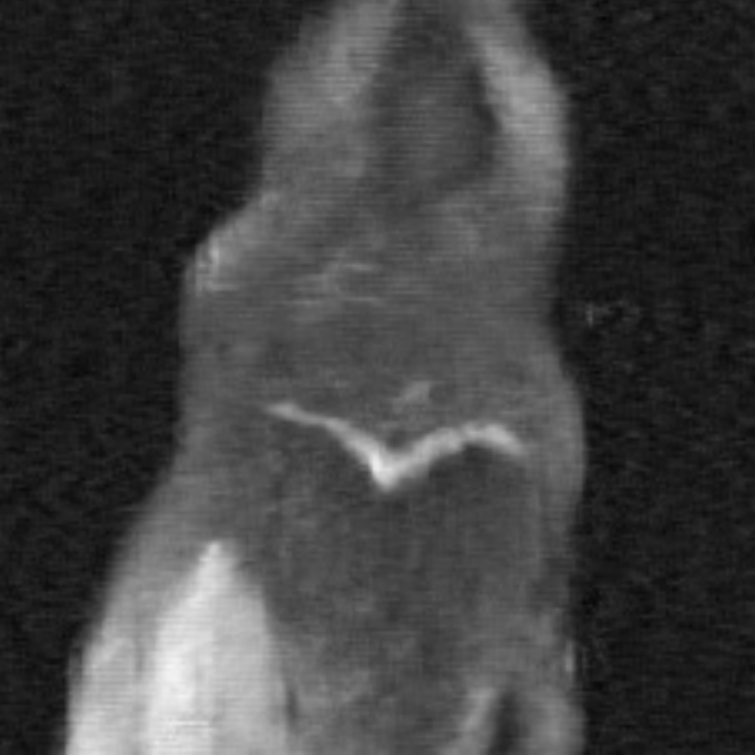}}
  \end{subfigure}
  \begin{subfigure}[t]{0.19\textwidth}
  \centering\scalebox{1}[-1]{\includegraphics[scale=0.36]{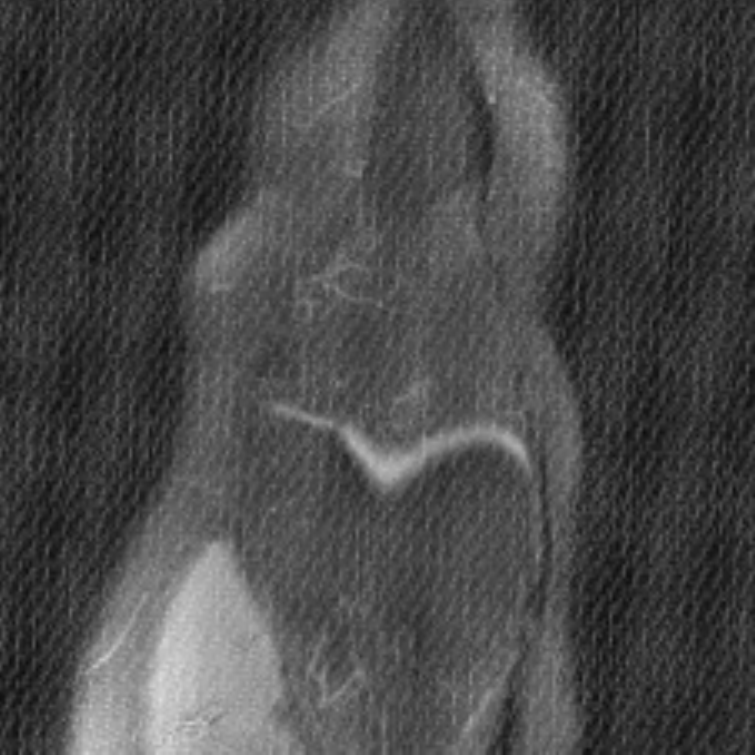}}
  \end{subfigure}
\caption{Samples from the fastMRI-A dataset and their corresponding reconstructions by $\ell_1$-norm minimization, ConvDecoder, U-net, and the end-to-end variational network (VarNet).
}
\label{fig:fastmriA_samples}
\end{figure*}

Finally, to give a visual impression of the fastMRI-A dataset, we provide a number of samples along with their 4x-accelerated reconstructions by the considered methods ($\ell_1$-norm minimization, ConvDecoder, U-net, and the end-to-end variational network (VarNet)), in Figure~\ref{fig:fastmriA_samples}. The figure shows that fastMRI-A contains samples that are naturally difficult to reconstruct. However, there are a few samples whose ground-truth image is corrupted with measurement noise. Consequently, all reconstruction methods yield low scores for such samples.



\section{Details on recovering small features in an image}\label{sec:small-feature-app-op}
In Section~\ref{sec:small-features}, we showed that (i) there is a location dependence associated with each reconstruction method when recovering small features, and (ii) reconstruction quality is correlated with small feature recovery error when evaluating different reconstruction methods on images which contain real-world features. In this section, we provide further details on recovering fine details based on the framework we proposed for determining the location dependence.

\subsection{Recovering small features via an un-trained network}
It was shown in Section~\ref{sec:small-features} that different reconstruction methods are sensitive to different regions in an image when recovering tiny details. 
According to Figure~\ref{fig:small_features}, the end-to-end variational network, U-net, and $\ell_1$-norm minimization all exhibit such location dependency. However, the un-trained network ConvDecoder shows a random dependence across the image when being optimized by the Adam optimizer~\citep{kingma2014adam}. In this section, we explain such behavior and provide empirical conditions under which ConvDecoder also shows dependency to different regions when recovering a small feature.

The choice of optimizer is critical for the performance of an un-trained neural network, as the interaction of optimizer with the network determines the prior~\cite{heckel2020denoising,heckel2020compressive}. 
For our setup, the Adam optimizer yields a significantly better reconstruction compared to Gradient Descent (GD), because Adam chooses learning rates associated with each layer in a way that imposes an effective prior, see~\citet{darestani2020can} for concrete experiments. 

Furthermore, when optimizing the network parameters over $n_{iter}$ number of iterations, we typically take the best-performing set of parameters.  The best-performing set is chosen based on the loss function which measures the closeness of network output and the under-sampled data.

These components of the method result in a potential randomness in recovering small feature at different locations in an image. To elaborate, the best-performing network at iteration $i$---according to the quality of the whole reconstructed image---may not have yet reconstructed the small feature. This can be mitigated by controlling the convergence smoothness, as we discuss next.

\begin{figure}[h!]
\begin{center}
\begin{tikzpicture}
\begin{groupplot}[
    group style={group size=3 by 1,
                group name = heat_plots,
                 xlabels at=edge bottom,
                 ylabels at=edge left,
                 yticklabels at=edge left,
                 xticklabels at= edge bottom,
                 horizontal sep=0.7cm, vertical sep=0.4cm,
                },
    width=4.3cm,
    height=5cm,
    colormap/hot2, 
    view={0}{270}]
\node[above left,label={[label distance=1pt]90:ground truth + feature}] (feature) at (9.6,-0.12) {\includegraphics[scale=0.36]{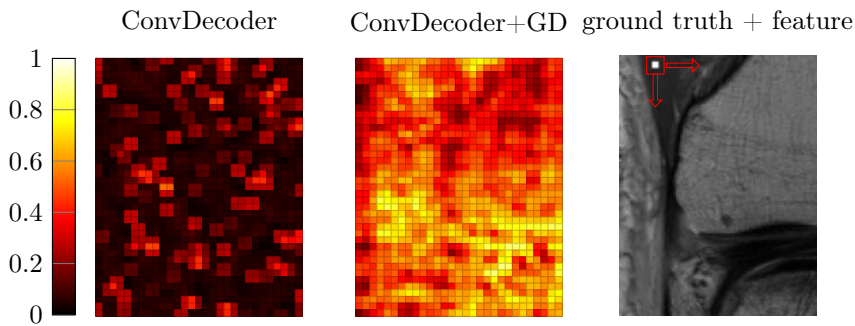}};
\nextgroupplot[colorbar left, colorbar style={at={(-0.1,1)},anchor=north east, width=0.3cm, height=3.4cm,}, yticklabels={,,},xticklabels={,,},tickwidth=0, title = ConvDecoder, title style={yshift=1.5pt}]
    \addplot3[surf] table {./files/heatc.txt};

\nextgroupplot[yticklabels={,,},xticklabels={,,},tickwidth=0,title = ConvDecoder+GD]
    \addplot3[surf] table {./files/heatc_gd.txt};
\end{groupplot}
\end{tikzpicture}
\caption{Smoother convergence brings location dependency at the cost losing average small-feature recovery performance. For Gradient Descent (GD), we use a linearly-increasing stepsize schedule.
}
\label{fig:small_features_app}
\end{center}
\end{figure}

There are numerous ways to achieve a smoother convergence for optimizing ConvDecoder. For example, one way is to choose a larger $\beta_1$ (we chose 0.98 instead of the default 0.9 value) for Adam. 
Alternatively, one can use GD instead of Adam for a smoother convergence. In this case, as shown in Figure~\ref{fig:small_features_app}, smoother convergence brings location dependency at the cost of losing small-feature-recovery performance (since the original Adam, on average performs significantly better than GD or Adam with larger $\beta_1$ for recovering small features).

\subsection{Quantitative analysis based on artificial features}
Based on our framework described in Section~\ref{sec:small-features}, we performed the following experiment to gain a quantitative understanding of artificial feature recovery, similar to the results in  Figure~\ref{fig:small-features-real} for real-world features. 

For all considered reconstruction methods, we put the fixed window of size $k$ on different locations and averaged the Mean-Squared Error (MSE) over 200 randomly-chosen images from the fastMRI knee validation set and also varied the window size in $\{2,3,4,5\}$.
Then, we randomly picked one of the images and put the fixed window on a random location and measured how much of it is lost during the reconstruction process. 
We performed this experiment 100 times for each reconstruction method to obtain an average information loss value as a metric for how reliable each method is in recovering fine details in an image.

\begin{figure}[ht!]
\begin{center}
\begin{tikzpicture}

\begin{groupplot}[
y tick label style={/pgf/number format/.cd,fixed,precision=4},
scaled y ticks = false, xticklabel style={
        /pgf/number format/fixed,
        /pgf/number format/precision=2
},
legend style={at={(1,1)} , nodes={scale=0.6}, draw={none}, fill = none, text opacity=1,
/tikz/every even column/.append style={column sep=-0.1cm}
 },
         group
         style={group size= 1 by 1, xlabels at=edge bottom, ylabels at=edge left,
         yticklabels at=edge left,
         xticklabels at=edge bottom,
         horizontal sep=0.5cm, vertical sep=1.5cm,
         }, 
         width=0.4\textwidth,height=0.26\textwidth,
         ylabel={error},
         scaled x ticks=false,
         ymax=0.14,
         ymin=0,
         xmax=5.4,
         legend cell align=left,
         ]
\nextgroupplot[title=Artificial features,xlabel={window size},]
	\addlegendentry{$\ell_1$}
	\addplot +[mark=none,fgreen,thick] table[x index=0,y expr=\thisrowno{1}/200]{./files/window.csv};
	
	\addlegendentry{VarNet}
	\addplot +[mark=none,black,thick] table[x index=0,y expr=\thisrowno{10}/200]{./files/window.csv};
	
	\addlegendentry{U-net}
	\addplot +[mark=none,red,thick] table[x index=0,y expr=\thisrowno{7}/200]{./files/window.csv};
	
	\addlegendentry{ConvDecoder}
	\addplot +[mark=none,blue,thick] table[x index=0,y expr=\thisrowno{4}/200]{./files/window.csv};
	
	\addplot +[name path=upper,draw=none, mark=none] table[x index=0,y expr=\thisrowno{2}/200] {./files/window.csv};
    \addplot +[name path=lower,draw=none,mark=none] table[x index=0,y expr=\thisrowno{3}/200] {./files/window.csv};
    \addplot +[fill=fgreen,opacity=0.3] fill between[of=upper and lower];
    
    \addplot +[name path=upper,draw=none, mark=none] table[x index=0,y expr=\thisrowno{5}/200] {./files/window.csv};
    \addplot +[name path=lower,draw=none,mark=none] table[x index=0,y expr=\thisrowno{6}/200] {./files/window.csv};
    \addplot +[fill,blue,opacity=0.3] fill between[of=upper and lower];
    
    \addplot +[name path=upper,draw=none, mark=none] table[x index=0,y expr=\thisrowno{8}/200] {./files/window.csv};
    \addplot +[name path=lower,draw=none,mark=none] table[x index=0,y expr=\thisrowno{9}/200] {./files/window.csv};
    \addplot +[fill=red,opacity=0.3] fill between[of=upper and lower];
    
    \addplot +[name path=upper,draw=none, mark=none] table[x index=0,y expr=\thisrowno{11}/200] {./files/window.csv};
    \addplot +[name path=lower,draw=none,mark=none] table[x index=0,y expr=\thisrowno{12}/200] {./files/window.csv};
    \addplot +[fill=black,opacity=0.3] fill between[of=upper and lower];
\end{groupplot}
\end{tikzpicture}
\caption{\textbf{Natural features are essential for evaluating robustness in recovering small features.} Normalized information recovery error based on feature size (note that window size = 2 represents a 2x2 feature). For all considered window sizes, the un-trained netowrk ConvDecoder performs best. The numbers are averaged over 100 runs on the 4x accelerated knee multi-coil measurements from the fastMRI dataset. 
}
\label{fig:small-features-app}
\end{center}
\end{figure}
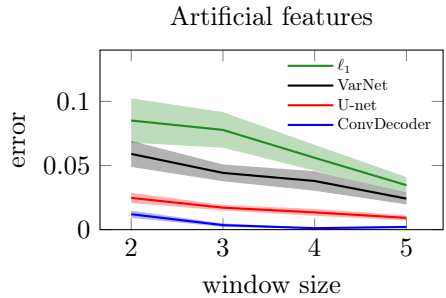

Figure~\ref{fig:small-features-app} shows the results for this experiment. By comparing this result to the one shown in Figure~\ref{fig:small-features-real} for real-world features, we see somewhat contradicting results: For example, VarNet performs much worse than ConvDecoder in recovering such details. 
This could be due to one of the following: Either artificial features are ineffective for evaluating robustness w.r.t. recovering fine details and real-world features are essential, or a more crafted procedure to synthesize these features is needed to generate real-world-like features. 

\subsection{Reconstruction examples for real-world small feature recovery}

As shown in Section~\ref{sec:small-features}, overall reconstruction quality correlates with accurate recovery of extremely small features in an image. In Figure~\ref{fig:path-sample-recs}, we provide reconstruction examples for three annotated images from the fastMRI validation set\footnote{Annotated slices are provided by \href{https://github.com/fcaliva/fastMRI_BB_abnormalities_annotation}{\color{black}https://github.com/fcaliva/fastMRI\_BB\_abnormalities\_annotation}}.

\begin{table*}[ht!]
\setlength{\tabcolsep}{1pt}
\centering
\begin{tabular}{cccccc}
  & $\ell_1$-minimization & U-net & VarNet & ConvDecoder & ground truth \\
  \rule{0pt}{8ex}
  &
  \scalebox{1}[1]{\includegraphics[width=0.16\textwidth,valign=m]{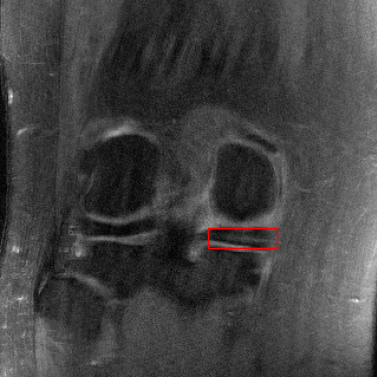}}&
  \scalebox{1}[1]{\includegraphics[width=0.16\textwidth,valign=m]{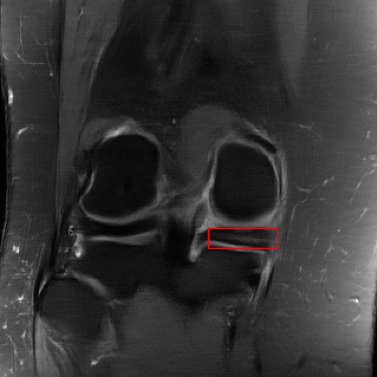}}&
  \scalebox{1}[1]{\includegraphics[width=0.16\textwidth,valign=m]{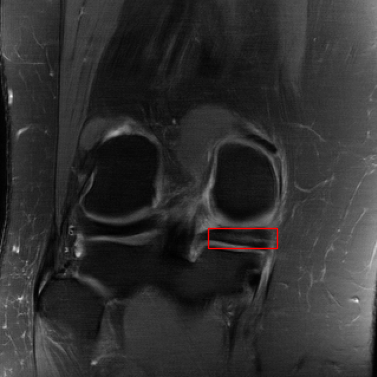}}&
  \scalebox{1}[1]{\includegraphics[width=0.16\textwidth,valign=m]{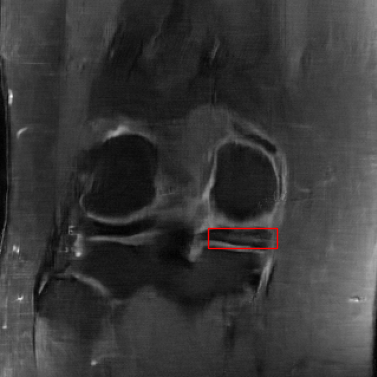}}&
  \scalebox{1}[1]{\includegraphics[width=0.16\textwidth,valign=m]{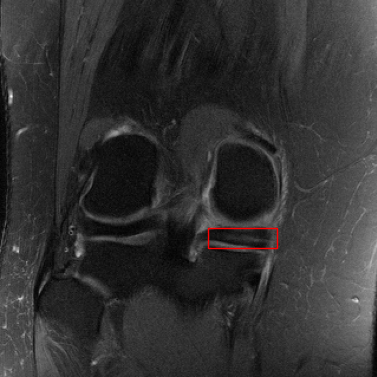}}\\
  \rule{0pt}{6.5ex}
  &
  \scalebox{1}[1]{\includegraphics[width=0.16\textwidth,valign=m]{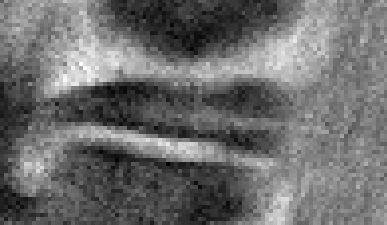}}&
  \scalebox{1}[1]{\includegraphics[width=0.16\textwidth,valign=m]{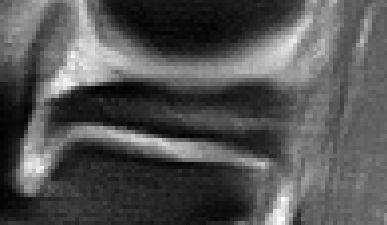}}&
  \scalebox{1}[1]{\includegraphics[width=0.16\textwidth,valign=m]{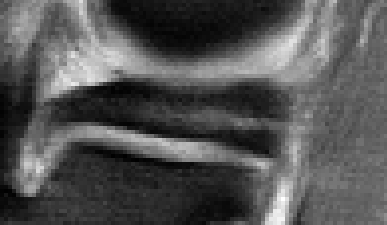}}&
  \scalebox{1}[1]{\includegraphics[width=0.16\textwidth,valign=m]{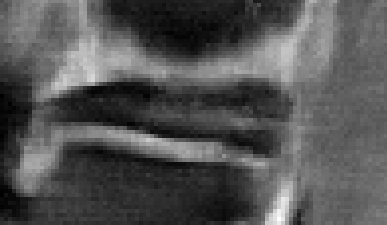}}&
  \scalebox{1}[1]{\includegraphics[width=0.16\textwidth,valign=m]{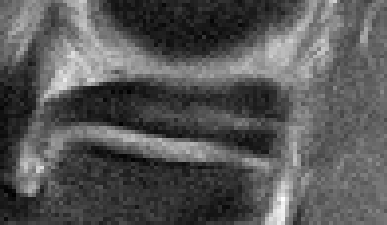}}\\
  \rule{0pt}{12ex}
  &
  \scalebox{1}[1]{\includegraphics[width=0.16\textwidth,valign=m]{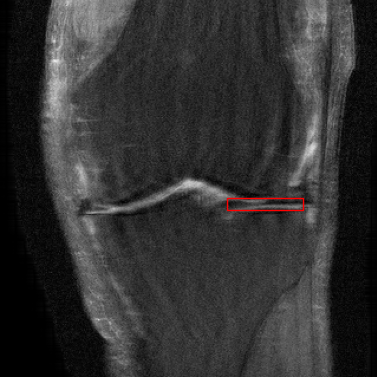}}&
  \scalebox{1}[1]{\includegraphics[width=0.16\textwidth,valign=m]{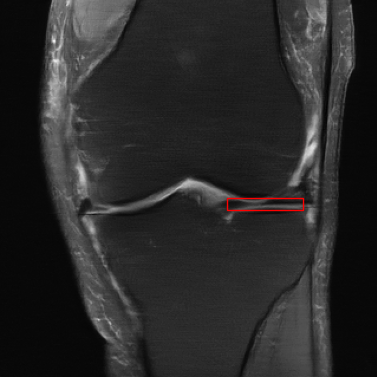}}&
  \scalebox{1}[1]{\includegraphics[width=0.16\textwidth,valign=m]{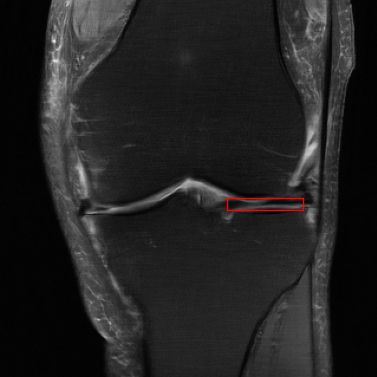}}&
  \scalebox{1}[1]{\includegraphics[width=0.16\textwidth,valign=m]{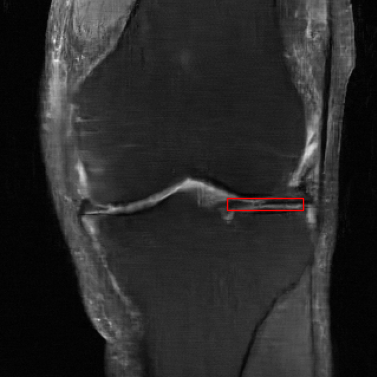}}&
  \scalebox{1}[1]{\includegraphics[width=0.16\textwidth,valign=m]{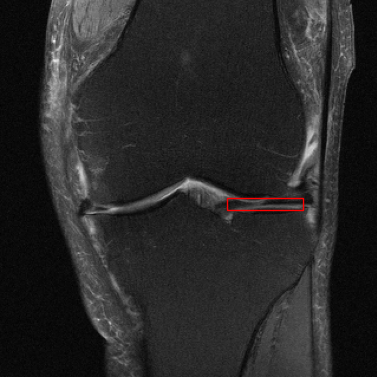}}\\
  \rule{0pt}{5.5ex}%
  &
  \scalebox{1}[1]{\includegraphics[width=0.16\textwidth,valign=m]{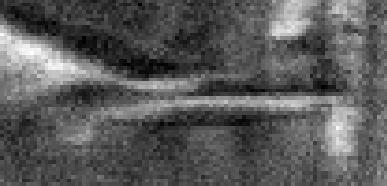}}&
  \scalebox{1}[1]{\includegraphics[width=0.16\textwidth,valign=m]{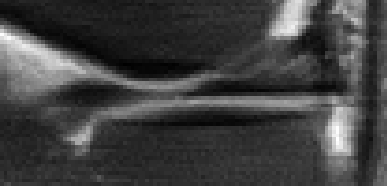}}&
  \scalebox{1}[1]{\includegraphics[width=0.16\textwidth,valign=m]{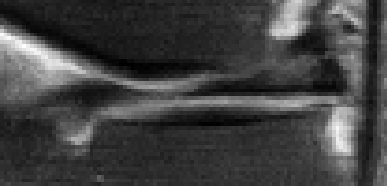}}&
  \scalebox{1}[1]{\includegraphics[width=0.16\textwidth,valign=m]{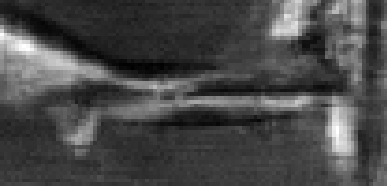}}&
  \scalebox{1}[1]{\includegraphics[width=0.16\textwidth,valign=m]{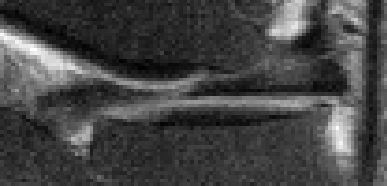}}\\
  \rule{0pt}{12ex}
  &
  \scalebox{1}[1]{\includegraphics[width=0.16\textwidth,valign=m]{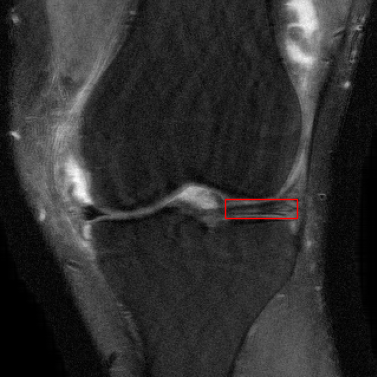}}&
  \scalebox{1}[1]{\includegraphics[width=0.16\textwidth,valign=m]{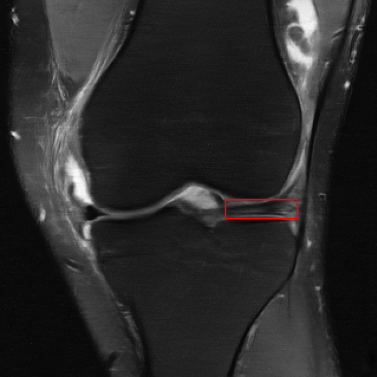}}&
  \scalebox{1}[1]{\includegraphics[width=0.16\textwidth,valign=m]{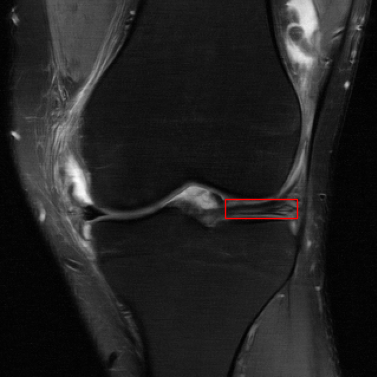}}&
  \scalebox{1}[1]{\includegraphics[width=0.16\textwidth,valign=m]{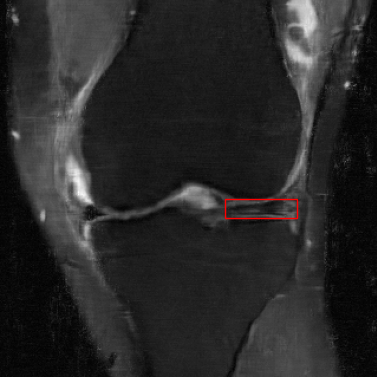}}&
  \scalebox{1}[1]{\includegraphics[width=0.16\textwidth,valign=m]{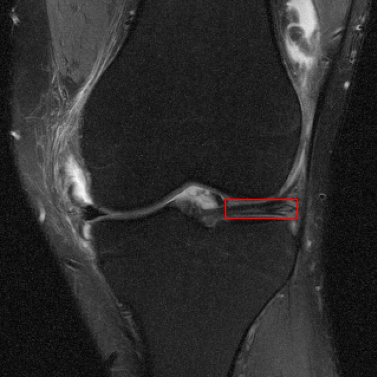}}\\
  \rule{0pt}{6ex}
  &
  \scalebox{1}[1]{\includegraphics[width=0.16\textwidth,valign=m]{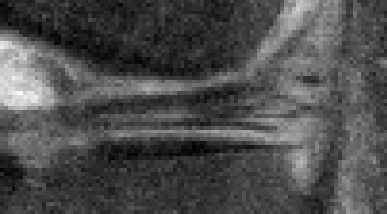}}&
  \scalebox{1}[1]{\includegraphics[width=0.16\textwidth,valign=m]{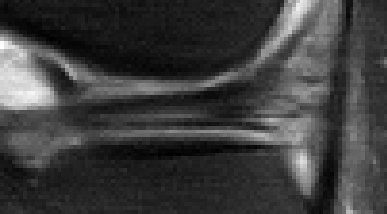}}&
  \scalebox{1}[1]{\includegraphics[width=0.16\textwidth,valign=m]{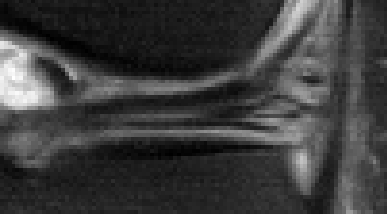}}&
  \scalebox{1}[1]{\includegraphics[width=0.16\textwidth,valign=m]{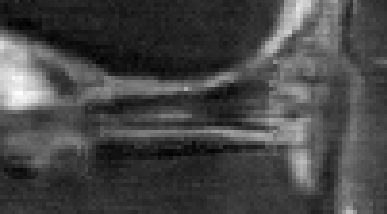}}&
  \scalebox{1}[1]{\includegraphics[width=0.16\textwidth,valign=m]{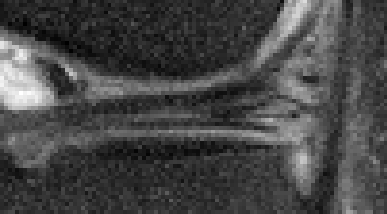}}\\
  \rule{0pt}{-8ex}
\end{tabular}
\captionsetup{skip=5pt}
\captionof{figure}{Reconstruction examples for real-world small feature recovery.
}
\label{fig:path-sample-recs}
\end{table*}


\section{Parameter setup}\label{sec:setup_app}

Throughout the paper, we have used four reconstruction methods: (i) the end-to-end variational network (VarNet), (ii) the U-net, (iii) the ConvDecoder, and (iv) $\ell_1$-norm minimization. The parameter setup we chose for VarNet and U-net are shown in Table~\ref{tab:vparams} and Table~\ref{tab:uparams}, respectively, and are as selected by the organizers of the fastMRI challenge\footnote{\href{https://github.com/facebookresearch/fastMRI/tree/master/models}{\color{black}https://github.com/facebookresearch/fastMRI/tree/master/models}} \citep{zbontar2018fastmri}.
The parameters of ConvDecoder (8 layers and 256 channels) and Deep Decoder (10 layers and 368 channels) are chosen according to the results discussed by \citet{darestani2020can}. 

For distribution shifts (Section~\ref{sec:data-shift}), in order to be able to study a larger variety of the methods, we studied differently sized networks. Specifically, we considered VarNet group (by altering the number of cascades and channels), U-net group (by changing the number of layers and channels), un-trained group (by considering Deep Decoder in addition to ConvDecoder), and $\ell_1$ group (by considering Wavelet, Fourier, and DCT bases). 

\begin{table*}[ht]
\centering
\begin{adjustbox}{max width=1\textwidth}
\begin{threeparttable}
\begin{tabular}{|c|c|c|c|c|c|c|}
\hline
Method & \#channels & convolutional  & \#pools & \#sens-pools &\#sens-channels & \#cascades\\
       &         (or width) & kernel size    &         &              &                &           \\
\hline
    VarNet-ref   & 18 & 3 & 4 & 4 & 8 & 12 \\
    VarNet2      & 9 & 3 & 4 & 4 & 4 & 12 \\
    VarNet3      & 9 & 3 & 4 & 4 & 4 & 6 \\
    VarNet4      & 4 & 3 & 4 & 4 & 2 & 12 \\
    VarNet5      & 4 & 3 & 4 & 4 & 2 & 6 \\
\hline
\end{tabular}
\end{threeparttable}
\end{adjustbox}
\caption{Model parameters for VarNet and its variants.
}
\label{tab:vparams}
\end{table*}

\begin{table*}[ht]
\centering
\begin{adjustbox}{max width=1\textwidth}
\begin{threeparttable}
\begin{tabular}{|c|c|c|c|c|}
\hline
Method & \#layers & \#channels & convolutional  & \#pools \\
       &          & (or width) & kernel size    &         \\
\hline
    U-net-ref    & 8  & 32  & 3 & 4 \\
    U-net2       & 8  & 16  & 3 & 4 \\
    U-net3       & 4  & 16  & 3 & 4 \\
    U-net4       & 8  & 8   & 3 & 4 \\
    U-net5       & 4  & 8   & 3 & 4 \\
\hline
\end{tabular}
\end{threeparttable}
\end{adjustbox}
\caption{Model parameters for U-net and its variants. \#channels denotes the width factor of each U-net (i.e., the number of channels for the first layer).
}
\label{tab:uparams}
\end{table*}



\end{document}